\documentclass[twocolumn,preprintnumbers,superscriptaddress,nofootinbib,aps,prd,floatfix]{revtex4-1}

\usepackage[english]{babel}
\usepackage{amsmath,amssymb}
\usepackage{hyperref}
\usepackage{graphicx}
\usepackage{amsfonts}
\usepackage[dvipsnames]{xcolor}
\usepackage{upgreek}
\usepackage[titletoc]{appendix}
\usepackage{setspace}
\usepackage{subfigure}
\usepackage{tcolorbox}

\newcommand{\newc}{\newcommand}
\newc{\fa}{f_{a}}
\newc{\Mp}{M_p}
\newc{\ga}{g_{a\gamma\gamma}}
\newc{\Tosc}{T_{\mathrm{osc}}}
\newc{\Tad}{T_{\mathrm{ad}}}
\newc{\rhotot}{\rho_{\mathrm{tot}}}
\newc{\TRH}{T_{\mathrm{RH}}}
\newc{\Tkin}{T_{\mathrm{kin}}}
\newc{\MeV}{\mathrm{MeV}}

\def\beq{\begin{eqnarray}}
\def\eeq{\end{eqnarray}}
\def\bea{\begin{eqnarray}}
\def\eea{\end{eqnarray}}

\begin{document}

\title{Dark Matter Targets for Axion-like Particle Searches}

\begin{abstract}
Many existing and proposed experiments targeting QCD axion dark matter (DM) can also search for a broad class of axion-like particles (ALPs). We analyze the experimental sensitivities to electromagnetically-coupled ALP DM in different cosmological scenarios with the relic abundance set by the misalignment mechanism. 
We obtain benchmark DM targets for the standard thermal cosmology, a pre-nucleosynthesis period of early matter domination, 
and a period of kination. These targets are theoretically simple and assume $\mathcal{O}(1)$ misalignment angles, avoiding 
 fine-tuning of the initial conditions.
We find that some experiments will have sensitivity to these ALP DM targets before they are sensitive to the QCD axion, and others can potentially reach interesting targets below the QCD band.
The ALP DM abundance also depends on the origin of the ALP mass. Temperature-dependent masses that are generated by strong dynamics (as for the QCD axion) correspond to DM candidates with smaller decay constants, resulting in even better detection prospects.
\end{abstract}

\author{Nikita Blinov}
\affiliation{Fermi National Accelerator Laboratory, Batavia, IL, USA\\[0.1cm]}

\author{Matthew J. Dolan}
\affiliation{ARC Centre of Excellence for Particle Physics at the
  Terascale, School of Physics, University of Melbourne, 3010, Australia\\[0.1cm]}

\author{Patrick Draper}
\affiliation{Department of Physics, University of Illinois, Urbana, IL 61801\\[0.1cm]}

\author{Jonathan Kozaczuk}
\affiliation{Department of Physics, University of Illinois, Urbana, IL 61801\\[0.1cm]}
\affiliation{Amherst Center for Fundamental Interactions, Department of Physics, University of Massachusetts, Amherst, MA 01003}

\preprint{FERMILAB-PUB-19-197-A-T}

\maketitle


\section{Introduction}

The particle nature of dark matter (DM) is unknown. One particularly well-motivated DM candidate is the QCD axion, which also provides a solution to the strong CP problem~\cite{Peccei:1977hh,Peccei:1977ur,Wilczek:1977pj,Weinberg:1977ma}. The QCD axion is a pseudoscalar boson with an approximate shift symmetry, and its mass and couplings are mostly controlled by a single parameter, the axion decay constant $f_a$. A relic cosmological abundance can be obtained from the misalignment mechanism~\cite{Abbott:1982af,Dine:1982ah,Preskill:1982cy}, and as a result, axion DM has phenomenological properties very different from thermally-produced weakly interacting massive particles.

More general light pseudoscalars are known as axion-like particles, or ALPs. ALPs arise as pseudo-Nambu Goldstone bosons (pNGBs) associated with the breaking of global $U(1)$ symmetries, or as zero modes of higher dimensional gauge fields that are generic in string theory~\cite{Arias:2012az,Svrcek:2006yi,Arvanitaki:2009fg,Cicoli:2012sz}. 
Unlike the QCD axion, ALPs do not have to interact via the strong force and 
therefore they are not associated with the strong CP problem.
As a result, they exhibit a wider range of couplings and masses and offer a compelling class of DM candidates. 
 For a review of ALP and axion model-building and cosmology see, e.g.,~\cite{Marsh:2015xka,Hook:2018dlk}.

Recent years have seen a resurgence of interest in searching for axions and ALPs, with a number of active experiments and new proposals under consideration (for reviews see e.g.~\cite{Graham:2015ouw,Irastorza:2018dyq}). These include resonant cavity experiments at various frequencies, such as ADMX~\cite{Asztalos:2009yp,Du:2018uak}, ORGAN~\cite{McAllister:2017lkb} and HAYSTAC~\cite{Brubaker:2016ktl,Zhong:2018rsr}, and also new ideas including dielectric haloscopes (MADMAX~\cite{TheMADMAXWorkingGroup:2016hpc,Brun:2019lyf} and photonic materials~\cite{Baryakhtar:2018doz}), resonant LC-circuits~\cite{Chaudhuri:2014dla,Silva-Feaver:2016qhh}, detection-induced magnetic flux oscillations (ABRACADABRA~\cite{Kahn:2016aff,Ouellet:2018beu}) and NMR-based techniques (ARIADNE~\cite{Arvanitaki:2014dfa,Geraci:2017bmq} and CASPEr~\cite{Graham:2013gfa,Budker:2013hfa,Garcon:2017ixh}).  Collectively these experiments cover many orders of magnitude of possible ALP mass, and are sensitive to ALP couplings to photons or nucleons depending on the experiment. In this work, we focus on cosmological relic populations of electromagnetically-interacting ALPs. The impact of resonant cavity searches on these ALPs has previously been considered in~\cite{Arias:2012az}.

Recently, the Physics Beyond Colliders Working Group has forecast the sensitivity of future experiments to axions and ALPs~\cite{Beacham:2019nyx}, building on the review~\cite{Irastorza:2018dyq}. For the QCD axion, a number of groups have developed models to expand the parameter space, classifying the possibilities for UV-complete theories~\cite{DiLuzio:2016sbl,DiLuzio:2017pfr,Agrawal:2017cmd}, model-building photophilic~\cite{Farina:2016tgd} and photophobic~\cite{Craig:2018kne} axions, and extending the standard misalignment mechanism~\cite{Agrawal:2017eqm,Co:2017mop}. 
These analyses highlight the breadth of viable QCD axion models extending beyond the canonical KSVZ and DFSZ scenarios, and motivate continued experimental exploration of the axion mass $m_a$ and the axion-photon coupling $\ga$ parameter space (the ``ALP plane").

Our aim is to map cosmological models onto the ALP plane, identifying regions where the correct relic abundance is obtained from simple assumptions about the expansion history, the ALP model, and the initial conditions. These regions of parameter space are therefore compelling targets for experiments searching for electromagnetically-coupled ALP DM.  Since ALPs do not necessarily couple to the strong interactions, and their relic density depends on the expansion rate at early times, these targets can differ significantly from the QCD axion with a standard radiation-dominated cosmological history.

In Section~\ref{sec:ALPDM} we consider an ALP with relic density  set by the misalignment mechanism. The final abundance 
   strongly depends on the expansion history of the universe before Big Bang Nucleosynthesis (BBN). We  study ALPs that begin to oscillate during 
   radiation domination (as in the standard cosmology), during an epoch of early matter domination (EMD), or during a kination phase. ALPs in these alternative cosmologies have also been considered recently in e.g.~\cite{Visinelli:2009kt, Ramberg:2019dgi,Visinelli:2018wza,Nelson:2018via,Draper:2018tmh}, which we build upon in our work. 
   In Section~\ref{sec:ALPmass} we study the impact of the origin of the ALP mass on the relic abundance.
   We determine the parameter space favored by ALP DM with a fixed mass during and after the onset of oscillations, a mass derived from higher-dimensional Planck-scale suppressed operators, and ALPs with a mass that changes with temperature.

At low masses, we find that experiments will be sensitive to ALPs with $\mathcal{O}(1)$ initial misalignment angles well before they are able to probe the QCD axion. 
At higher masses, the cosmological models motivate continuing ALP searches to couplings below the QCD region. In some cases, existing proposals will have the required sensitivity, while in other scenarios -- particularly EMD -- new search strategies may be required.  
We present the theoretical targets, existing constraints and experimental projections in Section~\ref{sec:results}, with the 
main results collected in Figs.~\ref{fig:masterplot_current},~\ref{fig:masterplot} and~\ref{fig:masterplot_Tdep}. Our findings are summarized in 
Section~\ref{sec:conc}. Appendices~\ref{appx:A} and~\ref{appx:B} contain details of the relic abundance calculations 
for different cosmologies and ALP mass temperature-dependence.

\section{ALP Dark Matter}
\label{sec:ALPDM}

We take the ALP Lagrangian to be
\begin{equation}
\mathcal{L}_{\rm{ALP}}= \frac{1}{2}\partial_\mu a \partial^\mu a - \frac{1}{2}m_a^2 a^2 - \frac{1}{4}\ga a F_{\mu\nu}\tilde{F}^{\mu\nu} \, , 
\end{equation}
where $\tilde{F}^{\mu\nu}$ is the dual electromagnetic field-strength tensor. The photon coupling $\ga=r\alpha/(2\pi f_a)$ is related to the ALP decay constant $f_a$. $r$ is a model-dependent constant generically expected to be ${\cal O}(1)$;  we set $r=1$ in this work. The free parameters are then the ALP mass $m_a$ and photon coupling $\ga$ (or equivalently $f_a $).

At early times, the ALP field is frozen. The relic abundance today depends on the distribution of initial values $a_0\equiv f_a \theta_0$ of the field before it begins to evolve. Here $\theta_0$ is the initial misalignment angle. 
  One possibility is that the angle $\theta_0 $ is uniform across all initially causally-disconnected regions that make up the observable universe today; this is the case if the ALP exists prior to inflation.
Typically we expect $\theta_0 \sim 1$, in which case saturating the observed dark matter density identifies favored regions of the ALP parameter.

An alternative initial condition is a stochastic distribution of $\theta_0$ across all causally-disconnected regions. This 
occurs in pNGB models where the global symmetry is broken after inflation.
This scenario can be modeled by considering an ALP with an effective average misalignment angle of $\langle \theta_0 \rangle \sim \pi/\sqrt{3}$~\cite{Kolb:1990vq}. 
In this case, topological defects and other large inhomogeneities formed during global symmetry-breaking also contribute to the present day dark matter density; 
however, the magnitude of these contributions is still a subject of debate.
According to different studies the inclusion of large fluctuations may increase~\cite{Hiramatsu:2012gg,Gorghetto:2018myk} or decrease~\cite{Klaer:2017ond} the relic density relative to that of the misalignment estimate. 
In what follows, we assume that the relic density is reasonably well-approximated by the misalignment calculation. 
Therefore, both the uniform and stochastic initial conditions can be studied if we vary $\theta_0$ over a sufficient range. 
 To avoid fine-tuning and to capture both possibilities, we take $\theta_0 \in [0.1,2]$ below.\footnote{Monodromy scenarios allow a much larger initial misalignment of the ALP field -- Refs.~\cite{Jaeckel:2016qjp,Berges:2019dgr} consider displacements of up to $10^3f_a$. This leads to larger possible values of $\ga$ for a given value of $m_a$.}
Accounting for topological defects should not significantly affect our conclusions, provided their contribution is at or below the same order of magnitude as the misalignment contribution.

The equation of motion for the ALP zero mode in the early universe is\footnote{The $m_a^2 a$ term should be replaced by $V^\prime(a)$ for field values larger than $\mathcal{O}(f_a)$. In what follows, we use the approximation above, noting that for larger initial misalignment angles, going beyond this approximation can have $\mathcal{O}(1)$ effects on the predicted relic density~\cite{Turner:1985si}.}  
\beq
\ddot{a} + 3H \dot{a} + m_a^2 a=0 \, ,
\eeq
where $H$ is the Hubble parameter
\beq
H^2 = \frac{\rhotot}{3\Mp^2},
\eeq
$\rhotot$ is the total energy density of the universe, and $\Mp$ is the reduced Planck mass.
At early times the ALP field is fixed. Oscillations begin when the Hubble parameter becomes comparable to $m_a$, 
\begin{align}\label{eq:qH}
m_a=qH,
\end{align} 
for some ${\cal O}(1)$ value of $q$. In Appendix~\ref{appx:A}, we give a detailed discussion of $q$ and list the values that give the best fits of the analytic formulae to the results of numerical integration. For temperature-independent ALP masses, we find $q\equiv q_0=1.6$ provides good precision across the various cosmological scenarios we consider.
At a given time, the ALP energy density is
\begin{equation}
  \rho_a = \frac{1}{2}\left(\dot{a}^2 +m_a^2a^2\right) 
\end{equation}
where again we keep only the quadratic part of the ALP potential as an approximation. The ALP number density at time $t$ can be defined as
\beq
n_a(t) = \rho_a(t) / m_a(t)
\eeq
where we have allowed for the possibility of a time-varying ALP mass.

\begin{figure}[!t]
  \includegraphics[width=0.49\textwidth]{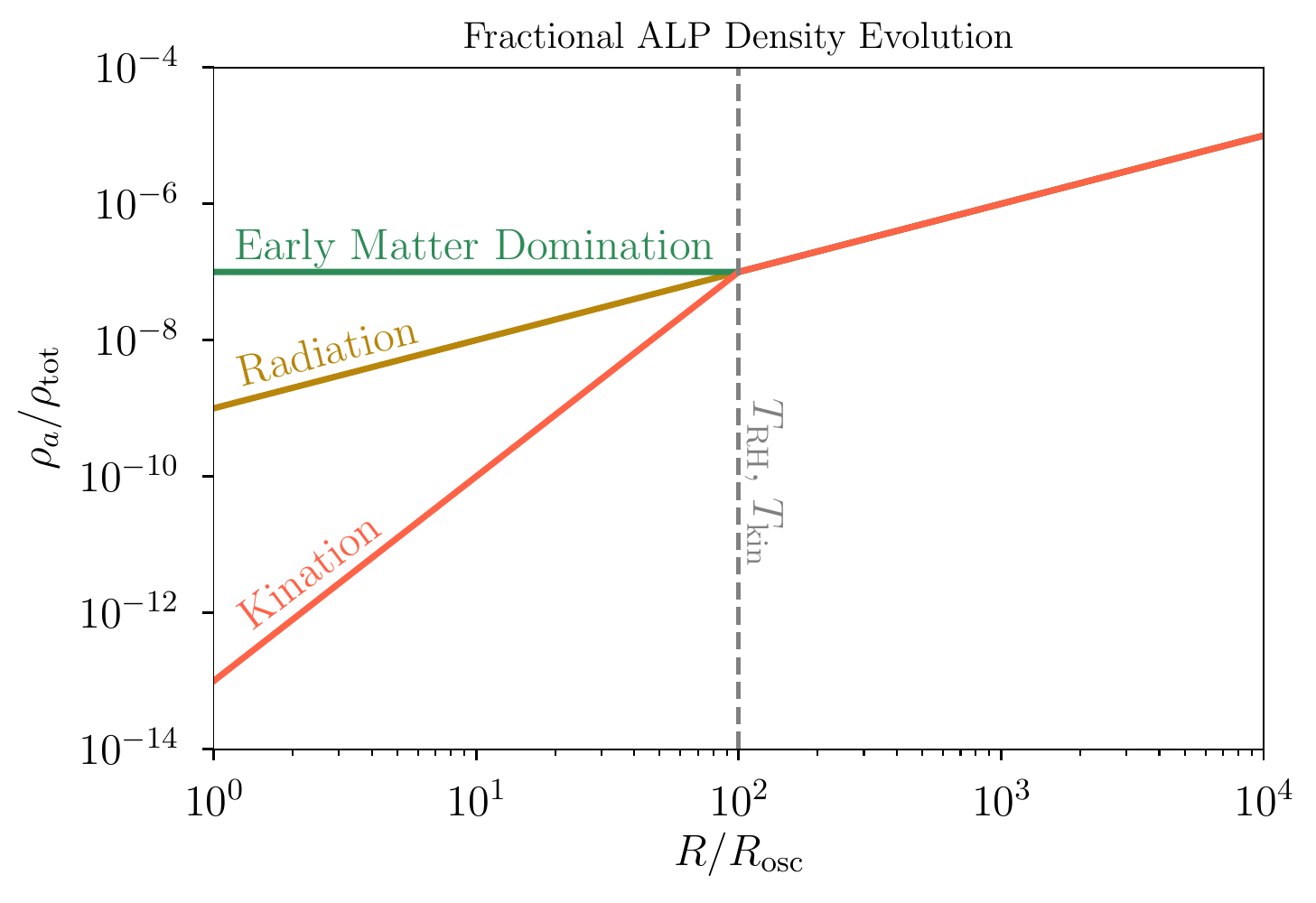}
  \caption{Schematic evolution of the ALP energy density relative to the 
      total energy as a function of the scale factor $R$ for different cosmologies.
      The scale factor is normalized to unity at the start of ALP oscillations.
      The lettuce, mustard and tomato lines correspond to 
      a universe with early matter (EMD), radiation, or kination domination before primordial nucleosynthesis, respectively.
      The transition from EMD or kination to radiation domination is denoted by the 
      vertical dashed line.
      The initial ALP density is fixed by requiring that ALP-radiation equality occurs 
      at the same value of $R/R_{\rm osc}$ for all three cases, such that these models have equal DM densities at late times.
      Since the initial value of the ALP energy density depends on $f_a \theta_0$, 
      cosmologies with an early period of early matter (kination) domination, require larger (smaller) values of $f_a \theta_0$ to saturate the observed dark matter relic density than in standard radiation domination, for fixed $m_a$.}
  \label{fig:cosmo}
  \end{figure}

Soon after oscillations begin, the ALP energy density redshifts as matter. Let us denote the corresponding temperature as $T_{\rm osc}$.  We also define a reference temperature $\Tad$ below which the evolution of the universe is adiabatic and the ratio of the ALP number density to entropy density $n/s$ is conserved. At the onset of oscillations, the ALP number density is
\beq
n_a(T_{\rm osc}) = \frac{1}{2} m_a^2 f_a^2 \theta_0^2.
\eeq
Its value at $\Tad$ is given by redshifting $n_a(T_{\rm osc})$ by $(R_{\rm osc}/R_{\rm ad})^3$ ($R$ is the FRW scale factor). The  present-day ALP density then depends on the cosmological evolution between $\Tosc$ and $\Tad$. The ALP relic density today can be written
\begin{equation}\label{eq:relic}
  \Omega_a = \frac{m_a n_a(\Tad)}{\rho_c} \left( \frac{T_0}{\Tad}\right)^3  \frac{g_{*S}(T_0)}{g_{*S}(\Tad)} \, .
  \end{equation}
Here $\rho_c \approx 10^{-5}\,h^2 \mathrm{GeV/cm}^3$, $h\approx 0.68$, and $T_0 \approx 2.7\,K$~\cite{Tanabashi:2018oca}. 
In these expressions we have assumed that the ALP mass is temperature independent. We will return to the temperature-dependent case in Sec.~\ref{sec:ALPmass}. 
Different cosmological scenarios correspond to different $\Tad$ and $n(\Tad)$. Below, we consider three well-motivated possibilities in which the ALP begins to oscillate during a ``standard'' period of radiation domination, during a period of early matter domination followed by reheating, or during kination. 
  The schematic evolution of the ALP energy density for these cosmologies is shown in Fig.~\ref{fig:cosmo}.

\subsection{Standard cosmology}
\label{sec:rd_cosmo}

In the conventional case, the ALP starts to oscillate during radiation domination (RD). The
total energy density is given by 
\beq
\rhotot = \frac{\pi^2}{30} g_*(T) T^4,
\eeq
where $g_*(T)$ is the effective number of relativistic degrees of freedom. 
Away from mass thresholds, $T\sim R^{-1}$ and so $\rhotot \propto R^{-4}$, where $R$ is the scale factor.
In this scenario, our approximate criterion for the onset of oscillations is
\begin{equation} \label{eq:Tosc_RD}
  m_a = q_0 H(\Tosc) = q_0\sqrt{\frac{\pi^2}{90}}\frac{g_*^{1/2}(\Tosc)\Tosc^2}{\Mp}
\end{equation}
where $q_0=1.6$, as discussed in Appendix~\ref{appx:A}.
Below $T_{\rm osc}$, the evolution is assumed to be adiabatic, so we can set $\Tad = T_{\rm osc}$. Using Eqs.~(\ref{eq:relic}) and~(\ref{eq:Tosc_RD}), we find an approximate expression for the ALP relic density today, assuming a temperature-independent mass during and after oscillations:
\begin{equation}
\begin{aligned} \label{eq:Omega_RD}
  \Omega_a h^2 & \simeq 0.12 \left(\frac{f_a \theta_0}{1.9\times 10^{13} \, {\rm GeV}} \right)^2
  \left( \frac{m_a}{1 \, \mu{\rm eV}} \right)^{1/2} \\ 
  & \times \left(\frac{90}{g_* (\Tosc)} \right)^{1/4}\,.
\end{aligned}
\end{equation}
Eq.~(\ref{eq:Omega_RD}) typically reproduces the results from numerical solutions of the ALP equation of motion (see Appendix~\ref{appx:B}) to within about 10-20\%.
The scaling with input parameters is straightforward to understand: at the onset of oscillations, the ALP constitutes a fraction $\rho_a/\rhotot\sim f_a^2/M_{\rm Pl}^2$ of the total energy density, which immediately starts growing since $\rho_a \propto R^{-3}$ redshifts more slowly than radiation. Correspondingly, larger $f_a$ leads to larger relic abundances. Similarly, increasing $m_a$ corresponds to earlier onset of oscillations and therefore a longer period over 
which $\rho_a/\rhotot$ grows, so the relic density also grows with $m_a$.

\subsection{Early Matter Domination}
\label{sec:emd_cosmo}
A period of early matter domination (EMD) modifies the conventional calculation of the axion relic density~\cite{Banks:1996ea} (for recent work, see Refs.~\cite{Ramberg:2019dgi,Visinelli:2018wza,Nelson:2018via,Draper:2018tmh}). EMD can be modeled by a heavy long-lived particle or an oscillating scalar field $\phi$ 
that dominates the energy density, such that $\rhotot \approx \rho_\phi \propto R^{-3}$. 
This scalar field can be a saxion or another scalar modulus with small couplings that lead to long lifetimes. 
The entropy injected by the decay of the scalar field dilutes the energy density of the ALP below the reheating scale, 
allowing for larger initial ALP energy densities and  reducing the tuning required in the misalignment angle for large $f_a$. 
This cosmology therefore favors a different region of the ALP parameter space compared to the RD case described above.

In the EMD scenario, the ALP is again initially displaced from the origin  and begins to oscillate when $m_a \sim q_0H=1.6H$ if the mass is independent of temperature. 
Assuming a reheating temperature  around 10 MeV (near the lower limit allowed by BBN~\cite{Kawasaki:2000en,Hannestad:2004px}), 
oscillation occurs during EMD for $m_a\gtrsim 10^{-13}$ eV, and during RD for smaller masses. 
The initial energy fraction in the ALP at oscillation is 
again of order $f_a^2 \theta_0^2/(M_{\rm pl}^2)$. The key difference in the EMD scenario is that $H^2\propto R^{-3}$ after the onset of ALP oscillations and prior to reheating, and so the ALP energy fraction $ \rho_a /(3M_{\rm Pl}^2 H^2)$ remains constant during this epoch. 
Accordingly, the ALP comes to dominate the energy density later than in the radiation-dominated case,  allowing for larger $f_a$ consistent with the observed dark matter relic density -- see Fig.~\ref{fig:cosmo}. 

Assuming adiabatic expansion below $T_{\rm RH}$, 
the present-day ALP density is given by Eq.~(\ref{eq:relic}) with $T_{\rm ad} = T_{\rm RH}$.
Since $\rho_a(t_{\rm RH})\simeq \rho_a(t_{\rm osc}) H^2(t_{\rm RH})/H^2(t_{\rm osc})$ and $\rho_a(t_{\rm osc}) \sim \theta_0^2 f_a^2 H^2(t_{\rm osc})$, $\rho_a(T_{\rm RH})$ entering Eq.~(\ref{eq:relic}) is independent of $H(t_{\rm osc})$ and $m_a$. The present-day ALP density is found to be
\beq \label{eq:Omega_EMD}
\Omega_a h^2 \, \simeq \, 0.12 \times \left(\frac{f_a \theta_0 }{9\times 10^{14} \, {\rm GeV}}\right)^2  \times \left(\frac{T_{\rm RH}}{10\, {\rm MeV}} \right)
\eeq
for temperature-independent ALP mass (see Appendix~\ref{appx:A} for more details).  In this case $\Omega_a$ is determined by $f_a$ and $T_{\rm RH}$, with larger $f_a$ and $T_{\rm RH}$ corresponding to larger $\Omega_a$. Note that this expression is only valid if $T_{\rm RH}$ is larger than $T_{\rm osc}$; otherwise, the standard radiation-dominated scenario is obtained. An expression for $T_{\rm osc}$ is given by Eq.~(\ref{eq:ToscEMD}) below, where a temperature-independent ALP mass corresponds to $b\rightarrow 0$.

We have also solved for the corresponding relic abundance numerically, by considering a three fluid model describing the modulus, the radiation energy density, and the ALP as described in Appendix.~\ref{appx:B}. These numerical solutions agree with Eq.~(\ref{eq:Omega_EMD}) to within about 20-25\% across the parameter space of interest and for the values of $T_{\rm RH}$ we have checked.

\subsection{Kination}
\label{sec:kin_cosmo}

The final cosmological scenario we consider is known as kination~\cite{Joyce:1996cp,Ferreira:1997hj}. ALP physics with an early period of kination has previously been studied in~\cite{Visinelli:2009kt,Visinelli:2018wza}. As for the EMD case, the energy density at early times is dominated by a long-lived scalar field $\phi$, but rolling in a steep potential such that its kinetic energy dominates $\rhotot$. For a polynomial potential $V(\phi) \propto \phi^N$, the energy density after some early time $t_0$ evolves as 
\beq \label{eq:kination}
\rho_\phi(t) \simeq \rho_{\phi} (t_0) \times \left(\frac{R(t_0)}{R(t)} \right)^{\frac{6N}{N+2}}.
\eeq
In the limit $N\rightarrow \infty$, $\rho_\phi$ dilutes as $ \sim R^{-6}$. We will consider this large $N$ limit in what follows.  
 Assuming that radiation comes to dominate when it reaches a temperature $T_{\rm kin}$, using Eq.~(\ref{eq:relic}) with $T_{\rm ad} = T_{\rm kin}$ we find that
\beq
\begin{aligned} \label{eq:Omega_kin}
  \Omega_a h^2 \simeq 0.12 &\times \left(\frac{f_a \theta_0}{2.4\times 10^{11} \, {\rm GeV}}\right)^2\times \left( \frac{m_a}{1 \, \mu{\rm eV}} \right)\\
& \times \left(\frac{11}{g_*(T_{\rm kin})}\right)^{\frac{1}{2}} \times \left(\frac{10\,{\rm MeV}}{T_{\rm kin}} \right).
\end{aligned}
\eeq
The relic density depends linearly on $m_a$ and inversely on $T_{\rm kin}$, which must be larger than about 5 MeV. The expression above only applies if $T_{\rm osc}>T_{\rm kin}$; otherwise, one reproduces the standard RD scenario given by Eq.~(\ref{eq:Omega_RD}). An expression for $T_{\rm osc}$ in the kination case is given in Eq.~(\ref{eq:Tosckin}), where $b\rightarrow 0$ corresponds to a temperature-independent ALP mass. Note that the fractional density $\rho_a/\rhotot \propto R^3$ grows rapidly during kination, allowing the ALP to 
saturate the DM relic abundance for smaller values of $f_a$ compared to RD and EMD scenarios considered above. Eq.~(\ref{eq:Omega_kin}) reproduces the numerically-obtained relic density (c.f.~Appendix~\ref{appx:B}) to within $\mathcal{O}(10)\%$.

As an illustration of the key differences between the three scenarios discussed so far, we sketch the evolution of the various relevant energy densities in Fig.~\ref{fig:cosmo} for the three cosmologies. Here  various parameters are fixed for illustrative purposes. 
The qualitative picture is clear: 
the faster the dilution of the dominant energy component in the pre-BBN era, the larger the final ALP abundance for fixed $f_a\theta_0$.
In the kination cosmology, for example, the ALP energy fraction rises more rapidly than in radiation domination. 
In contrast, in the EMD case this energy fraction remains constant until reheating.
Since -- in order to saturate the observed DM relic density -- ALP-radiation equality must occur around $T\simeq 1$ eV, 
the EMD and kination cases require a larger and smaller initial energy fraction, respectively, than in radiation domination, corresponding to larger and smaller preferred values of $f_a \theta_0$ for a given mass. From the experimental standpoint, this means that the kination scenario will provide a compelling and more easy-to-reach target than in the standard ALP cosmology, while a period of early matter domination will make the ALP more difficult to access with terrestrial experiments. However, all relic density-preferred bands can lie above the QCD band (i.e. at stronger coupling) for sufficiently small ALP masses. We will detail this picture further in Sec.~\ref{sec:results}.

\section{Origin of the ALP mass}
\label{sec:ALPmass}

In the previous section, we assumed that the ALP mass is independent of temperature at the onset of oscillations. This is the simplest class of models, and in general it seems reasonable to remain agnostic about the origin of the ALP mass. However, motivated by the QCD axion, we consider two further variations.

Famously, the QCD axion appears to conflict with the straightforward application of effective field theory principles and the expectation that quantum gravity violates 
global symmetries~\cite{Barr:1992qq,Holman:1992us,Holman:1992va,Kamionkowski:1992ax,Kamionkowski:1992mf}. Adding Planck-suppressed PQ-violating higher-dimension operators to the action, one finds that the axion solution to the strong CP problem is inoperative unless the Wilson coefficients are strongly suppressed up to operator dimension $d\sim12$.  Solutions to this problem are known; it might be the case that all quantum gravity-induced PQ-violation is exponentially small~\cite{Svrcek:2006yi,Arvanitaki:2009fg}.  In the ALP case, it is also of interest to compare the masses and couplings for which a viable dark matter candidate is obtained with  the typical mass generated by Planck-suppressed operators.

Secondly, the QCD axion relic abundance is non-trivially affected by the strong temperature dependence of the topological susceptibility of QCD. 
Similarly, it is imaginable that the ALP mass is controlled by infrared physics (e.g., a new strongly coupled gauge theory) that introduces temperature dependence. As in QCD this dependence can have important implications for the preferred regions of the mass-coupling parameter space.

\subsection{ALP mass from UV physics}

We consider the typical contribution to the ALP mass from a dimension-$d$ operator,
\begin{equation}
  \mathcal{L}\supset \frac{c\Phi^d}{\Mp^{d-4}} + \rm{h.c.} \,,
  \label{eq:mpbreaking}
  \end{equation}
parametrizing $\Phi=f_a \exp(ia/f_a)$.
For simplicity, we suppose that the Wilson coefficient $c$ is real and that the full potential is minimized at $a=0$. 
The contribution to the ALP mass from Eq.~(\ref{eq:mpbreaking}) is
\begin{equation}
m_a^2 = 2cd^2 f_a^2 \left( \frac{f_a}{\Mp }\right)^{d-4} \, .
\end{equation}
We relate the scale $f_a$ to the ALP-photon coupling $\ga$ by assuming~\cite{Agrawal:2017cmd}
\begin{equation}
  \ga = r\frac{\alpha}{2\pi f_a} \, ,
  \label{eq:agg}
  \end{equation}
where $r$ is an anomaly coefficient that we expect to be ${\cal O}(1)$.\footnote{Ref.~\cite{Agrawal:2017cmd}  constructs models with $r \gg 1$, leading to a large enhancement of $\ga$.}
Combining Eqs.~(\ref{eq:mpbreaking}) and~(\ref{eq:agg})  we obtain
\begin{equation}
\ga = \frac{r\alpha}{2\pi\Mp} \left( \frac{m_a^2}{2cd^2\Mp^2} \right)^{-1/(d-2)} \, .
  \end{equation}

In the results presented in Sec.~\ref{sec:results}, we set $c=r=1$. The resulting mass-coupling relation for $d=8,10,12$ is shown along with the preferred DM regions and the experimental limits and projections in Figs.~\ref{fig:masterplot_current} and~\ref{fig:masterplot}.
To summarize, we will find that Planck-suppressed operators below dimension 8 must be absent across all of the parameter space we consider. 
In the high-$f_a$ region, even more suppression is required. 
For example, almost all of the viable ALP parameter space in the standard RD scenario requires that the Planck-suppressed contributions to the ALP potential start at dimension 12. The viability of this possibility depends on the specific UV model. We will discuss the implications of these results further below.

\subsection{\texorpdfstring{$T$}{T}-dependent ALP masses: general considerations}
\label{sec:general}

We now turn to the complementary case where the ALP mass is set by $T$-dependent  infrared (IR) physics. 
First, we outline generic properties, constraints, and requirements on these scenarios. We then define a simple family of $T$-dependent masses and compute the relic density in the different cosmological scenarios, providing simple analytic expressions that reproduce the results of a more complete numerical treatment to within a few tens of percent.

First, we note that the temperature controlling the ALP mass does not need to equal the temperature of the SM bath.
This is generically the case if the ALP mass is generated by couplings to a hidden sector (HS) that is not in kinetic equilibrium with the SM. 
For a given SM temperature $T$ we parametrize the temperature of the hidden sector, $T_{\rm HS}$, as
\beq
T_{\rm HS} \equiv \xi(T) \, T
\eeq 
In what follows, all temperatures will correspond to temperatures of the SM photon bath, unless otherwise stated, and factors of $\xi$ will be used to convert to hidden sector temperatures.

We assume that $m_a$ is primarily sensitive to the temperature above a scale $\Lambda$, corresponding to a Standard Model (SM) bath temperature $T_{\Lambda}$. 
The ALP zero mode is initially frozen at $f_a \theta_0$ and starts to oscillate when $T = \Tosc$. 
In order for $T$-dependence to have an effect on $\Omega_a$, 
we  require $\xi_{\rm osc} \, T_{\rm osc} > \Lambda$ where $\xi_{\rm osc} \equiv \xi(T_{\rm osc})$.

The scale  $\Lambda$ cannot be arbitrarily low. 
In order for $m_a$ to vary significantly with temperature, 
there must exist a population of relativistic degrees of freedom in the HS. 
The presence of additional relativistic degrees of freedom modifies the expansion rate of the Universe, and
which can alter the predictions of light element abundances and the CMB power spectrum.
These constraints can be avoided if $T_\Lambda \gtrsim T_{\rm BBN}$, where $T_{\rm BBN}\sim 5$ MeV is the temperature of the SM bath around the onset of BBN. 
Otherwise, we must ensure that the effects from radiation in the HS at temperatures above $T_\Lambda$ are consistent  
with the measurements of the primordial abundances and CMB. 
Modifications of the expansion rate are typically parametrized by the effective number of neutrino species, $N_{\rm eff}$. 
For the parameter space of interest, $T_\Lambda$ is always above the temperature of recombination, so the BBN limit is most relevant. 
These constraints, detailed in e.g.~Ref.~\cite{Aghanim:2018eyx}, can be satisfied at $\sim 2\sigma$ confidence level 
provided $\Delta N_{\rm eff} \lesssim 0.5$. 
In terms of the effective number of relativistic degrees of freedom $g_{*{\rm HS}}(T_{\rm BBN})$ in the HS at $T_{\rm BBN}$, we have
\beq
\Delta N_{\rm eff} = \frac{4}{7}\left(\frac{11}{4}\right)^{4/3} g_{*{\rm HS}}(T_{\rm BBN}) \, \xi_{\rm BBN}^4
\eeq
where we use the shorthand $\xi_{\rm BBN} \equiv \xi(T_{\rm BBN})$. From this we see that BBN $N_{\rm eff}$ constraints can be avoided if $\xi_{\rm BBN} \lesssim 0.1$ for $g_{*{\rm HS}}(T_{\rm BBN}) \lesssim 10^3$ . 

While the considerations above are quite general, 
there may be additional model-dependent constraints in concrete realizations.  
For example, one must also ensure that the relic abundance of any heavy states in the HS makes up a small component of the matter density today. 
These can decay or annihilate into HS radiation; however, one must then verify that $\xi$ remains small.
Heavy hidden sector states can also decay or annihilate to the SM, but this may require connector particles between the HS and SM that may again increase $\xi(T_{\rm BBN})$. 
Furthermore, their decays to the SM must not significantly disrupt BBN or the CMB. In an effort to be as model-agnostic as possible we will not consider these issues further, although we emphasize that they will likely be important in concrete ALP scenarios with $T$-dependent masses. For related discussions in specific strongly coupled hidden sector models, see, e.g., Refs.~\cite{Feng:2011ik, Cline:2013zca, Boddy:2014yra, Hochberg:2014dra, Forestell:2017wov, Berlin:2018tvf, Draper:2018tmh}. 

Summarizing, in order for $T$-dependence to affect the ALP relic density and be in agreement with $N_{\rm eff}$ constraints, we require either
\beq\label{eq:req1}
\xi_{\rm osc}T_{\rm osc} > \Lambda, \quad T_{\Lambda} > T_{\rm BBN} 
\eeq
or
\beq\label{eq:req2}
\xi_{\rm osc} T_{\rm osc} > \Lambda, \quad \xi_{\rm BBN} \lesssim 0.1.
\eeq
If  $\xi_{\rm osc} T_{\rm osc} < \Lambda$, the onset of ALP oscillations proceeds as in the $T$-independent case discussed earlier. 
The relationship between $T_{\rm osc}$, $m_a$,  and $f_a$ depends on the particular cosmological scenario, as we discuss below. 
EMD and kination cosmologies will have additional requirements in order for $T$-dependence to be relevant.

Nontrivial temperature dependence enhances $\Omega_a$ relative to the $T$-independent prediction for a given $m_a$, $f_a$, and $\theta_0$. The resulting abundance can be computed numerically (as discussed in Appendix~\ref{appx:B}), but simple analytic estimates can again be used to reproduce the full results to within $\mathcal{O}(10\%)$ in most cases. 
 The size of the enhancement for the different cosmological scenarios can be estimated as follows (see also Appendix~\ref{appx:A} for more details). Let us define the enhancement factor
\beq
\gamma \equiv \frac{\Omega_a}{\Omega_{a}^{T-{\rm ind}}}
\eeq
where $\Omega_a$ is the ALP relic density assuming a $T$-dependent ALP mass and $\Omega_{a}^{T-{\rm ind}}$ is the corresponding $T$-independent result 
as computed in Sec.~\ref{sec:ALPDM}.
Both quantities are evaluated for the same set of $m_a(T=0)$, $f_a\theta_0$. We model the different cosmological scenarios by assuming that the Hubble parameter for temperatures above some scale $T_*$ evolves as
\beq
H^2 \propto R^{-3(w+1)}
\eeq
where $w$ is the equation of state parameter, $\rho = w p$. Early matter domination, radiation domination, and kination correspond to $w=0$, 1/3, and 1, respectively. Below $T_*$, the evolution is assumed to be adiabatic and follows that of a standard radiation-dominated cosmology.  We approximate the transition to radiation domination at $T_*$ (if it occurs) as instantaneous. As in Eq.~(\ref{eq:qH}), the ALP begins to oscillate when 
\beq
m_{\rm osc} = q_T H_{\rm osc}.
\eeq
Here the subscript $T$ indicates that the value of $q$ for the $T$-dependent case can differ from $q_0 = 1.6$. Given these assumptions and provided the ALP begins oscillating while its mass is changing with temperature, a straightforward calculation discussed further in Appendix~\ref{appx:A} shows that the enhancement factor is given approximately by
\beq \label{eq:enhancement}
\gamma_T \simeq \left(\frac{q_T}{q_{0}}\right)^{\frac{2}{w+1}} \left(\frac{m_a}{m_{\rm osc}}\right)^{\frac{2}{w+1}-1}.
\eeq
The subscript in $\gamma_T$ is indicates that this expression applies if the mass at the onset of oscillations, $m_{\rm osc}$, differs from  $m_a$, the low-temperature ALP mass. If $m_{\rm osc}\ll m_a$, the relic density can be significantly enhanced in the RD and EMD cosmologies. 
The scaling with $m_a/m_{\rm osc}$ is a product of two counter-acting effects: the delay in the start of oscillations and the 
growth of ALP mass with time. This is made explicit in Eq.~\ref{eq:gamma_app} below.
In the kination case, $2/(w+1)-1=0$ and these effects nearly cancel, so the relic density is only enhanced if $q_T\neq q_0$. Given our assumptions about the origins of $m_a(T)$, discussed below, this enhancement is milder than in RD and EMD, and is at most an $\mathcal{O}(1)$ effect.

To proceed further, we focus on a class of models with ALP mass $T$-dependence similar to that of the QCD axion.
We will assume that, for $T_{\rm HS}=\xi(T) T>\Lambda$, the ALP mass is given by
\beq \label{eq:TdepM}
m_a(T)=m_{a} \left(\frac{\Lambda}{\xi(T)T}\right)^b
\eeq
where $m_{a}$ is the zero-temperature mass, taken to be of the form
\beq 
m_{a}=\frac{\Lambda^2}{f_a}.
\label{eq:zero_temp_mass}
\eeq
For $\xi(T) T<\Lambda$, $m_a(T)=m_{a}$. In Eq.~(\ref{eq:TdepM}), $b$ is a positive exponent. In QCD-like theories, $b$ is related to the $\beta$-function of the gauge group and can be obtained analytically from the dilute instanton gas approximation (DIGA). DIGA predicts $b=\left( 11 N_c -2N_f \right)/6 + N_{f}/2 -2 $, where $N_c$ and $N_f$ are the number of colors and light flavors, respectively~\cite{Gross:1980br}. For QCD, $b=4$, and the semiclassical approximation is in reasonable agreement with lattice results at high temperatures~\cite{borsanyinature,borsanyia,Dine:2017swf}. In these simulations the scaling predicted by DIGA appears to hold down to $\Lambda_{QCD}$, where $m_a$ saturates to near its zero-temperature value and remains approximately constant at lower temperatures~\cite{borsanyinature,borsanyia}. In this sense our model of the temperature dependence mimics QCD and generalizes it to arbitrary $\Lambda$, $b$, and $\xi(T)$. In our plots we will take $b=4$ as an illustrative example, corresponding to the QCD-like case. As such, we assume
\beq \label{eq:gstar_for_plots}
g_{* {\rm HS}}(T) = \frac{52}{2}\left( 1+\tanh\left[10\left(1-\frac{\Lambda}{ \xi(T)T} \right)\right]\right)
\eeq
where the factor of 52 corresponds to the number of relativistic degrees of freedom for $SU(3)$ with three light flavors and the temperature $T$ is understood to be that of the SM radiation bath; the $\tanh$ function smoothly decouples these degrees of freedom at the transition temperature.
The total number of relativistic degrees of freedom at a temperature $T$ is then $g_*(T) = g_{* {\rm SM}}(T) + \xi(T)^4 g_{* {\rm HS}}(T)$. Note that for the $T$-independent predictions we take $g_{* {\rm HS}}=0$, as the mass $m_a$ can be set by physics in the ultraviolet and does not necessarily require new light degrees of freedom present near the onset of oscillations.

With the form of $T$-dependence specified, one can show (c.f.~Appendix~\ref{appx:A}) that there is a maximum allowed enhancement factor, $\gamma_T \leq \gamma_{\rm max}$. Defining $T_\Lambda$ such that $T_{\Lambda} = \Lambda/\xi(T_{\Lambda})$ (the SM temperature at which the ALP mass saturates to its low-temperature value), if the ALP has not started oscillating by $T_\Lambda$ and $m_a > q_0 H(T_\Lambda)$, oscillations will begin suddenly at $T_{\Lambda}$ and so $q_{\rm max} = m_a/H(T_\Lambda)$. Thus,
\beq
\label{eq:max_enhancement}
\gamma_{\rm max} =  \left(\frac{m_a}{H(T_{\Lambda}) q_{0}}\right)^{\frac{2}{w+1}}.
\eeq 
In this case the oscillation temperature is simply $T_{\rm osc} = T_{\Lambda}=\Lambda/\xi_{\rm osc}$. Note that the opposite limit in which the ALP is still frozen at $T_\Lambda$ and $m_a < q_0 H(T_\Lambda)$ corresponds to $\xi_{\rm osc} T_{\rm osc} < \Lambda$, which reproduces the $T$-independent case. One can show that this occurs when $\operatorname{min}(\gamma_T,\gamma_{\rm max})<1$.

Summarizing these considerations, the enhancement factor can be written compactly as 
\beq \label{eq:gamma_master}
\gamma = \operatorname{max}\left\{ 1, \, \operatorname{min}\left\{\gamma_T, \gamma_{\rm max} \right\} \right\},
\eeq
where $\gamma_T$ and $\gamma_{\rm max}$ are defined in Eqs.~(\ref{eq:enhancement}) and~(\ref{eq:max_enhancement}), respectively. 
Explicit expressions for these quantities in concrete cosmological scenarios are given below.
Denoting the predicted oscillation temperature for a given exponent $b$ in Eq.~(\ref{eq:TdepM}) as $T_{\rm osc}(b)$, the true oscillation temperature is given by
\beq\label{eq:Tosc}
T_{\rm osc} = \left\{\begin{array}{c c} 
\vspace{.1cm} T_{\rm osc} (b=0), & \gamma=1\\

\vspace{.1cm} T_{\rm osc} (b), & \gamma = \gamma_T \\ 

\vspace{.1cm}  \frac{\Lambda}{\xi_{\rm osc}}, &  \gamma=\gamma_{\rm max}. 
\end{array}
\right. 
\eeq
Further details can be found in Appendix~\ref{appx:A}. Finally, in the EMD and kination cosmologies, if $T_{\rm osc}$ predicted by Eq.~(\ref{eq:Tosc}) is smaller than $T_{\rm RH}$ or $T_{\rm kin}$, the results for RD should be used.

\subsection{Radiation domination with \texorpdfstring{$T$}{T}-dependence}

Let us first apply these results to determine the effects of $T$-dependence on the standard calculation of the ALP relic abundance, where the ALP is assumed to oscillate during radiation domination. A related discussion can be found in Ref.~\cite{Arias:2012az}, which we generalize to allow for a HS at a different temperature than the SM. In general, $\xi(T)$ changes across mass thresholds as particles in both the HS and SM annihilate. This heating typically changes $\xi$ by at most $\mathcal{O}(1)$ factors unless the change in number of degrees of freedom is very large. Since the precise form of $\xi(T)$ is model-dependent, in the remainder of this study we assume for simplicity that $\xi(T) \approx \xi_{\rm osc}$ for temperatures of interest and treat $\xi_{\rm osc}$ as a free parameter. In concrete models $\xi_{\rm osc}$ can be set by, e.g., the branching ratio of the inflaton into the HS relative to the SM~\cite{Adshead:2016xxj}, or be equal to one for a HS in kinetic equilibrium with the SM.

\begin{figure*}[!t]
  \centering
    \includegraphics[width=0.49\textwidth]{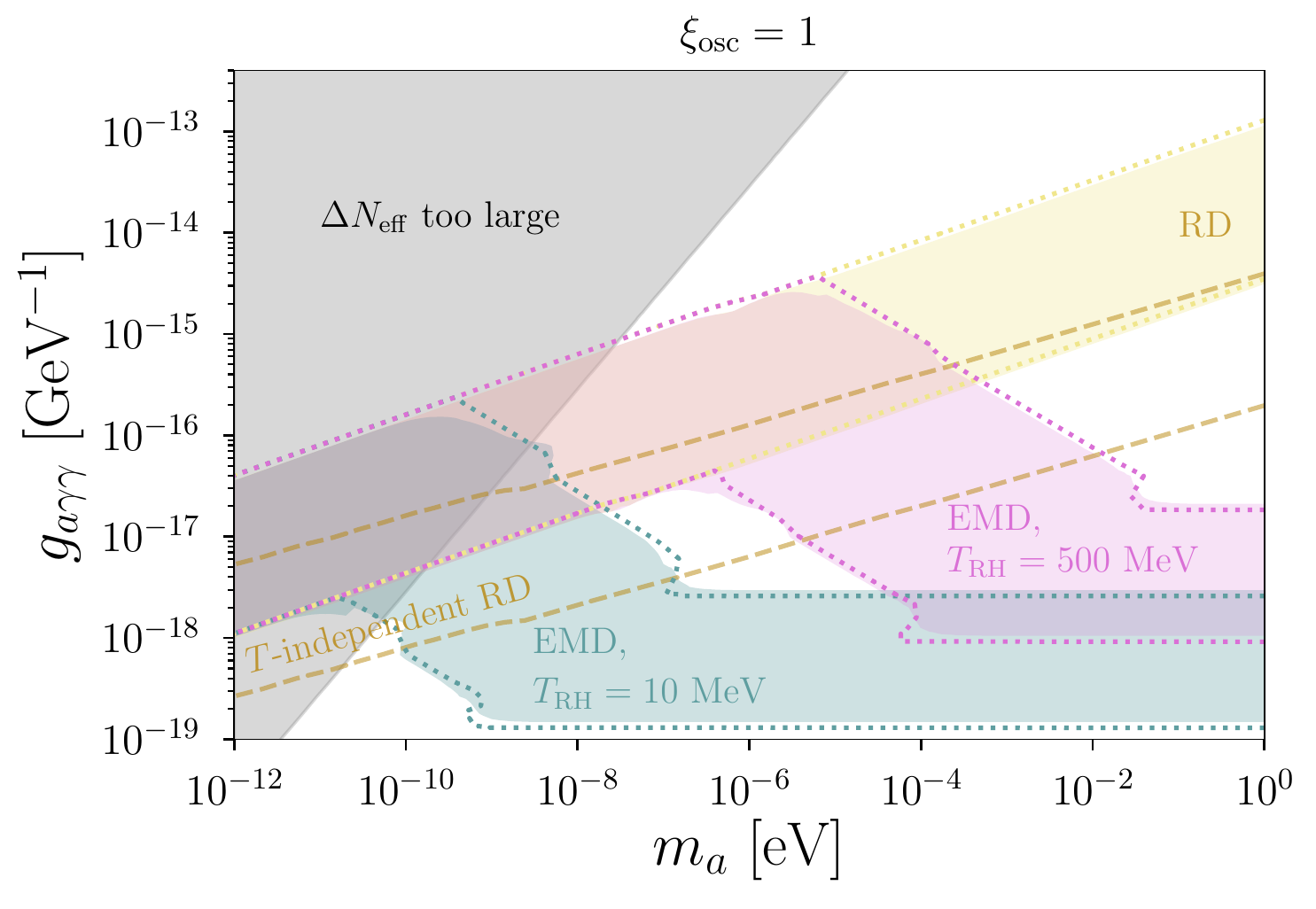} \,  \includegraphics[width=0.49\textwidth]{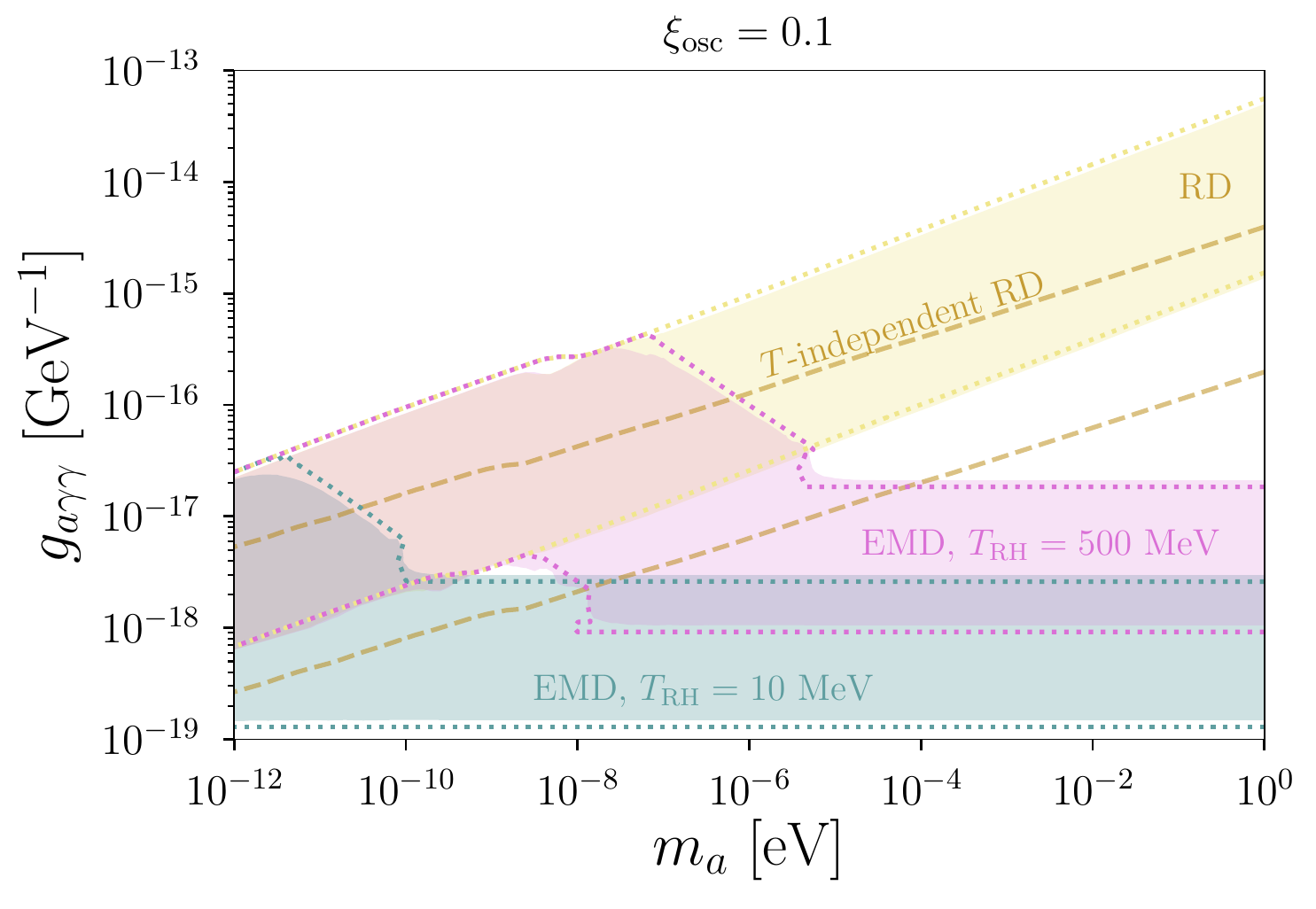}
    \caption{Preferred regions in the ALP parameter space allowing for a temperature-dependent ALP mass given by Eq.~(\ref{eq:TdepM}) with $b =4$.
    The left (right) panel corresponds to a hidden sector with temperature ratio $\xi_{\rm osc} =1$ ($\xi_{\rm osc} =0.1$) relative to the SM. The pastel shaded regions feature an ALP that saturates the observed dark matter relic density with $\theta_0 \in [0.1,2]$ for radiation domination (gold) and early matter domination with $T_{\rm RH}=10$ MeV (green) and 500 MeV (purple), obtained via numerical solution of the evolution equations. The dotted contours show the analytic predictions given in the text, which we find are a good match to the full numerical solutions. 
   For reference, we also indicate the preferred region for the RD scenario with a $T$-independent ALP mass between the yellow dotted contours.
In both the RD and EMD cosmologies, the relic density can be substantially increased for a fixed $m_a$ and $f_a$ if the ALP mass is temperature-dependent during the onset of oscillations. 
Note that $T$-dependence in the EMD cosmology interpolates between the $T$-independent EMD and $T$-dependent RD scenarios. 
In the left panel, the gray shaded region is excluded for the $T$-dependent case by the value of $N_{\rm eff}$ during BBN. 
These constraints are avoided in the right panel due to the lower hidden sector temperature, at the price of a smaller enhancement of the relic abundance. Note that the scales of the vertical axes are  different in the left and right panels.}
  \label{fig:Tdep}
  \end{figure*}

In Appendix~\ref{appx:A}, we find that the predicted oscillation temperature $T_{\rm osc}(b)$ and enhancement in this case can be estimated by
\beq \label{eq:ToscRD}
T_{\rm osc}(b) \simeq  \frac{\Lambda}{\xi_{\rm osc}}
\left(\frac{1125 \,M_{\rm Pl}^2 \,\xi_{\rm osc}^{4}}{8\, (2+b)^2 \pi^2  \, g_*(T_{\rm osc})\, f_a^2 } \right)^{\frac{1}{4+2b}}
\eeq
\beq
\gamma_T \simeq \left(\frac{2+b}{2}\right)^{\frac{3+b}{2+b}} \left(\frac{5 \sqrt{90} M_{\rm Pl} \xi_{\rm osc}^2}{8\pi \sqrt{g_*(T_{\rm osc}) } f_a}\right)^{\frac{b}{4+2b}}.
\label{eq:gammaRD} 
\eeq
 Meanwhile, the maximum enhancement factor $\gamma_{\rm max}$ is approximately
\beq\label{eq:gmax_RD}
\gamma_{\rm max} \simeq \left(\frac{5\sqrt{90} M_{\rm Pl}\xi^2_{\rm osc}}{8 \pi \sqrt{g_*(T_\Lambda)} f_a}\right)^{\frac{3}{2}}.
\eeq
With these expressions, Eqs.~(\ref{eq:Omega_RD}), (\ref{eq:gamma_master}) and (\ref{eq:Tosc}) can then be used to estimate $\Omega_a$ across the ALP plane accounting for $T$-dependence in the ALP mass.  In the parameter space we consider, Eq.~(\ref{eq:gamma_master}) typically yields $\gamma = \gamma_T$ for RD. 
 
The preferred ALP dark matter region in the $T$-dependent case for $\xi_{\rm osc}=1$ is shown on the left in Fig.~\ref{fig:Tdep} for $b = 4$ and $g_{*\mathrm{HS}}(T)$ given by Eq.~(\ref{eq:gstar_for_plots}) with $\xi(T) = \xi_{\rm osc}$. The region shaded gold features an ALP with $h^2 \Omega_a =0.12$ for natural values of the initial misalignment angle, $\theta_0 \in [0.1, 2]$, obtained by solving the ALP equation of motion (EOM) numerically (c.f.~Appendix~\ref{appx:B}). The gold dotted contours correspond to the analytic estimates given above. The corresponding $T$-independent preferred ALP DM region, obtained by numerically solving the ALP EOM to late times, lies between the dashed gold contours. $\xi_{\rm osc}=1$ illustrates the maximum allowed enhancement of the relic abundance in a RD cosmology. Since the HS is at the same temperature as the SM, there are strong bounds from $N_{\rm eff}$ for $\xi_{\rm osc}=1$, and the shaded gray region on the left in Fig.~\ref{fig:Tdep} is excluded by requiring $\Delta N_{\rm eff} < 0.5$ (corresponding to $T_{\Lambda} > T_{\rm BBN} \sim 5$ MeV). It is likely that in concrete models the lower bound on $T_{\Lambda}$ will need to be somewhat higher than 5 MeV to avoid BBN constraints, and so the results shown should be understood to correspond to the most optimistic case. 
 
 On the right in Fig.~\ref{fig:Tdep} we show corresponding results assuming a decoupled hidden sector with $\xi_{\rm osc}=0.1$. The enhancement of $h^2 \Omega_a$ is smaller, however the cooler HS in principle allows for $T_{\Lambda} < T_{\rm BBN}$, since $\xi_{\rm BBN} < 0.1$ (assuming the number of relativistic degrees of freedom in the hidden sector does not change significantly between oscillation and the onset of BBN). Again, one must be mindful of additional model-dependent constraints on small-$\Lambda$ scenarios, as well as those with decoupled hidden sectors with dark radiation or significant late-time abundances of stable relics. 
 
 Larger $\xi_{\rm osc}$ and $b$ can in principle increase $\Omega_a$ further, however this often comes at the cost of additional entropy injection after oscillation in simple models. For example, it could be that $\xi_{\rm osc}>1$, however the large corresponding amount of HS entropy needs to be transferred to the SM before BBN, erasing the resulting enhancement for $\xi_{\rm osc}$ much larger than 1. A hidden sector predicting $b>4$ could also increase $h^2 \Omega_a$ somewhat, however as $b$ increases one also expects $g_{*{\rm HS}} (T_{\rm osc})$ to increase in a QCD-like theory, and so the resulting enhancement again gets washed out by the requisite HS entropy dump for large $b$ before BBN. These effects are encapsulated in the $g_*$ dependence of $\gamma_T$ in Eq.~(\ref{eq:gammaRD}).

\subsection{Early Matter Domination with \texorpdfstring{$T$}{T}-dependence}

We proceed similarly for the case of early matter domination, deriving a set of analytic expressions that can be used to estimate the relic abundance.
We again allow the HS to be at a different temperature than the SM bath and parametrize 
ALP mass temperature dependence as in Eq.~(\ref{eq:TdepM}).
The evolution of the energy densities in $\phi$ (the field responsible for EMD), SM and HS radiation can be modeled by
\begin{align}
\dot \rho_\phi + 3H \rho_\phi  & = -\Gamma_\phi \rho_\phi \\
\dot\rho_{\rm SM} + 4 H \rho_{\rm SM} & =+\Gamma_{\phi \to {\rm SM}} \rho_\phi\\
\dot\rho_{\rm HS} + 4 H \rho_{\rm HS} & =+\Gamma_{\phi \to {\rm HS}} \rho_\phi
\end{align}
where the dot indicates a derivative with respect to time, 
$\Gamma_\phi = \Gamma_{\phi \to {\rm SM}} + \Gamma_{\phi \to {\rm HS}}$ and 
$\Gamma_{\phi \to {\rm SM}}$, $\Gamma_{\phi \to {\rm HS}}$ are the partial widths of $\phi$ into to SM and HS radiation, respectively. 
These equations can be solved during $\phi$ domination (i.e. while $\Gamma_\phi/H\ll 1$ and $\rho_\phi R^3 \approx \mathrm{const}$) and yield
\beq \label{eq:EMDdensitysol}
\rho_i = \frac{6}{5} H_{\rm MD} M_{\rm Pl}^2 \left(\frac{R_{\rm MD}}{R}\right)^{3/2} \Gamma_{\phi \to i}, 
\eeq
where we assumed that the initial energy densities are negligible compared to those produced by $\phi$ decays.
Here $i$ corresponds to either HS or SM subscripts, and $H_{\rm MD}$ and $R_{\rm MD}$ are the Hubble parameter and FRW scale factor at the onset of EMD, respectively. 
We can use Eq.~(\ref{eq:EMDdensitysol}) to compute the temperatures of the HS and SM radiation baths during the epoch of matter domination. 
Defining the reheating temperature as that for which $\rho_\phi = \rho_{\rm SM} + \rho_{\rm HS}$ and assuming Eq.~(\ref{eq:EMDdensitysol}) holds down to that temperature allows one to relate $\Gamma_{\rm SM}$ and $\Gamma_{\rm HS}$ to $T_{\rm RH}$ and $\xi_{\rm osc}$ (again neglecting relative heating effects). We then find that the oscillation temperature and abundance enhancement factor can be approximated by
\beq
 \label{eq:ToscEMD}
 T_{\rm osc}(b) \simeq \frac{\Lambda}{\xi_{\rm osc}}  \left(\frac{28125 \, M_{\rm Pl}^2 \,T_{\rm RH}^4  \, g_*(T_{\rm RH})\,\xi_{\rm osc}^{8}}{32\, (4+b)^2 \pi^2 \, g^2_*(T_{\rm osc})\, m_a^2 \, f_a^4} \right)^{\frac{1}{8+2b}}
 \eeq
 \beq
\gamma_T \simeq \left(\frac{4+b}{4}\right)^{\frac{8+b}{4+b}} \left(\frac{1125 M_{\rm Pl}\sqrt{g_*(T_{\rm RH})} T_{\rm RH}^2 \xi_{\rm osc}^4}{16 \pi \sqrt{90} \,g_*(T_{\rm osc}) \, m_a f_a^2}\right)^{\frac{b}{4+b}}\label{eq:gammaEMD}
\eeq
(see Appendix~\ref{appx:A} for more details).  $\gamma_{\rm max}$ in EMD is given by
\beq\label{eq:gmax_EMD}
\gamma_{\rm max} \simeq \left(\frac{225 M_{\rm Pl}\sqrt{g_*(T_{\rm RH})} T_{\rm RH}^2 \xi_{\rm osc}^4}{4 \pi \sqrt{90}g_*(T_\Lambda) m_a f_a^2}\right)^2
\eeq
These expressions can be inserted into Eqs.~(\ref{eq:gamma_master}) -- (\ref{eq:Tosc}) and used along with Eq.~(\ref{eq:Omega_EMD}) to estimate $\Omega_a$ in the $T$-dependent case. Again if $T_{\rm osc} < T_{\rm RH}$, the RD expressions should be used. The resulting predictions agree well with full numerical solutions of the three-fluid system of equations, discussed in Appendix~\ref{appx:B}. 

The preferred regions of the ALP parameter space in the EMD scenario are illustrated in Fig.~\ref{fig:Tdep} for $\xi_{\rm osc} = 1$ and $\xi_{\rm osc}=0.1$ with $b=4$. We show results  for $T_{\rm RH}=10$ and 500 MeV. The blue and purple shaded regions feature an ALP with $\Omega_a h^2 \simeq 0.12$ for $\theta_0\in [0.1, 2]$ as obtained from the numerical solution. The corresponding dotted contours show the analytic predictions of Eqs.~(\ref{eq:ToscEMD}) -- (\ref{eq:gmax_EMD}) and are a good fit to the numerical results. For $\xi_{\rm osc}=1$, the gray shaded region is excluded by the measured value of $N_{\rm eff}$ at BBN. 
This constraint is alleviated for $\xi_{\rm osc}=0.1$, however the enhancement factor $\gamma$ is reduced as a result. 
Other model-dependent constraints are likely to apply in the region where $T_{\Lambda} < T_{\rm BBN}$ as discussed in Sec.~\ref{sec:general}.

The behavior illustrated in Fig.~\ref{fig:Tdep} is straightforward to understand. 
First, note that the oscillation temperature is reduced as $m_a$ and $f_a$ are increased. 
If the oscillations begin below $T_{\Lambda}$, temperature-dependent effects are unimportant 
and the preferred regions are the same as discussed in Sec.~\ref{sec:emd_cosmo} (i.e.~$\gamma = 1$). 
This occurs for larger ALP masses, and the preferred value of $\ga$ is independent of $m_a$. 
For small enough masses, $T_{\rm osc}>T_{\Lambda}$ so that $T$-dependence enhances the relic density for a fixed $m_a$, $f_a$ relative to the $T$-independent case. 
Here, the ALP DM regions pick up dependence on $f_a$, increasing the preferred values of $\ga$ in Fig.~\ref{fig:Tdep}.
 At even lower values of $m_a$, oscillation occurs after reheating ($T_{\rm osc} < T_{\rm RH}$), 
and the predictions reduce to those of the $T$-dependent RD scenario of Sec.~\ref{sec:rd_cosmo} (the gold shaded region). 

Fig.~\ref{fig:Tdep} shows that allowing for a $T$-dependent ALP mass interpolates between the $T$-independent EMD and $T$-dependent RD scenarios. Smaller values for $b$ tilt the interpolating region towards the left, while larger values steepen it. Increasing $T_{\rm RH}$ causes the EMD band to match onto RD predictions at larger $m_a$. In all cases, the preferred ALP DM regions are bounded by the $T$-dependent RD and $T$-independent EMD contours for a given $\theta_0$. 

\subsection{Kination with \texorpdfstring{$T$}{T}-dependence}

\begin{figure*}[t!]
  \centering
  \includegraphics[width=1.0\textwidth]{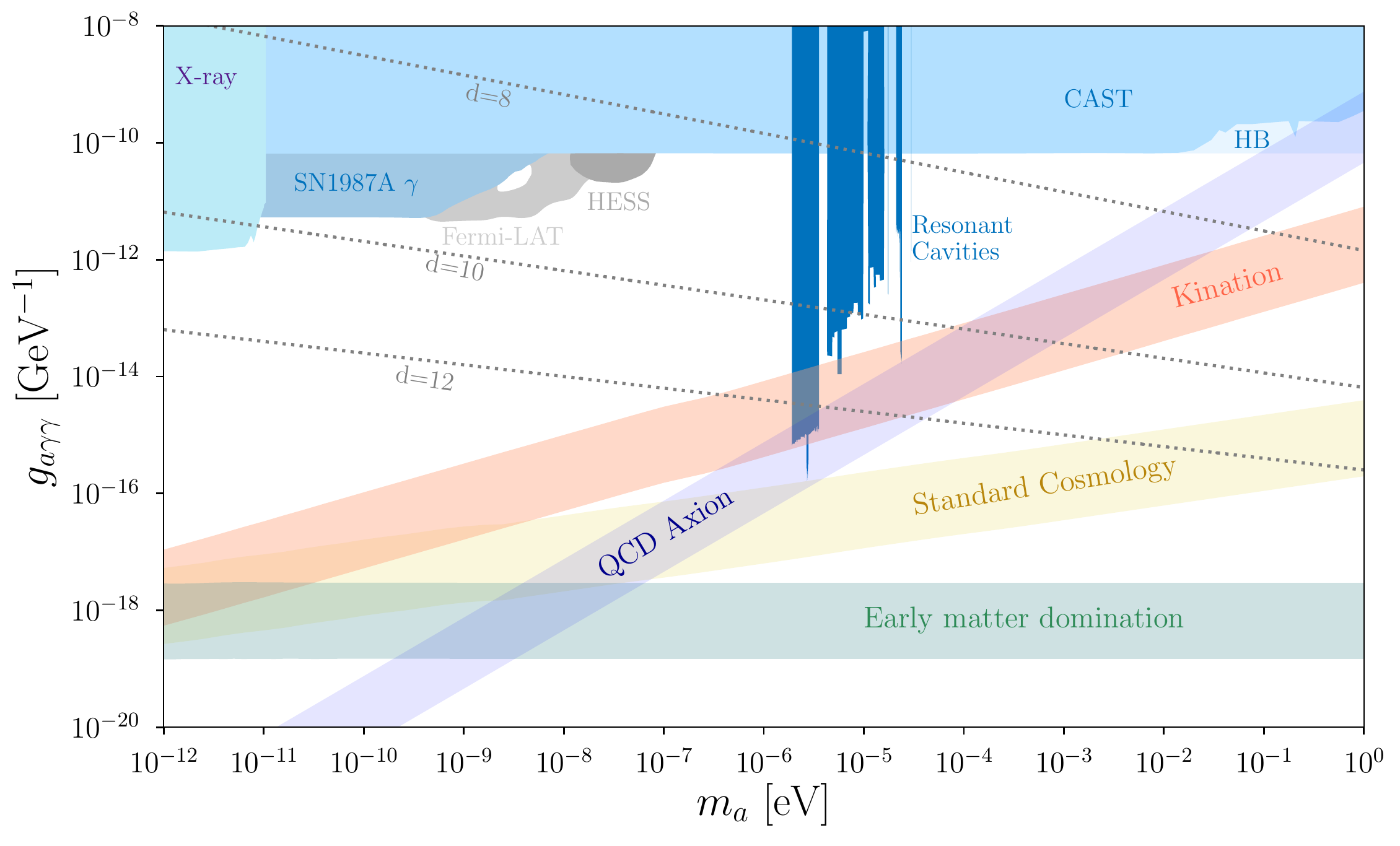}
  \caption{ Theoretical targets (colored bands) and current experimental constraints (filled regions) on the ALP-photon coupling $\ga$ as a function of the ALP mass $m_a$. The shaded bands show regions where the ALP saturates the observed DM relic abundance for the standard (yellow), early matter-dominated (green) and kination (red) cosmologies for initial misalignment angles of $\theta_0 \in [0.1,2]$.
  For the latter cosmologies, we take the reheating/kination temperature to be $10$ MeV.
  We also show the  QCD axion band, which does not have a relic density requirement imposed, in blue.
The gray dotted diagonal lines correspond to ALPs which get their mass from dimension 8, 10 and 12 Planck-suppressed operators. 
Further discussion can be found in Section~\ref{sec:results}.
}
  \label{fig:masterplot_current}
  \end{figure*}

Finally, we comment on the kination cosmology with a $T$-dependent ALP mass near the onset of oscillations. From Eq.~(\ref{eq:enhancement}), the only enhancement comes from the slightly different values of $q$ defining the oscillation time. In other words, the gain in energy from the growth of the mass is almost completely canceled by the loss in energy from starting to oscillate later. As explained in Appendix~\ref{appx:A}, we find
\beq \label{eq:Tosckin}
T_{\rm osc}(b) \simeq \frac{\Lambda} {\xi_{\rm osc}} \,\left(\frac{10125 \,M_{\rm Pl}^2 \,T_{\rm kin}^{2}  \, g_*(T_{\rm kin}) \,\xi_{\rm osc}^{6}}{32 \,(3+b)^2 \pi^2  \, g^2_*(T_{\rm osc})\, m_a \, f_a^{3} } \right)^{\frac{1}{6+2b}},
\eeq
\beq\label{eq:gammakin}
\gamma_T \simeq \frac{1}{3}(3+b).
\eeq
  Meanwhile, 
 \beq\label{eq:gmax_kin}
 \gamma_{\rm max} \simeq \frac{5 \sqrt{90} M_{\rm Pl} \sqrt{g_*(T_{\rm kin})} T_{\rm kin} \xi_{\rm osc}^3}{8 \pi \sqrt{m_a f_a^3} g_*(T_\Lambda)}.
 \eeq
 In most realistic models, one expects $b\sim \mathcal{O}(1)$, and so typically $\gamma=\gamma_T$. The enhancement is milder than in RD and EMD as it only depends on the exponent $b$.
Nevertheless, allowing for $m_{\rm osc} \neq m_a$ changes the ALP oscillation temperature. Since $T_{\rm osc} > T_{\rm kin}$ in order for the period of kination to modify the ALP evolution, $T$-dependence will change the regions of the ALP plane where kination is relevant for fixed $T_{\rm kin}$.

In our results below we will take $T_{\rm kin} = 10$ MeV. Since the preferred regions on the ALP parameter space assuming kination with and without $T$-dependence are similar for $T_{\rm kin}=10$ MeV, we do not show predictions for kination in Fig.~\ref{fig:Tdep}. However, we provide the corresponding $T$-dependent predictions in Fig.~\ref{fig:masterplot_Tdep}. Again we find that the analytic estimates above provide a good fit to the numerics (c.f.~Appendix~\ref{appx:B}) across the parameter space considered.

\section{Projections and Results}
\label{sec:results}
\begin{figure*}[t!]
  \centering
  \includegraphics[width=1.0\textwidth]{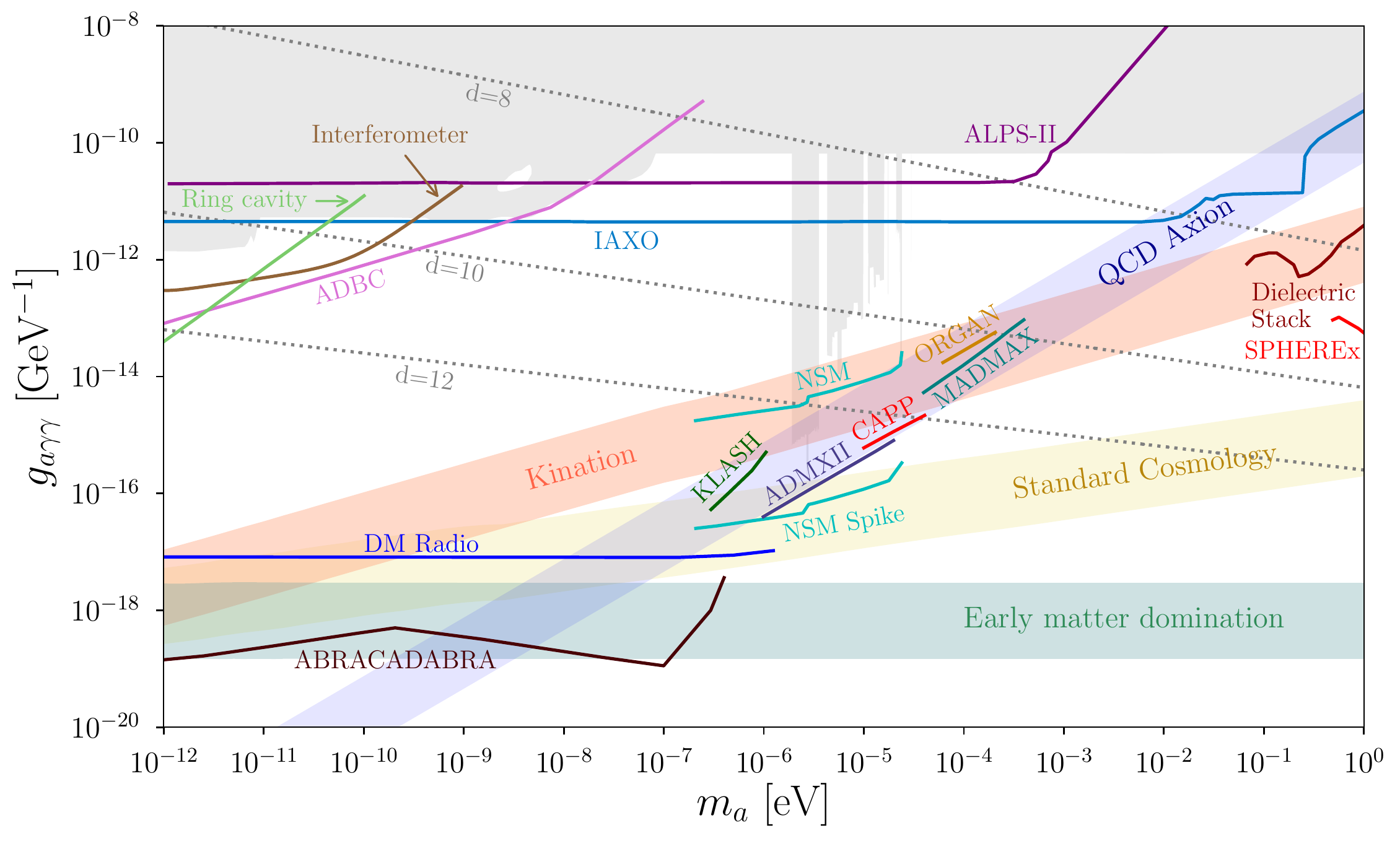}
  \caption{ Theoretical targets (colored bands) and projected experimental reach (colored lines) in the ALP-photon coupling $\ga$ as a function of the ALP mass $m_a$. The shaded bands show regions where the ALP saturates the observed DM relic abundance for the standard, early matter-dominated, and kination cosmologies for initial misalignment angles of $\theta_0 \in [0.1,2]$. For the latter cosmologies, we take the reheating temperature to be $10$ MeV. We also show the standard QCD axion target in blue and existing experimental constraints in solid gray. The gray dotted diagonal lines correspond to ALPs which obtain mass from dimension 8, 10 or 12 Planck scale suppressed operators. Further discussion can be found in Section~\ref{sec:results}.}
  \label{fig:masterplot}
  \end{figure*}

\begin{figure*}[t!]
  \centering
  \includegraphics[width=.5\textwidth]{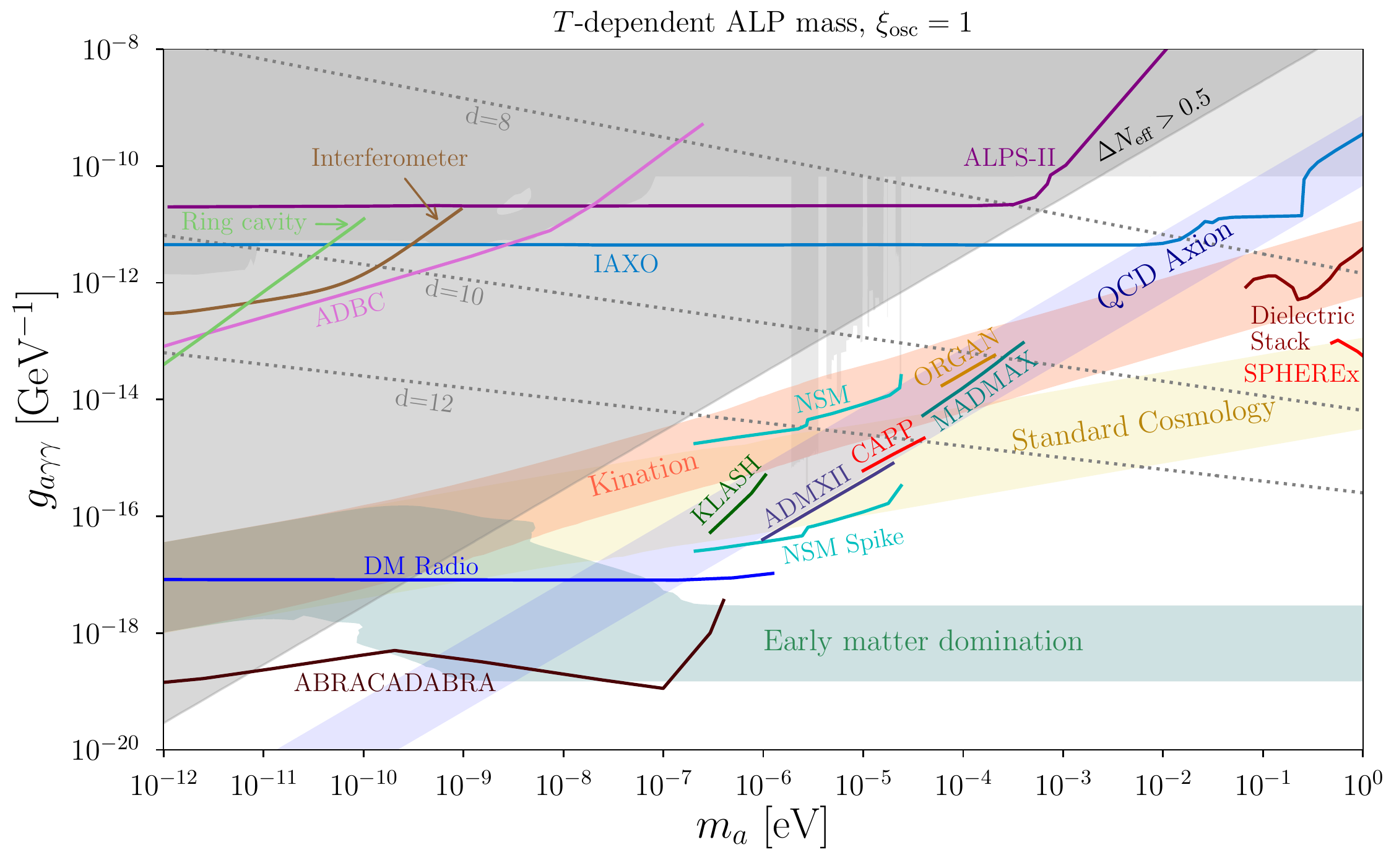}\,\,\includegraphics[width=.5\textwidth]{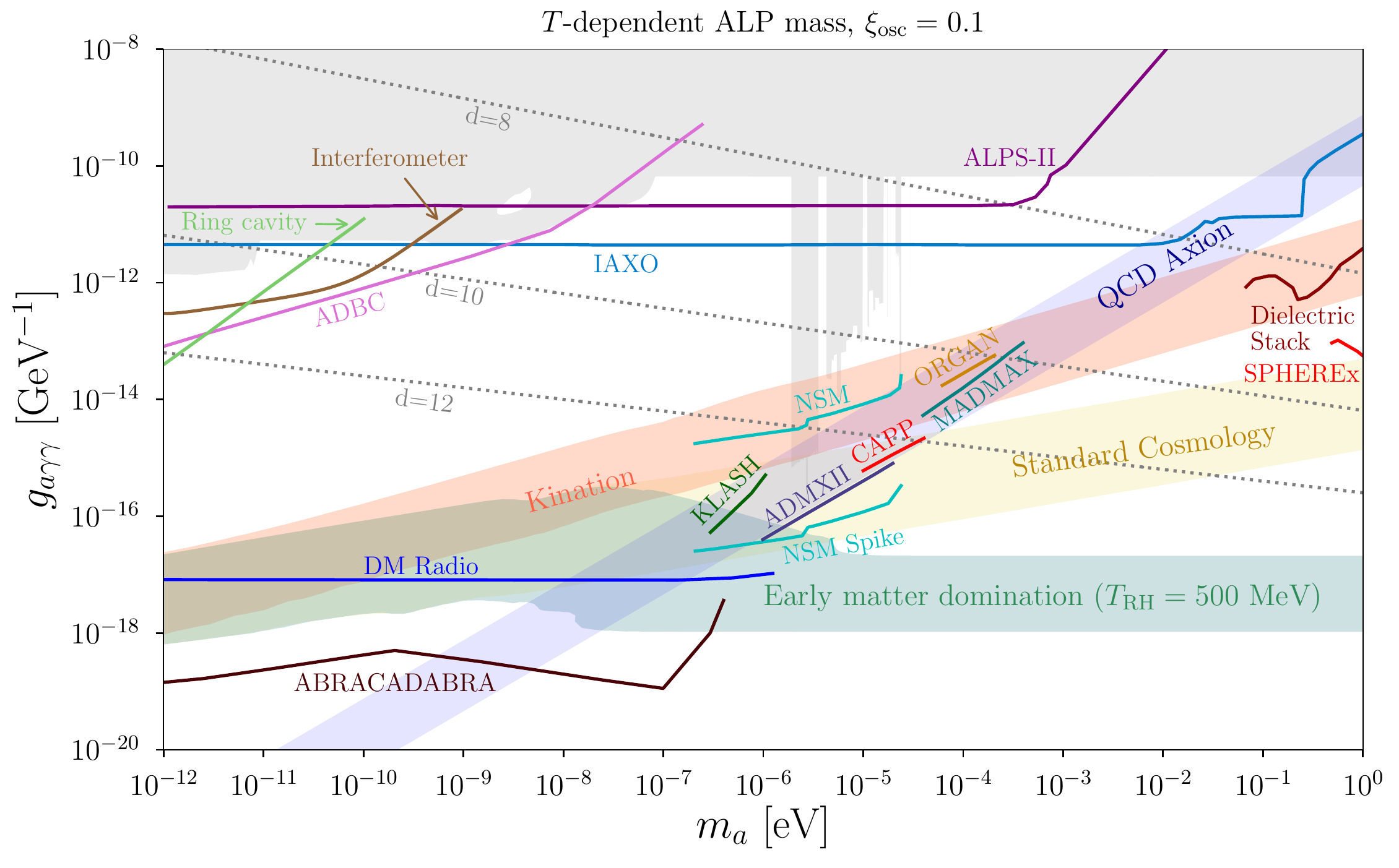}
  \caption{As in Fig.~\ref{fig:masterplot}, but for a $T$-dependent ALP mass. In the left panel, the hidden sector responsible for generating the ALP potential is assumed to be in thermal equilibrium with the SM, while in the right panel we assume the hidden sector is decoupled with temperature given by $0.1 \times T_{\rm SM}$. In both cases $b=4$ was assumed in Eq.~(\ref{eq:TdepM}). Note that there may be additional important constraints on the hidden sector, as discussed further in the text. On the left, the EMD band assumes $T_{\rm RH} = 10$ MeV, while on the right $T_{\rm RH} = 500$ MeV. In both cases $T_{\rm kin} = 10$ MeV. The parameter space on the left is constrained by $\Delta N_{\rm eff}$ at BBN, since there are necessarily new HS states in equilibrium with the SM bath at $T_{\rm \Lambda}$. The ALP target regions assume that $g_{*{\rm HS}}=52$ above $T_{\Lambda}$, corresponding to the expected value for a $SU(3)$ hidden sector with 3 light flavors. }
  \label{fig:masterplot_Tdep}
  \end{figure*}

We now investigate the potential for current and future ALP direct detection experiment and astrophysical observations to explore these natural ALP dark matter targets. 
The present status is summarized in Fig.~\ref{fig:masterplot_current}, and future prospects are shown in in Figs.~\ref{fig:masterplot} (for temperature-independent masses) and~\ref{fig:masterplot_Tdep} (for temperature-dependent masses).
These figures also show the preferred regions in the three 
cosmological histories considered in Secs.~\ref{sec:ALPDM} and~\ref{sec:ALPmass}: the standard cosmology,
early matter domination (EMD) with $\TRH = 10\,\MeV$, and kination with $\Tkin = 10\,\MeV$.  $\TRH$ and $\Tkin$ are the temperatures at which the universe transitions to standard radiation-dominated 
evolution;  temperatures of $5-10\,\MeV$ correspond to the lowest values compatible with BBN. In each case the bands are obtained by varying the initial misalignment angle $\theta_0$ between $0.1$ (bottom edge of each band) and $2$ (upper edge of each band). The ALP regions can be extended to smaller values of $\ga$ at the cost of fine-tuning $\theta_0 < 0.1$. The gray dotted lines in Figs.~\ref{fig:masterplot_current},~\ref{fig:masterplot} and~\ref{fig:masterplot_Tdep} show the ALP mass-coupling relation if the masses are generated by Planck-suppressed operators of various dimensions as discussed in Sec~\ref{sec:ALPDM}. For the temperature-dependent results in Fig.~\ref{fig:masterplot_Tdep} we have assumed $b=4$ in Eq.~(\ref{eq:TdepM}) and taken $g_{*{\rm HS}}$ as in Eq.~(\ref{eq:gstar_for_plots}) with $\xi=\xi_{\rm osc}$. The left hand-plot in Fig.~\ref{fig:masterplot_Tdep} shows the case where the hidden sector is in thermal equilibrium the SM, and the right-hand plot the case where the HS is  decoupled from the SM with a lower temperature, $0.1\times T_{\rm{SM}}$.

Axion-like particles can be constrained by a variety of astrophysical measurements. 
These limits include the results from CAST~\cite{Anastassopoulos:2017ftl}; 
cooling of Horizontal Branch (``HB'' in Fig.~\ref{fig:masterplot_current}) stars, massive stars~\cite{Cadamuro:2011fd,Friedland:2012hj}, 
and SN1987A~\cite{Raffelt:1996wa,Dolan:2017osp,Lee:2018lcj}; 
non-observation of a $\gamma$-ray excess from SN1987A~\cite{Payez:2014xsa}; 
the extragalactic background light~\cite{Masso:1997ru,Overduin:2004sz}; 
searches for spectral irregularities in $\gamma$ rays with HESS~\cite{Abramowski:2013oea} and Fermi-LAT~\cite{TheFermi-LAT:2016zue}, 
and in X-rays with Chandra~\cite{Chen:2017mjf}.\footnote{ALPs can also be constrained by observations of near-extremal black holes and the resulting constraints on superradiance~\cite{Arvanitaki:2014wva}. However, these constraints are strongly model-dependent in that they are sensitive to the properties of the ALP self-interactions. Accordingly we omit them from our plots, noting that they impact the region of parameter space $m_a \lesssim 10^{-11}$~eV.} 
These limits are shown in Fig.~\ref{fig:masterplot_current} as pastel-colored shaded regions.

The ALP parameter space is also constrained by a number of resonant cavity experiments. 
We show the regions excluded by ADMX~\cite{Asztalos:2009yp,Du:2018uak} and ADMX Sidecar~\cite{Boutan:2018uoc}, Phase 1 of HAYSTAC~\cite{Zhong:2018rsr}, the ORGAN Pathfinder~\cite{McAllister:2017lkb} and the older UF~\cite{Hagmann:1990tj} and RBF~\cite{DePanfilis:1987dk,Wuensch:1989sa} experiments in dark blue in Fig.~\ref{fig:masterplot_current}. These experiments target the classical QCD axion DM window for $m_a$ between $10^{-6}$ and $10^{-4}$~eV. 
We see in Fig.~\ref{fig:masterplot_current} that the resonant cavity experiments are already probing significant regions 
of the kination-favored parameter space and are just beginning to extend into the QCD axion window.

We  turn now to near-term prospects for direct detection in the ALP parameter space.
The past few years have seen a renaissance in ideas for searching very light DM, including coherent bosonic candidates like ALPs. We show in Fig.~\ref{fig:masterplot} a summary of the impact these new experiments will have on the ALP parameter space for temperature-independent ALP masses, and in Fig.~\ref{fig:masterplot_Tdep} a similar summary for temperature-dependent ALP masses.  In many cases, allowing for $T$-dependence the experimental prospects are even more promising, although constraints on new relativistic degrees of freedom generating the ALP potential can exclude some of the parameter space. We emphasize that these considerations are model-dependent and that specific scenarios could feature even more stringent constraints on the hidden sector than those considered.
Each experiment is capable of ruling out the region above the corresponding solid line.

  Some future experiments are extensions of resonant microwave cavities technique, as in upgrades to ADMX~\cite{Shokair:2014rna}, CAPP~\cite{Petrakou:2017epq}, KLASH~\cite{Alesini:2017ifp,Gatti:2018ojx}, and, at higher frequencies, ORGAN~\cite{McAllister:2017lkb}. These experiments provide a broader sensitivity in the QCD axion region $m_a\sim 10^{-6}-10^{-5}$~eV extending to lower values of $\ga$. More recently, new ideas based on dielectric stacks have appeared which are sensitive to higher mass ALPs, as in MADMAX~\cite{TheMADMAXWorkingGroup:2016hpc} and photonic materials~\cite{Baryakhtar:2018doz} (``Dielectric Stack'' in Figs.~\ref{fig:masterplot} and~\ref{fig:masterplot_Tdep}). 
  We also show the sensitivity of the proposal for a large-scale helioscope, IAXO~\cite{Irastorza:2013dav}, which will extend the reach of CAST, and the projections for the ALPS-II light-shining-through-walls experiment~\cite{Bahre:2013ywa}, which is currently under construction at DESY and will have sensitivity above $\ga \sim 10^{-11}~\rm{GeV}^{-1}$. 
  It is also possible that future measurements of radio emission lines from the magnetospheres of neutron stars could lead to constraints in the $\mu$eV mass-range~\cite{Hook:2018iia}. We show the limits that could be obtained with 100 hours of observation of the magnetar SGR J1745-2900 under the assumptions of an NFW dark matter density profile (``NSM'' in Figs.~\ref{fig:masterplot} and~\ref{fig:masterplot_Tdep}), and also a spike profile (``NSM Spike'') which would lead to stronger bounds. At large ALP masses the intensity line-mapping experiment SPHEREx~\cite{Creque-Sarbinowski:2018ebl} will be able to probe to the bottom of the kination region.

At very low masses, the ABRACADABRA suite of experiments (the region we show is the union of the broadband and resonant searches) and DM-Radio promise to cover a large amount of parameter space down to very small values of $\ga$\footnote{We also note that there is the BEAST proposal~\cite{McAllister:2018ndu} which could be relevant at low masses. However, since the BEAST projections are a topic of current discussion in the literature~\cite{Ouellet:2018nfr,Beutter:2018xfx,Kim:2018sci} we do not show them on our plot. We also do not show other limits from other as-yet-unpublished proposals, such as~\cite{Goryachev:2018vjt,Marsh:2018dlj,Bogorad:2019pbu,Janish:2019dpr,Edwards:2019tzf}.}. Other recent proposals at low mass make use 
of birefringence in the presence of an ALP background and include the interferometer concept~\cite{DeRocco:2018jwe}, ADBC~\cite{Liu:2018icu}, and an experiment based on optical ring cavities~\cite{Obata:2018vvr}.

While these will be able to explore new parts of parameter space, we find that they will not be sensitive to the kinds of ALP dark matter we study in this paper. We find that DM-Radio  will be able to probe ALP dark matter up to $m_a\sim 10^{-6}$~eV assuming a standard cosmology or a period of kination in the early Universe. ABRACADABRA will be able to discover (or rule out) ALP dark matter in all of the cosmological scenarios we have considered with masses below $\sim 4\times 10^{-7}$~eV. If the ALP mass is generated by a new strongly coupled gauge sector, the signal at ABRACADABRA for a given mass is likely to be even larger.

\section{Summary and Conclusions}
\label{sec:conc}

We have investigated the implications of current and future direct detection experiments for ALP dark matter with mass $10^{-12}\leq m_a \leq 1$~eV in a variety of well-motivated cosmological scenarios. We have presented simple analytic expressions for the corresponding relic density from misalignment in the standard cosmological scenario with radiation domination (RD), as well as allowing for a period of early matter domination (EMD) and kination, in Eqs.~(\ref{eq:Omega_RD}), (\ref{eq:Omega_EMD}), and (\ref{eq:Omega_kin}), respectively. These results apply to ALPs for which the mass is independent of the temperature between the onset of oscillations and today. A $T$-dependent ALP mass of the form in Eq.~(\ref{eq:TdepM}) enhances the relic abundance relative to these predictions so that $\Omega_a = \gamma \Omega_a^{T-{\rm ind}}$, with the enhancement factor $\gamma$ given by Eq.~(\ref{eq:gamma_master}) and Eqs.~(\ref{eq:gammaRD}) -- (\ref{eq:gmax_RD}) for RD,  Eqs.~(\ref{eq:gammaEMD}) -- (\ref{eq:gmax_EMD}) for EMD, and Eqs.~(\ref{eq:gammakin}) -- (\ref{eq:gmax_kin}) for kination, respectively.

While ALP dark matter is currently relatively unconstrained, future experiments have the ability to probe much of the well-motivated ALP parameter space. ALPs that obtain their masses from Planck-suppressed operators will be thoroughly tested by future experiments (provided they can saturate the observed dark matter relic abundance). The amount of suppression required for a viable ALP dark matter candidate depends on the cosmological scenario under consideration. It is possible for an ALP associated with $d=12$ operators to be consistent with the DM relic density in a standard cosmological scenario for masses above $10^{-2}$~eV. In other cases, such as a period of kination down to temperatures of a few MeV, an ALP with mass set by $d=8$ Planck-suppressed operators can provide a viable dark matter candidate, a significantly less stringent requirement than that for the QCD axion (which requires Planck-suppressed PQ-breaking operators to arise at $d=12$ or higher). 

ALP dark matter can be easier to detect than the QCD axion. For low masses (below the standard QCD axion window for a fixed $f_a$) experiments such as ABRACADABRA and DM-Radio will have sensitivity to ALP dark matter before they are sensitive to the QCD axion. In particular, the ABRACADABRA experiments can constrain the existence of ALPs in the various cosmologies we have considered for $m_a \lesssim 4\times 10^{-7}$~eV down to $10^{-12}$~eV. Below $10^{-11}$~eV black-hole super-radiance complements the ABRACADABRA and DM-Radio sensitivity, although the precise details are model-dependent. If there was a period of kination in the early Universe down to temperatures near the BBN scale, or if the ALP mass at the onset of oscillations is smaller than its present day value, ABRACADABRA and DM-Radio can be more sensitive to ALP dark matter than to the QCD axion across their entire mass sensitivity ranges. At higher $m_a$, the experimental prospects are more positive in the kination and $T$-dependent RD cases as well. We find that resonant cavity experiments (such as ADMX, CAPP and ORGAN), as well as MADMAX, can also probe ALPs in these more optimistic scenarios before they reach the QCD axion window.

For even larger masses, $m_a \gtrsim 10^{-4}$ eV, ALP dark matter becomes more difficult to detect than the QCD axion in all of the scenarios we have considered. However, some proposed experiments using terahertz frequency resonators, dielectric stacks, or line-intensity mapping targeting the QCD axion in this mass range can also probe ALP DM that begins oscillating during kination (for low $T_{\rm kin}$) and come close to the standard ALP prediction with $\mathcal{O}(1)$ initial misalignment angles and $T$-dependent masses for $m_a$ up to an eV. For ALPs in the standard RD and EMD cosmologies with masses set in the UV, this high-$m_a$ region will be difficult to access with existing experimental proposals. However, other probes of this parameter space beyond direct detection experiments may exist in some cases. For example, a period of early matter domination can also lead to the formation of ALP miniclusters, which can have interesting astrophysical consequences~\cite{ALPminiclusters,Nelson:2018via,Visinelli:2018wza}. Future inquiry along these lines, and new ideas to access this region experimentally, are worth continued investigation.

ALPs can provide a compelling and viable dark matter candidate, behaving much like the QCD axion in the early Universe, 
but in many cases allowing for larger couplings to photons. ALP dark matter, therefore, can be easier to detect than the QCD axion, especially at low masses. More generally we emphasize the importance of vigorously pursuing the axion direct detection program, targeting a wide range of masses and exploring the ALP parameter space beyond the canonical QCD axion window.

\vskip 1\baselineskip
\noindent {\emph{Acknowledgements}} 
We thank Aaron Chou, Yonatan Kahn, Jeff Filippini, and Manuel Meyer for useful discussions.
MJD is supported by the Australian Research Council. The work of PD and JK was supported by NSF grant PHY-1719642.
This manuscript has been authored by Fermi Research Alliance, LLC under 
Contract No. DE-AC02-07CH11359 with the U.S. Department of Energy, Office of Science, Office of High Energy Physics.

\appendix
\begin{widetext}
\section{Analytic Estimates for the Relic Density}
\label{appx:A}

In this Appendix we derive analytic estimates of the ALP DM number density.

\subsection{Temperature-Independent ALP Mass}

First we consider cases where $m_a$ is fixed to its zero-temperature value at times before the onset of ALP oscillations. The relic density is then given by Eq.~(\ref{eq:relic}), reproduced here for convenience:
\beq
  \Omega_a = \frac{m_a n_a(\Tad)}{\rho_c} \left( \frac{T_0}{\Tad}\right)^3  \frac{g_{*S}(T_0)}{g_{*S}(\Tad)} \, .
  \label{eq:relic_app}
\eeq
$T_{ad}$ is a reference temperature below which the evolution of the universe is adiabatic and $T_0$ is the temperature today. $T_{\rm osc}$ is the SM temperature at the onset of ALP oscillations, and we have $\dot{\theta}(t_{\rm osc}) \simeq 0$ and $n_a(T_{\rm osc}) = 1/2 m_a f_a^2 \theta_0^2$. Below $T_{\rm osc}$ the comoving ALP number density is assumed to be conserved, so that
\beq 
n_a(T_{\rm ad}) = \frac{1}{2} m_a f_a^2 \theta_0^2 \left(\frac{R_{\rm osc}}{R_{\rm ad}}\right)^3.
\label{eq:na_app}
\eeq 
Combining Eqs.~(\ref{eq:relic_app}) -- (\ref{eq:na_app}) yields a general expression for the relic density,
\beq
\Omega_a = \frac{1}{2}\frac{m_a^2 f_a^2 \theta_0^2 }{\rho_c}\left(\frac{R_{\rm osc}}{R_{\rm ad}}\right)^3 \left( \frac{T_0}{\Tad}\right)^3  \frac{g_{*S}(T_0)}{g_{*S}(\Tad)}.
\label{eq:relic_general}
\eeq
To obtain $\Omega_a$ we therefore need to determine $T_{\rm osc}/T_{\rm ad}$ and/or $R_{\rm osc}/R_{\rm ad}$ in the various cases. 

We parametrize the relationship at oscillation between the Hubble parameter and ALP mass as
\beq
H_{\rm osc} &= m_a/q_0~~~{\rm (T-independent)}.
 \label{eq:q0}
\eeq
Here $q_0$ is a positive number that should be chosen to accurately reproduce the numerical predictions, described in Appendix~\ref{appx:B}. In our final estimates we will take $q_0=1.6$; the reasoning behind this choice is explained below. At the time when oscillations begin, it is assumed that the universe is dominated by a fluid with equation of state $p=w\rho$ (for which $H^2 \propto R^{-3(w+1)}$), and then later transitions instantaneously to radiation domination at some $T=T_*$:
\beq
H_{\rm osc}^2 \left(\frac{R_{\rm osc}}{R_*}\right)^{3(w+1)}  \simeq \frac{\pi^2}{90 M^2_{\rm Pl}} g_*(T_*) T_*^4\;.
\label{eq:Hosc}
\eeq
In all cases we consider (RD, EMD, and kination), we can take 
\begin{align}
\Tad\rightarrow T_*.
\end{align} 
Henceforth we will refer only to $T_*$. In RD, we can also set $T_*=T_{\rm osc}$. In EMD, $T_* = T_{\rm RH}$, and in kination $T_* = T_{\rm kin}$. 
Combining Eqs.~(\ref{eq:q0}) -- (\ref{eq:Hosc}), we obtain $R_{\rm osc}/R_{\rm *}$  in terms of $T_*$ and $m_a$. Plugging in to Eq.~(\ref{eq:relic_general}) and setting $q_0=1.6$ yields the final result for the relic density for each case, given in the main text as Eqs.~(\ref{eq:Omega_RD}),~(\ref{eq:Omega_EMD}), and (\ref{eq:Omega_kin}). Note that in the kination case we could have instead taken $T_{\rm ad} = T_{\rm osc}$ and used the conservation of comoving entropy to relate $T_{\rm osc}$  to $T_{\rm kin}$. This yields the same result.

We now discuss $q_0$ in more detail, which will be particularly relevant when the ALP mass is $T$-dependent. First we inspect the form of solutions to the ALP EOM. Consider the EOM during a period with equation of state $p=w\rho$, such that $R \sim t^{\frac{2}{3(w+1)}}$. With $\theta(t)\equiv a(t)/f_a$ we have
\beq
\ddot{\theta} + \frac{2}{(w+1)\,t} \dot\theta + m_a^2 \,\theta = 0
\eeq
which has solutions
\beq
\theta(t) =t^{r} \left[ c_1 J_{r}(m_a t) + c_2 Y_{r}(m_a t) \right].
\label{eq:ALPsol1}
\eeq
Here $r=\frac{1}{2}-\frac{1}{1+w}$, $c_{1,2}$ are integration constants, and $J_{r}(x)$, $Y_{r}(x)$ are Bessel functions of the first and second kind. Given that the Bessel functions only exhibit oscillatory behavior when their arguments are $\mathcal{O}(1)$ or larger, we see that ALP oscillations begin when 
\begin{align}
m_a \simeq A/t,
\end{align}
 with $A$ an $\mathcal{O}(1)$ number. Since $H\simeq 2/(3t(1+w)$, we  define the onset of oscillations as
\beq
m_a \simeq \frac{3A(1+w)}{2} H_{\rm osc}.~~~{\rm (T-independent~ALP~mass)}
\label{eq:mosc1}
\eeq
For EMD ($w=0$), RD ($w=1/3$), and kination ($w=1$), we obtain $m_a \simeq \left\{3/2,2, 3\right\}\times A H_{\rm osc}$, respectively. Comparing our analytic and numerical solutions, we find that choosing $A$ such that $m_a \simeq 1.6 H_{\rm osc}$ reproduces the numerical results to within a few tens of percent across the parameter space considered in the various cosmologies. (This appears consistent with the discussion of Ref.~\cite{Marsh:2015xka}, which found $m_a\sim 2 H_{\rm osc}$ is a better choice than $m_a\sim 3 H_{\rm osc}$ in the RD scenario.) This value of $q_0$ can be adjusted to yield slightly better agreement in each cosmology, but for simplicity we take a common value. Therefore, introducing the parametrization~(\ref{eq:mosc1}) was not really necessary in this case; however, a similar parametrization is useful when considering temperature-dependent masses, so we keep it for comparison. Summarizing, we take $q_0 = 1.6$ for all cosmologies, or in the parametrization~(\ref{eq:mosc1}), 
\beq\label{eq:Avals}
A = \left\{\begin{array}{c c} 
\vspace{.1cm}\frac{16}{15}, & {\rm EMD} \\

\vspace{.1cm} \frac{4}{5}, & {\rm RD}\\

\vspace{.1cm}\frac{8}{15}, & {\rm kination}.

\end{array}
\right.
\eeq

\subsection{Temperature-Dependent ALP Mass}
We can proceed similarly when $m_a(T)$ varies near the onset of oscillations.  
 As discussed in the main text, the temperature controlling the ALP mass does not need to equal the temperature of the SM bath.
We parametrize the temperature of the hidden sector $T_{\rm HS}$ as
\beq
T_{\rm HS} \equiv \xi(T) \, T
\eeq 
where $T$ is the temperature of the SM photon bath. We take an instanton-motivated class of models in which $m_a(T_{HS}) = (\Lambda/T_{HS})^b m_a$ for $T>\Lambda$.
Subsequently, all temperatures will correspond to SM temperatures, unless otherwise stated, and factors of $\xi$ will be used to convert to hidden sector temperatures. 

Denoting the ALP mass at the onset of oscillations as $m_{\rm osc}$ and the mass today as $m_a$, the relic density can be expressed as:
\beq\label{eq:relic_2_app}
\Omega_a \simeq\frac{1}{2\rho_c} \left(\frac{R_{\rm osc}}{R_0}\right)^3 m_a m_{\rm osc} f_a^2 \theta_0^2 \simeq \frac{1}{2\rho_c}\left(\frac{R_{\rm osc}}{R_*}\right)^3 \frac{g_{*s}(T_0) T_0^3}{g_{*s}(T_*) T_*^3} \,m_a \,m_{\rm osc} \,f_a^2 \, \theta_0^2\;.
\label{eq:rho0}
\eeq
Here we have again assumed that at the time when oscillations begin, the universe is dominated by a fluid with equation of state $p=w\rho$ (for which $H^2 \propto R^{-3(w+1)}$), and then later transitions instantaneously to radiation domination at some $T=T_*$ so that Eq.~(\ref{eq:Hosc}) applies. We parametrize the oscillation time in the case of temperature-dependent ALP masses via
\beq
H_{\rm osc} &= m_{\rm osc}/q_T~~~{\rm (T-dependent)}
\eeq
generalizing Eq.~(\ref{eq:q0}). $q_T$ should again be a positive number chosen so that the analytic formulae provide a good approximation to the numerical results. Note that $q_T$ and $q_0$ can be different.

For a fixed $m_a$, $f_a$ and $\theta_0$, we can compare the relic densities in the $T$-dependent and $T$-independent cases. From Eqs.~(\ref{eq:relic_general}), (\ref{eq:Hosc}), and (\ref{eq:relic_2_app}) we have
\beq \label{eq:gamma_app}
\gamma = \frac{\Omega_{a}}{\Omega_{a}^{\rm T-ind}} \simeq \frac{m_{\rm osc}}{m_a} \left(\frac{H_{\rm osc}^{\rm T-ind}}{H_{\rm osc}} \right)^{\frac{2}{w+1}} \simeq \frac{m_{\rm osc}}{m_a} \left(\frac{q_T m_a}{q_0 m_{\rm osc}} \right)^{\frac{2}{w+1}} = \left(\frac{q_T}{q_{0}}\right)^{\frac{2}{w+1}} \left(\frac{m_a}{m_{\rm osc}}\right)^{\frac{2}{w+1}-1}.
\eeq
Thus, to determine $\Omega_a$ in the $T$-dependent case, we can use the $T$-independent results of Eqs.~(\ref{eq:Omega_RD}), (\ref{eq:Omega_EMD}), and (\ref{eq:Omega_kin}), multiplying by $\gamma$ (assuming that $T$-dependence is relevant at the time of oscillations) and using the value of $T_{\rm osc}$ predicted for $T$-dependent masses. To do so, we must determine $m_{\rm osc}$ and appropriate choices of $q_T$ in each case. 

How should $q_T$ be chosen? Consider the ALP EOM for times when $m_a(T) = m_a (\Lambda/\xi(T)T)^b$. During a period of adiabatic evolution dominated by a fluid with equation of state $\rho = wp$, the temperature evolves as
\beq
T\propto R^{-1} \propto t^{-\frac{2}{3(w+1)}}
\eeq
away from mass thresholds. As a result, the $T$-dependent mass term in the ALP EOM can be written in terms of $t$ so that the EOM becomes
\beq
\ddot{\theta} + \frac{2}{(w+1)\,t}  \dot{\theta} + m_a^2 \left(\frac{\Lambda}{T_i} \right)^b t^{\frac{4b}{3(w+1)}} \theta = 0
\eeq
where $T_i$ is a constant (neglecting the $T$-dependence of $g_*$ and $\xi$). The solutions of this equation are sightly more complicated than in the $T$-independent case, but can still be written in terms of Bessel functions with arguments
\beq
\frac{3(1+w)\, t}{3(1+w)+2b} \, m_a t^{\frac{2b}{3(1+w)}} \left(\frac{\Lambda}{T_i}\right)^b .
\eeq
The solution starts to oscillate when the above quantity $\sim \mathcal{O}(1)\equiv A$. Using $H=2/(3(1+w)t)$ and $m_a(t) = m_a t^{\frac{2b}{3(1+w)}} (\Lambda/T_i)^b$, oscillations begin when
\beq
m_{\rm osc}\simeq \frac{A\left(3+3w+2b\right)}{2} H_{\rm osc} \quad {\rm (RD, kination)}.
\label{eq:mosc_adiabatic}
\eeq  
 Since we have assumed adiabatic evolution, this result applies to our radiation-dominated and kination scenarios. 
 
 For the EMD cosmology, entropy is injected into the bath from decays. However, we can still derive an approximate time-temperature relation from Eq.~(\ref{eq:EMDdensitysol}) (again neglecting $T$-dependence in $g_*$ and $\xi$).  The ALP EOM in this case is approximately
 \beq
 \ddot{\theta} + \frac{2}{t}  \dot{\theta} + m_a^2 \left(\frac{\Lambda}{T_i} \right)^b t^{\frac{b}{2}} \theta = 0.
 \eeq
 Proceeding as before, we find that the onset of oscillations occurs when
\beq
m_{\rm osc}\simeq \frac{ 3A\left(4+b\right)}{8} H_{\rm osc} \quad {\rm (EMD)}.
\label{eq:mosc_EMD}
\eeq
Eqs.~(\ref{eq:mosc_adiabatic}) and (\ref{eq:mosc_EMD}) reduce to the results from Eq.~(\ref{eq:mosc1}) when $b=0$, as they should. Taking the same values for $A$ as in Eq.~(\ref{eq:Avals}), we obtain
\beq
q_T = \left\{\begin{array}{c c} 
\vspace{.1cm} \frac{2}{5}\left(4+b \right), & {\rm EMD} \\

\vspace{.1cm} \frac{4}{5}(2+b), & {\rm RD}\\

\vspace{.1cm}\frac{8}{15}\left(3+b\right), & {\rm kination}

\end{array}
\right.
\label{eq:q1}
\eeq
When comparing to our numerical solutions, we find that neglecting the $b$-dependence in $q_T$ for $b=4$ typically results in $\sim 70-100\%$ discrepancies from the numerics. The disagreement becomes worse for larger values of $b$, in which case choosing the correct value of $q_T$ becomes particularly important.

Finally, we obtain the oscillation temperature from $m_{\rm osc} = q_T H_{\rm osc}$, assuming that $m_a(T) = m_a (\Lambda/\xi T)^b$ at oscillation. We refer to this oscillation temperature as $T_{\rm osc}(b)$ to distinguish it from the $T$-independent prediction. To obtain $T_{\rm osc}(b)$, we must specify $H(T)$. In RD this is simple. In kination, the temperature of the radiation bath is given by conservation of the comoving entropy density once $T_{\rm kin}$ is specified, and the scale factor evolution from $T_{\rm kin}$ to $T_{\rm osc}$ is determined by Eq.~(\ref{eq:Hosc}). For EMD, one can use the late-time solution for the radiation energy density, Eq.~(\ref{eq:EMDdensitysol}), along with Eq.~(\ref{eq:Hosc}), to write the Hubble parameter in terms of the temperature. Solving for $T_{\rm osc}(b)$ in this way yields Eqs.~(\ref{eq:ToscRD}), (\ref{eq:ToscEMD}), and (\ref{eq:Tosckin}). Inserting these temperatures into $m_a(T)$ and Eq.~(\ref{eq:gamma_app}) yields Eqs.~(\ref{eq:gammaRD}), (\ref{eq:gammaEMD}) and (\ref{eq:gammakin}) for $\gamma_T$.

This is not quite the end of the story for the temperature-dependent scenarios. Our results for $q_T$, $T_{\rm osc}(b)$, and $\gamma_T$ followed from inspecting solutions to the ALP EOM with $m_a(T) = m_a(\Lambda/\xi T)^b $. Therefore, they only apply if $q_T H_{\Lambda} \leq m_a$, where $H_{\Lambda}$ is the Hubble parameter at $T_\Lambda = \Lambda/\xi(T_\Lambda)$, when the mass saturates to its $T$-independent value.  If  instead $q_T H_{\Lambda} > m_a$, oscillations begin after the mass has already saturated to $m_a$, and the ALP EOM solution is given by Eq.~(\ref{eq:ALPsol1}). Then there are two possibilities we must consider. First, if $q_0 H_{\Lambda} > m_a$, we recover the $T$-independent case. Second, if $q_0 H_{\Lambda} < m_a$, at $T_{\Lambda}$ the ALP will start to rapidly oscillate across this threshold, and $H_{\rm osc}$ is given by
\beq
m_a(T_{\rm osc} )= m_a, \quad H_{\rm osc} \simeq H_\Lambda, \quad q= q_{\rm max} \equiv  \frac{m_a}{H_\Lambda}.
\eeq
In this sudden-oscillations case, $T_{\rm osc} = T_\Lambda = \Lambda/\xi_{\rm osc}$ and $\gamma$ is given by
\beq
\gamma_{\rm max} = \left(\frac{m_a}{H_\Lambda q_{0}}\right)^{\frac{2}{w+1}},
\eeq
which yields Eqs.~(\ref{eq:gmax_RD}), (\ref{eq:gmax_EMD}), and (\ref{eq:gmax_kin}).

Summarizing, we have derived the enhancement factor relevant for $T$-dependent ALP masses of the form $m_a(T) = m_a(\Lambda/\xi T)^b $ for three qualitatively different regimes. If $q_T H_{\Lambda} \leq m_a$, $\gamma=\gamma_T$. If instead, $q_T H_{\Lambda} > m_a > q_0 H_{\Lambda}$, oscillations begin suddenly at $T_\Lambda$, and $\gamma$ should instead be taken as $\gamma_{\rm max}$. The condition $q_T H_{\Lambda} > m_a > q_0 H_{\Lambda}$ corresponds to $\gamma_T > \gamma_{\rm max}$. Finally, if $q_T H_{\Lambda}, \,  q_0 H_{\Lambda} > m_a $, one reproduces the $T$-independent case, and $\gamma=1$. This case occurs when both $\gamma_T,\gamma_{\rm max}<1$. Therefore, the correct value of $\gamma$ is given compactly by
\beq
\gamma = \operatorname{max}\left\{ 1, \, \operatorname{min}\left\{\gamma_b, \gamma_{\rm max} \right\} \right\}
\eeq
The temperature $T_{\rm osc}$ should be specified as in Eq.~(\ref{eq:Tosc}). 
 If the resulting $T_{\rm osc}< T_{\rm RH}$ or $T_{\rm kin}$,  the results for radiation domination should be used. Using the above expressions typically reproduces the numerical results to within a few tens of percent across the parameter space. 

\section{Numerics}\label{appx:B}

To fit  coefficients in some of the analytical expressions, and to determine their accuracy, we solve the ALP EOM
\beq\label{eq:ALPeom}
\ddot{\theta} + 3H \dot{\theta} + m_a^2(t) \theta = 0
\eeq
numerically for the EMD, RD, and kination scenarios. We consider times between $t_0$ and $t_f$ with $\theta(t_0) = \theta_0$. The initial time $t_0$ is chosen sufficiently above $t_{\rm osc}$ (defined implicitly by Eq.~(\ref{eq:Tosc})) to capture the start of ALP evolution, while $t_f$ is taken large enough so that we can subsequently approximate the ALP number density as redshifting like nonrelativistic matter. To obtain the relic density today, we  use
\beq\label{eq:numOmega}
\Omega_a = \frac{1}{\rho_c} \left(\frac{T_0}{T(t_f)}\right)^3 \frac{g_{*s}(T_0)}{g_{*s}(T(t_f))} n_a(t_f)
\eeq
where $n_a(t_f)$ is the ALP number density at the time $t_f$:
\beq
n_a(t_f) = \frac{f_a^2}{2m_a^2(t_f)}\left[\dot{\theta}^2(t_f)+m_a^2(t_f) \theta^2(t_f)\right].
\eeq

The various cosmologies predict different relations for $H(t)$ and $m_a(t)$ entering Eq.~(\ref{eq:ALPeom}). The ALP contribution to $H$ is negligible at times near oscillation in the parameter space of interest. We therefore neglect it and model the matter, radiation, and kination components as perfect fluids with equation of state parameters $w=0, \frac{1}{3}$, and $1$, respectively. Furthermore, we treat the SM and HS radiation (if any) as a single fluid with temperatures related by $\xi(T)$, which is assumed to be approximately constant and set by $\xi_{\rm osc}$, taken as an input parameter.  In the standard RD cosmology, we can simply use 
\beq
H(t)=\frac{1}{2t} = \frac{\pi}{\sqrt{90}M_{\rm Pl}} \sqrt{g_*(T)} T^2
\eeq
to determine $T(t)$. In the kination scenario, we have 
\beq\label{eq:Hkin}
H(r) = \frac{\pi}{\sqrt{90} M_{\rm Pl}}\sqrt{g_*(T_{\rm kin})} T^2_{\rm kin} \,r^{-2}  \sqrt{1+r^{-2} }
\eeq
where $r\equiv R/R_{\rm kin}$ is the ratio of the FRW scale factor to its value at $T_{\rm kin}$, defined as the temperature for which the radiation and kination field energy densities are equal. The parameter $r$ is related to $t$ through
\beq
t = t_0 + \int_{r_0}^r \frac{dr'}{r' H(r')} 
\eeq 
where $t_0$ should be taken before ALP oscillation and during the kination phase so that $r_0$ can be defined via $H(r_0) = 1/(3t_0)$. The resulting $r(t)$ yields $H(t)$ through Eq.~(\ref{eq:Hkin}), while $T(t)$ is defined from
\beq
g_*(T) T^4 = g_*(T_{\rm kin}) T_{\rm kin}^4 r^{-4}(t).
\eeq
In modeling a period of early matter domination, the radiation and matter densities are coupled and given as a function of time by numerically solving the system 
\begin{align}
\dot \rho_\phi + 3H \rho_\phi  & = -\Gamma_\phi \rho_\phi \\
\dot\rho_{\rm R} + 4 H \rho_{\rm R} & =+\Gamma_{\phi} \rho_\phi
\end{align}
with $H=1/(\sqrt{3}M_{\rm Pl}) \sqrt{\rho_\phi + \rho_{\rm R} }$ and $\Gamma_{\phi}$ chosen so that $\rho_R = \rho_\phi$ at a temperature $T_{\rm RH}$. The temperature is defined from the radiation energy density $\rho_R  = \pi^2/30 g_*(T) T^4$ in the usual way. The final time is chosen to correspond to temperatures below $T_{\rm RH}$ so that Eq.~(\ref{eq:numOmega}) can be used.

It is worth noting that the numerical treatment of the radiation bath outlined above is technically not entirely correct: near the QCD phase transition (or when the relativistic HS DOFs annihilate) the radiation bath equation of state can deviate somewhat from $\rho = 1/3 p$. Conservation of comoving entropy density can instead be used in this case to track the evolution of $\rho_{\rm R}$ with $t$. However, we find that the corresponding effects are rather small and so we neglect them in our analysis.

\end{widetext}
\bibliography{ALPDD}

\begin{thebibliography}{113}%
\makeatletter
\providecommand \@ifxundefined [1]{%
 \@ifx{#1\undefined}
}%
\providecommand \@ifnum [1]{%
 \ifnum #1\expandafter \@firstoftwo
 \else \expandafter \@secondoftwo
 \fi
}%
\providecommand \@ifx [1]{%
 \ifx #1\expandafter \@firstoftwo
 \else \expandafter \@secondoftwo
 \fi
}%
\providecommand \natexlab [1]{#1}%
\providecommand \enquote  [1]{``#1''}%
\providecommand \bibnamefont  [1]{#1}%
\providecommand \bibfnamefont [1]{#1}%
\providecommand \citenamefont [1]{#1}%
\providecommand \href@noop [0]{\@secondoftwo}%
\providecommand \href [0]{\begingroup \@sanitize@url \@href}%
\providecommand \@href[1]{\@@startlink{#1}\@@href}%
\providecommand \@@href[1]{\endgroup#1\@@endlink}%
\providecommand \@sanitize@url [0]{\catcode `\\12\catcode `\$12\catcode
  `\&12\catcode `\#12\catcode `\^12\catcode `\_12\catcode `\%12\relax}%
\providecommand \@@startlink[1]{}%
\providecommand \@@endlink[0]{}%
\providecommand \url  [0]{\begingroup\@sanitize@url \@url }%
\providecommand \@url [1]{\endgroup\@href {#1}{\urlprefix }}%
\providecommand \urlprefix  [0]{URL }%
\providecommand \Eprint [0]{\href }%
\providecommand \doibase [0]{http://dx.doi.org/}%
\providecommand \selectlanguage [0]{\@gobble}%
\providecommand \bibinfo  [0]{\@secondoftwo}%
\providecommand \bibfield  [0]{\@secondoftwo}%
\providecommand \translation [1]{[#1]}%
\providecommand \BibitemOpen [0]{}%
\providecommand \bibitemStop [0]{}%
\providecommand \bibitemNoStop [0]{.\EOS\space}%
\providecommand \EOS [0]{\spacefactor3000\relax}%
\providecommand \BibitemShut  [1]{\csname bibitem#1\endcsname}%
\let\auto@bib@innerbib\@empty
\bibitem [{\citenamefont {Peccei}\ and\ \citenamefont
  {Quinn}(1977{\natexlab{a}})}]{Peccei:1977hh}%
  \BibitemOpen
  \bibfield  {author} {\bibinfo {author} {\bibfnamefont {R.~D.}\ \bibnamefont
  {Peccei}}\ and\ \bibinfo {author} {\bibfnamefont {H.~R.}\ \bibnamefont
  {Quinn}},\ }\href {\doibase 10.1103/PhysRevLett.38.1440} {\bibfield
  {journal} {\bibinfo  {journal} {Phys. Rev. Lett.}\ }\textbf {\bibinfo
  {volume} {38}},\ \bibinfo {pages} {1440} (\bibinfo {year}
  {1977}{\natexlab{a}})},\ \bibinfo {note} {[,328(1977)]}\BibitemShut {NoStop}%
\bibitem [{\citenamefont {Peccei}\ and\ \citenamefont
  {Quinn}(1977{\natexlab{b}})}]{Peccei:1977ur}%
  \BibitemOpen
  \bibfield  {author} {\bibinfo {author} {\bibfnamefont {R.~D.}\ \bibnamefont
  {Peccei}}\ and\ \bibinfo {author} {\bibfnamefont {H.~R.}\ \bibnamefont
  {Quinn}},\ }\href {\doibase 10.1103/PhysRevD.16.1791} {\bibfield  {journal}
  {\bibinfo  {journal} {Phys. Rev.}\ }\textbf {\bibinfo {volume} {D16}},\
  \bibinfo {pages} {1791} (\bibinfo {year} {1977}{\natexlab{b}})}\BibitemShut
  {NoStop}%
\bibitem [{\citenamefont {Wilczek}(1978)}]{Wilczek:1977pj}%
  \BibitemOpen
  \bibfield  {author} {\bibinfo {author} {\bibfnamefont {F.}~\bibnamefont
  {Wilczek}},\ }\href {\doibase 10.1103/PhysRevLett.40.279} {\bibfield
  {journal} {\bibinfo  {journal} {Phys. Rev. Lett.}\ }\textbf {\bibinfo
  {volume} {40}},\ \bibinfo {pages} {279} (\bibinfo {year} {1978})}\BibitemShut
  {NoStop}%
\bibitem [{\citenamefont {Weinberg}(1978)}]{Weinberg:1977ma}%
  \BibitemOpen
  \bibfield  {author} {\bibinfo {author} {\bibfnamefont {S.}~\bibnamefont
  {Weinberg}},\ }\href {\doibase 10.1103/PhysRevLett.40.223} {\bibfield
  {journal} {\bibinfo  {journal} {Phys. Rev. Lett.}\ }\textbf {\bibinfo
  {volume} {40}},\ \bibinfo {pages} {223} (\bibinfo {year} {1978})}\BibitemShut
  {NoStop}%
\bibitem [{\citenamefont {Abbott}\ and\ \citenamefont
  {Sikivie}(1983)}]{Abbott:1982af}%
  \BibitemOpen
  \bibfield  {author} {\bibinfo {author} {\bibfnamefont {L.~F.}\ \bibnamefont
  {Abbott}}\ and\ \bibinfo {author} {\bibfnamefont {P.}~\bibnamefont
  {Sikivie}},\ }\href {\doibase 10.1016/0370-2693(83)90638-X} {\bibfield
  {journal} {\bibinfo  {journal} {Phys. Lett.}\ }\textbf {\bibinfo {volume}
  {B120}},\ \bibinfo {pages} {133} (\bibinfo {year} {1983})},\ \bibinfo {note}
  {[,URL(1982)]}\BibitemShut {NoStop}%
\bibitem [{\citenamefont {Dine}\ and\ \citenamefont
  {Fischler}(1983)}]{Dine:1982ah}%
  \BibitemOpen
  \bibfield  {author} {\bibinfo {author} {\bibfnamefont {M.}~\bibnamefont
  {Dine}}\ and\ \bibinfo {author} {\bibfnamefont {W.}~\bibnamefont
  {Fischler}},\ }\href {\doibase 10.1016/0370-2693(83)90639-1} {\bibfield
  {journal} {\bibinfo  {journal} {Phys. Lett.}\ }\textbf {\bibinfo {volume}
  {B120}},\ \bibinfo {pages} {137} (\bibinfo {year} {1983})},\ \bibinfo {note}
  {[,URL(1982)]}\BibitemShut {NoStop}%
\bibitem [{\citenamefont {Preskill}\ \emph {et~al.}(1983)\citenamefont
  {Preskill}, \citenamefont {Wise},\ and\ \citenamefont
  {Wilczek}}]{Preskill:1982cy}%
  \BibitemOpen
  \bibfield  {author} {\bibinfo {author} {\bibfnamefont {J.}~\bibnamefont
  {Preskill}}, \bibinfo {author} {\bibfnamefont {M.~B.}\ \bibnamefont {Wise}},
  \ and\ \bibinfo {author} {\bibfnamefont {F.}~\bibnamefont {Wilczek}},\ }\href
  {\doibase 10.1016/0370-2693(83)90637-8} {\bibfield  {journal} {\bibinfo
  {journal} {Phys. Lett.}\ }\textbf {\bibinfo {volume} {B120}},\ \bibinfo
  {pages} {127} (\bibinfo {year} {1983})},\ \bibinfo {note}
  {[,URL(1982)]}\BibitemShut {NoStop}%
\bibitem [{\citenamefont {Arias}\ \emph {et~al.}(2012)\citenamefont {Arias},
  \citenamefont {Cadamuro}, \citenamefont {Goodsell}, \citenamefont {Jaeckel},
  \citenamefont {Redondo},\ and\ \citenamefont {Ringwald}}]{Arias:2012az}%
  \BibitemOpen
  \bibfield  {author} {\bibinfo {author} {\bibfnamefont {P.}~\bibnamefont
  {Arias}}, \bibinfo {author} {\bibfnamefont {D.}~\bibnamefont {Cadamuro}},
  \bibinfo {author} {\bibfnamefont {M.}~\bibnamefont {Goodsell}}, \bibinfo
  {author} {\bibfnamefont {J.}~\bibnamefont {Jaeckel}}, \bibinfo {author}
  {\bibfnamefont {J.}~\bibnamefont {Redondo}}, \ and\ \bibinfo {author}
  {\bibfnamefont {A.}~\bibnamefont {Ringwald}},\ }\href {\doibase
  10.1088/1475-7516/2012/06/013} {\bibfield  {journal} {\bibinfo  {journal}
  {JCAP}\ }\textbf {\bibinfo {volume} {1206}},\ \bibinfo {pages} {013}
  (\bibinfo {year} {2012})},\ \Eprint {http://arxiv.org/abs/1201.5902}
  {arXiv:1201.5902 [hep-ph]} \BibitemShut {NoStop}%
\bibitem [{\citenamefont {Svrcek}\ and\ \citenamefont
  {Witten}(2006)}]{Svrcek:2006yi}%
  \BibitemOpen
  \bibfield  {author} {\bibinfo {author} {\bibfnamefont {P.}~\bibnamefont
  {Svrcek}}\ and\ \bibinfo {author} {\bibfnamefont {E.}~\bibnamefont
  {Witten}},\ }\href {\doibase 10.1088/1126-6708/2006/06/051} {\bibfield
  {journal} {\bibinfo  {journal} {JHEP}\ }\textbf {\bibinfo {volume} {06}},\
  \bibinfo {pages} {051} (\bibinfo {year} {2006})},\ \Eprint
  {http://arxiv.org/abs/hep-th/0605206} {arXiv:hep-th/0605206 [hep-th]}
  \BibitemShut {NoStop}%
\bibitem [{\citenamefont {Arvanitaki}\ \emph {et~al.}(2010)\citenamefont
  {Arvanitaki}, \citenamefont {Dimopoulos}, \citenamefont {Dubovsky},
  \citenamefont {Kaloper},\ and\ \citenamefont
  {March-Russell}}]{Arvanitaki:2009fg}%
  \BibitemOpen
  \bibfield  {author} {\bibinfo {author} {\bibfnamefont {A.}~\bibnamefont
  {Arvanitaki}}, \bibinfo {author} {\bibfnamefont {S.}~\bibnamefont
  {Dimopoulos}}, \bibinfo {author} {\bibfnamefont {S.}~\bibnamefont
  {Dubovsky}}, \bibinfo {author} {\bibfnamefont {N.}~\bibnamefont {Kaloper}}, \
  and\ \bibinfo {author} {\bibfnamefont {J.}~\bibnamefont {March-Russell}},\
  }\href {\doibase 10.1103/PhysRevD.81.123530} {\bibfield  {journal} {\bibinfo
  {journal} {Phys. Rev.}\ }\textbf {\bibinfo {volume} {D81}},\ \bibinfo {pages}
  {123530} (\bibinfo {year} {2010})},\ \Eprint {http://arxiv.org/abs/0905.4720}
  {arXiv:0905.4720 [hep-th]} \BibitemShut {NoStop}%
\bibitem [{\citenamefont {Cicoli}\ \emph {et~al.}(2012)\citenamefont {Cicoli},
  \citenamefont {Goodsell},\ and\ \citenamefont {Ringwald}}]{Cicoli:2012sz}%
  \BibitemOpen
  \bibfield  {author} {\bibinfo {author} {\bibfnamefont {M.}~\bibnamefont
  {Cicoli}}, \bibinfo {author} {\bibfnamefont {M.}~\bibnamefont {Goodsell}}, \
  and\ \bibinfo {author} {\bibfnamefont {A.}~\bibnamefont {Ringwald}},\ }\href
  {\doibase 10.1007/JHEP10(2012)146} {\bibfield  {journal} {\bibinfo  {journal}
  {JHEP}\ }\textbf {\bibinfo {volume} {10}},\ \bibinfo {pages} {146} (\bibinfo
  {year} {2012})},\ \Eprint {http://arxiv.org/abs/1206.0819} {arXiv:1206.0819
  [hep-th]} \BibitemShut {NoStop}%
\bibitem [{\citenamefont {Marsh}(2016)}]{Marsh:2015xka}%
  \BibitemOpen
  \bibfield  {author} {\bibinfo {author} {\bibfnamefont {D.~J.~E.}\
  \bibnamefont {Marsh}},\ }\href {\doibase 10.1016/j.physrep.2016.06.005}
  {\bibfield  {journal} {\bibinfo  {journal} {Phys. Rept.}\ }\textbf {\bibinfo
  {volume} {643}},\ \bibinfo {pages} {1} (\bibinfo {year} {2016})},\ \Eprint
  {http://arxiv.org/abs/1510.07633} {arXiv:1510.07633 [astro-ph.CO]}
  \BibitemShut {NoStop}%
\bibitem [{\citenamefont {Hook}(2018)}]{Hook:2018dlk}%
  \BibitemOpen
  \bibfield  {author} {\bibinfo {author} {\bibfnamefont {A.}~\bibnamefont
  {Hook}},\ }\href@noop {} {\  (\bibinfo {year} {2018})},\ \Eprint
  {http://arxiv.org/abs/1812.02669} {arXiv:1812.02669 [hep-ph]} \BibitemShut
  {NoStop}%
\bibitem [{\citenamefont {Graham}\ \emph {et~al.}(2015)\citenamefont {Graham},
  \citenamefont {Irastorza}, \citenamefont {Lamoreaux}, \citenamefont
  {Lindner},\ and\ \citenamefont {van Bibber}}]{Graham:2015ouw}%
  \BibitemOpen
  \bibfield  {author} {\bibinfo {author} {\bibfnamefont {P.~W.}\ \bibnamefont
  {Graham}}, \bibinfo {author} {\bibfnamefont {I.~G.}\ \bibnamefont
  {Irastorza}}, \bibinfo {author} {\bibfnamefont {S.~K.}\ \bibnamefont
  {Lamoreaux}}, \bibinfo {author} {\bibfnamefont {A.}~\bibnamefont {Lindner}},
  \ and\ \bibinfo {author} {\bibfnamefont {K.~A.}\ \bibnamefont {van Bibber}},\
  }\href {\doibase 10.1146/annurev-nucl-102014-022120} {\bibfield  {journal}
  {\bibinfo  {journal} {Ann. Rev. Nucl. Part. Sci.}\ }\textbf {\bibinfo
  {volume} {65}},\ \bibinfo {pages} {485} (\bibinfo {year} {2015})},\ \Eprint
  {http://arxiv.org/abs/1602.00039} {arXiv:1602.00039 [hep-ex]} \BibitemShut
  {NoStop}%
\bibitem [{\citenamefont {Irastorza}\ and\ \citenamefont
  {Redondo}(2018)}]{Irastorza:2018dyq}%
  \BibitemOpen
  \bibfield  {author} {\bibinfo {author} {\bibfnamefont {I.~G.}\ \bibnamefont
  {Irastorza}}\ and\ \bibinfo {author} {\bibfnamefont {J.}~\bibnamefont
  {Redondo}},\ }\href {\doibase 10.1016/j.ppnp.2018.05.003} {\bibfield
  {journal} {\bibinfo  {journal} {Prog. Part. Nucl. Phys.}\ }\textbf {\bibinfo
  {volume} {102}},\ \bibinfo {pages} {89} (\bibinfo {year} {2018})},\ \Eprint
  {http://arxiv.org/abs/1801.08127} {arXiv:1801.08127 [hep-ph]} \BibitemShut
  {NoStop}%
\bibitem [{\citenamefont {Asztalos}\ \emph {et~al.}(2010)\citenamefont
  {Asztalos} \emph {et~al.}}]{Asztalos:2009yp}%
  \BibitemOpen
  \bibfield  {author} {\bibinfo {author} {\bibfnamefont {S.~J.}\ \bibnamefont
  {Asztalos}} \emph {et~al.} (\bibinfo {collaboration} {ADMX}),\ }\href
  {\doibase 10.1103/PhysRevLett.104.041301} {\bibfield  {journal} {\bibinfo
  {journal} {Phys. Rev. Lett.}\ }\textbf {\bibinfo {volume} {104}},\ \bibinfo
  {pages} {041301} (\bibinfo {year} {2010})},\ \Eprint
  {http://arxiv.org/abs/0910.5914} {arXiv:0910.5914 [astro-ph.CO]} \BibitemShut
  {NoStop}%
\bibitem [{\citenamefont {Du}\ \emph {et~al.}(2018)\citenamefont {Du} \emph
  {et~al.}}]{Du:2018uak}%
  \BibitemOpen
  \bibfield  {author} {\bibinfo {author} {\bibfnamefont {N.}~\bibnamefont {Du}}
  \emph {et~al.} (\bibinfo {collaboration} {ADMX}),\ }\href {\doibase
  10.1103/PhysRevLett.120.151301} {\bibfield  {journal} {\bibinfo  {journal}
  {Phys. Rev. Lett.}\ }\textbf {\bibinfo {volume} {120}},\ \bibinfo {pages}
  {151301} (\bibinfo {year} {2018})},\ \Eprint
  {http://arxiv.org/abs/1804.05750} {arXiv:1804.05750 [hep-ex]} \BibitemShut
  {NoStop}%
\bibitem [{\citenamefont {McAllister}\ \emph {et~al.}(2017)\citenamefont
  {McAllister}, \citenamefont {Flower}, \citenamefont {Ivanov}, \citenamefont
  {Goryachev}, \citenamefont {Bourhill},\ and\ \citenamefont
  {Tobar}}]{McAllister:2017lkb}%
  \BibitemOpen
  \bibfield  {author} {\bibinfo {author} {\bibfnamefont {B.~T.}\ \bibnamefont
  {McAllister}}, \bibinfo {author} {\bibfnamefont {G.}~\bibnamefont {Flower}},
  \bibinfo {author} {\bibfnamefont {E.~N.}\ \bibnamefont {Ivanov}}, \bibinfo
  {author} {\bibfnamefont {M.}~\bibnamefont {Goryachev}}, \bibinfo {author}
  {\bibfnamefont {J.}~\bibnamefont {Bourhill}}, \ and\ \bibinfo {author}
  {\bibfnamefont {M.~E.}\ \bibnamefont {Tobar}},\ }\href {\doibase
  10.1016/j.dark.2017.09.010} {\bibfield  {journal} {\bibinfo  {journal} {Phys.
  Dark Univ.}\ }\textbf {\bibinfo {volume} {18}},\ \bibinfo {pages} {67}
  (\bibinfo {year} {2017})},\ \Eprint {http://arxiv.org/abs/1706.00209}
  {arXiv:1706.00209 [physics.ins-det]} \BibitemShut {NoStop}%
\bibitem [{\citenamefont {Brubaker}\ \emph {et~al.}(2017)\citenamefont
  {Brubaker} \emph {et~al.}}]{Brubaker:2016ktl}%
  \BibitemOpen
  \bibfield  {author} {\bibinfo {author} {\bibfnamefont {B.~M.}\ \bibnamefont
  {Brubaker}} \emph {et~al.},\ }\href {\doibase 10.1103/PhysRevLett.118.061302}
  {\bibfield  {journal} {\bibinfo  {journal} {Phys. Rev. Lett.}\ }\textbf
  {\bibinfo {volume} {118}},\ \bibinfo {pages} {061302} (\bibinfo {year}
  {2017})},\ \Eprint {http://arxiv.org/abs/1610.02580} {arXiv:1610.02580
  [astro-ph.CO]} \BibitemShut {NoStop}%
\bibitem [{\citenamefont {Zhong}\ \emph {et~al.}(2018)\citenamefont {Zhong}
  \emph {et~al.}}]{Zhong:2018rsr}%
  \BibitemOpen
  \bibfield  {author} {\bibinfo {author} {\bibfnamefont {L.}~\bibnamefont
  {Zhong}} \emph {et~al.} (\bibinfo {collaboration} {HAYSTAC}),\ }\href
  {\doibase 10.1103/PhysRevD.97.092001} {\bibfield  {journal} {\bibinfo
  {journal} {Phys. Rev.}\ }\textbf {\bibinfo {volume} {D97}},\ \bibinfo {pages}
  {092001} (\bibinfo {year} {2018})},\ \Eprint
  {http://arxiv.org/abs/1803.03690} {arXiv:1803.03690 [hep-ex]} \BibitemShut
  {NoStop}%
\bibitem [{\citenamefont {Caldwell}\ \emph {et~al.}(2017)\citenamefont
  {Caldwell}, \citenamefont {Dvali}, \citenamefont {Majorovits}, \citenamefont
  {Millar}, \citenamefont {Raffelt}, \citenamefont {Redondo}, \citenamefont
  {Reimann}, \citenamefont {Simon},\ and\ \citenamefont
  {Steffen}}]{TheMADMAXWorkingGroup:2016hpc}%
  \BibitemOpen
  \bibfield  {author} {\bibinfo {author} {\bibfnamefont {A.}~\bibnamefont
  {Caldwell}}, \bibinfo {author} {\bibfnamefont {G.}~\bibnamefont {Dvali}},
  \bibinfo {author} {\bibfnamefont {B.}~\bibnamefont {Majorovits}}, \bibinfo
  {author} {\bibfnamefont {A.}~\bibnamefont {Millar}}, \bibinfo {author}
  {\bibfnamefont {G.}~\bibnamefont {Raffelt}}, \bibinfo {author} {\bibfnamefont
  {J.}~\bibnamefont {Redondo}}, \bibinfo {author} {\bibfnamefont
  {O.}~\bibnamefont {Reimann}}, \bibinfo {author} {\bibfnamefont
  {F.}~\bibnamefont {Simon}}, \ and\ \bibinfo {author} {\bibfnamefont
  {F.}~\bibnamefont {Steffen}} (\bibinfo {collaboration} {MADMAX Working
  Group}),\ }\href {\doibase 10.1103/PhysRevLett.118.091801} {\bibfield
  {journal} {\bibinfo  {journal} {Phys. Rev. Lett.}\ }\textbf {\bibinfo
  {volume} {118}},\ \bibinfo {pages} {091801} (\bibinfo {year} {2017})},\
  \Eprint {http://arxiv.org/abs/1611.05865} {arXiv:1611.05865
  [physics.ins-det]} \BibitemShut {NoStop}%
\bibitem [{\citenamefont {Brun}\ \emph {et~al.}(2019)\citenamefont {Brun} \emph
  {et~al.}}]{Brun:2019lyf}%
  \BibitemOpen
  \bibfield  {author} {\bibinfo {author} {\bibfnamefont {P.}~\bibnamefont
  {Brun}} \emph {et~al.} (\bibinfo {collaboration} {MADMAX}),\ }\href@noop {}
  {\  (\bibinfo {year} {2019})},\ \Eprint {http://arxiv.org/abs/1901.07401}
  {arXiv:1901.07401 [physics.ins-det]} \BibitemShut {NoStop}%
\bibitem [{\citenamefont {Baryakhtar}\ \emph {et~al.}(2018)\citenamefont
  {Baryakhtar}, \citenamefont {Huang},\ and\ \citenamefont
  {Lasenby}}]{Baryakhtar:2018doz}%
  \BibitemOpen
  \bibfield  {author} {\bibinfo {author} {\bibfnamefont {M.}~\bibnamefont
  {Baryakhtar}}, \bibinfo {author} {\bibfnamefont {J.}~\bibnamefont {Huang}}, \
  and\ \bibinfo {author} {\bibfnamefont {R.}~\bibnamefont {Lasenby}},\ }\href
  {\doibase 10.1103/PhysRevD.98.035006} {\bibfield  {journal} {\bibinfo
  {journal} {Phys. Rev.}\ }\textbf {\bibinfo {volume} {D98}},\ \bibinfo {pages}
  {035006} (\bibinfo {year} {2018})},\ \Eprint
  {http://arxiv.org/abs/1803.11455} {arXiv:1803.11455 [hep-ph]} \BibitemShut
  {NoStop}%
\bibitem [{\citenamefont {Chaudhuri}\ \emph {et~al.}(2015)\citenamefont
  {Chaudhuri}, \citenamefont {Graham}, \citenamefont {Irwin}, \citenamefont
  {Mardon}, \citenamefont {Rajendran},\ and\ \citenamefont
  {Zhao}}]{Chaudhuri:2014dla}%
  \BibitemOpen
  \bibfield  {author} {\bibinfo {author} {\bibfnamefont {S.}~\bibnamefont
  {Chaudhuri}}, \bibinfo {author} {\bibfnamefont {P.~W.}\ \bibnamefont
  {Graham}}, \bibinfo {author} {\bibfnamefont {K.}~\bibnamefont {Irwin}},
  \bibinfo {author} {\bibfnamefont {J.}~\bibnamefont {Mardon}}, \bibinfo
  {author} {\bibfnamefont {S.}~\bibnamefont {Rajendran}}, \ and\ \bibinfo
  {author} {\bibfnamefont {Y.}~\bibnamefont {Zhao}},\ }\href {\doibase
  10.1103/PhysRevD.92.075012} {\bibfield  {journal} {\bibinfo  {journal} {Phys.
  Rev.}\ }\textbf {\bibinfo {volume} {D92}},\ \bibinfo {pages} {075012}
  (\bibinfo {year} {2015})},\ \Eprint {http://arxiv.org/abs/1411.7382}
  {arXiv:1411.7382 [hep-ph]} \BibitemShut {NoStop}%
\bibitem [{\citenamefont {Silva-Feaver}\ \emph {et~al.}(2017)\citenamefont
  {Silva-Feaver} \emph {et~al.}}]{Silva-Feaver:2016qhh}%
  \BibitemOpen
  \bibfield  {author} {\bibinfo {author} {\bibfnamefont {M.}~\bibnamefont
  {Silva-Feaver}} \emph {et~al.},\ }\bibfield  {booktitle} {\emph {\bibinfo
  {booktitle} {{Proceedings, Applied Superconductivity Conference (ASC 2016):
  Denver, Colorado, September 4-9, 2016}}},\ }\href {\doibase
  10.1109/TASC.2016.2631425} {\bibfield  {journal} {\bibinfo  {journal} {IEEE
  Trans. Appl. Supercond.}\ }\textbf {\bibinfo {volume} {27}},\ \bibinfo
  {pages} {1400204} (\bibinfo {year} {2017})},\ \Eprint
  {http://arxiv.org/abs/1610.09344} {arXiv:1610.09344 [astro-ph.IM]}
  \BibitemShut {NoStop}%
\bibitem [{\citenamefont {Kahn}\ \emph {et~al.}(2016)\citenamefont {Kahn},
  \citenamefont {Safdi},\ and\ \citenamefont {Thaler}}]{Kahn:2016aff}%
  \BibitemOpen
  \bibfield  {author} {\bibinfo {author} {\bibfnamefont {Y.}~\bibnamefont
  {Kahn}}, \bibinfo {author} {\bibfnamefont {B.~R.}\ \bibnamefont {Safdi}}, \
  and\ \bibinfo {author} {\bibfnamefont {J.}~\bibnamefont {Thaler}},\ }\href
  {\doibase 10.1103/PhysRevLett.117.141801} {\bibfield  {journal} {\bibinfo
  {journal} {Phys. Rev. Lett.}\ }\textbf {\bibinfo {volume} {117}},\ \bibinfo
  {pages} {141801} (\bibinfo {year} {2016})},\ \Eprint
  {http://arxiv.org/abs/1602.01086} {arXiv:1602.01086 [hep-ph]} \BibitemShut
  {NoStop}%
\bibitem [{\citenamefont {Ouellet}\ \emph {et~al.}(2018)\citenamefont {Ouellet}
  \emph {et~al.}}]{Ouellet:2018beu}%
  \BibitemOpen
  \bibfield  {author} {\bibinfo {author} {\bibfnamefont {J.~L.}\ \bibnamefont
  {Ouellet}} \emph {et~al.},\ }\href@noop {} {\  (\bibinfo {year} {2018})},\
  \Eprint {http://arxiv.org/abs/1810.12257} {arXiv:1810.12257 [hep-ex]}
  \BibitemShut {NoStop}%
\bibitem [{\citenamefont {Arvanitaki}\ and\ \citenamefont
  {Geraci}(2014)}]{Arvanitaki:2014dfa}%
  \BibitemOpen
  \bibfield  {author} {\bibinfo {author} {\bibfnamefont {A.}~\bibnamefont
  {Arvanitaki}}\ and\ \bibinfo {author} {\bibfnamefont {A.~A.}\ \bibnamefont
  {Geraci}},\ }\href {\doibase 10.1103/PhysRevLett.113.161801} {\bibfield
  {journal} {\bibinfo  {journal} {Phys. Rev. Lett.}\ }\textbf {\bibinfo
  {volume} {113}},\ \bibinfo {pages} {161801} (\bibinfo {year} {2014})},\
  \Eprint {http://arxiv.org/abs/1403.1290} {arXiv:1403.1290 [hep-ph]}
  \BibitemShut {NoStop}%
\bibitem [{\citenamefont {Geraci}\ \emph {et~al.}(2018)\citenamefont {Geraci}
  \emph {et~al.}}]{Geraci:2017bmq}%
  \BibitemOpen
  \bibfield  {author} {\bibinfo {author} {\bibfnamefont {A.~A.}\ \bibnamefont
  {Geraci}} \emph {et~al.} (\bibinfo {collaboration} {ARIADNE}),\ }\bibfield
  {booktitle} {\emph {\bibinfo {booktitle} {{Proceedings, 2nd Workshop on
  Microwave Cavities and Detectors for Axion Research: Livermore, California,
  USA, January 10-13, 2017}}},\ }\href {\doibase 10.1007/978-3-319-92726-8_18}
  {\bibfield  {journal} {\bibinfo  {journal} {Springer Proc. Phys.}\ }\textbf
  {\bibinfo {volume} {211}},\ \bibinfo {pages} {151} (\bibinfo {year}
  {2018})},\ \Eprint {http://arxiv.org/abs/1710.05413} {arXiv:1710.05413
  [astro-ph.IM]} \BibitemShut {NoStop}%
\bibitem [{\citenamefont {Graham}\ and\ \citenamefont
  {Rajendran}(2013)}]{Graham:2013gfa}%
  \BibitemOpen
  \bibfield  {author} {\bibinfo {author} {\bibfnamefont {P.~W.}\ \bibnamefont
  {Graham}}\ and\ \bibinfo {author} {\bibfnamefont {S.}~\bibnamefont
  {Rajendran}},\ }\href {\doibase 10.1103/PhysRevD.88.035023} {\bibfield
  {journal} {\bibinfo  {journal} {Phys. Rev.}\ }\textbf {\bibinfo {volume}
  {D88}},\ \bibinfo {pages} {035023} (\bibinfo {year} {2013})},\ \Eprint
  {http://arxiv.org/abs/1306.6088} {arXiv:1306.6088 [hep-ph]} \BibitemShut
  {NoStop}%
\bibitem [{\citenamefont {Budker}\ \emph {et~al.}(2014)\citenamefont {Budker},
  \citenamefont {Graham}, \citenamefont {Ledbetter}, \citenamefont
  {Rajendran},\ and\ \citenamefont {Sushkov}}]{Budker:2013hfa}%
  \BibitemOpen
  \bibfield  {author} {\bibinfo {author} {\bibfnamefont {D.}~\bibnamefont
  {Budker}}, \bibinfo {author} {\bibfnamefont {P.~W.}\ \bibnamefont {Graham}},
  \bibinfo {author} {\bibfnamefont {M.}~\bibnamefont {Ledbetter}}, \bibinfo
  {author} {\bibfnamefont {S.}~\bibnamefont {Rajendran}}, \ and\ \bibinfo
  {author} {\bibfnamefont {A.}~\bibnamefont {Sushkov}},\ }\href {\doibase
  10.1103/PhysRevX.4.021030} {\bibfield  {journal} {\bibinfo  {journal} {Phys.
  Rev.}\ }\textbf {\bibinfo {volume} {X4}},\ \bibinfo {pages} {021030}
  (\bibinfo {year} {2014})},\ \Eprint {http://arxiv.org/abs/1306.6089}
  {arXiv:1306.6089 [hep-ph]} \BibitemShut {NoStop}%
\bibitem [{\citenamefont {Garcon}\ \emph {et~al.}(2017)\citenamefont {Garcon}
  \emph {et~al.}}]{Garcon:2017ixh}%
  \BibitemOpen
  \bibfield  {author} {\bibinfo {author} {\bibfnamefont {A.}~\bibnamefont
  {Garcon}} \emph {et~al.},\ }\href {\doibase 10.1088/2058-9565/aa9861} {\
  (\bibinfo {year} {2017}),\ 10.1088/2058-9565/aa9861},\ \Eprint
  {http://arxiv.org/abs/1707.05312} {arXiv:1707.05312 [physics.ins-det]}
  \BibitemShut {NoStop}%
\bibitem [{\citenamefont {Beacham}\ \emph {et~al.}(2019)\citenamefont {Beacham}
  \emph {et~al.}}]{Beacham:2019nyx}%
  \BibitemOpen
  \bibfield  {author} {\bibinfo {author} {\bibfnamefont {J.}~\bibnamefont
  {Beacham}} \emph {et~al.},\ }\href@noop {} {\  (\bibinfo {year} {2019})},\
  \Eprint {http://arxiv.org/abs/1901.09966} {arXiv:1901.09966 [hep-ex]}
  \BibitemShut {NoStop}%
\bibitem [{\citenamefont {Di~Luzio}\ \emph
  {et~al.}(2017{\natexlab{a}})\citenamefont {Di~Luzio}, \citenamefont
  {Mescia},\ and\ \citenamefont {Nardi}}]{DiLuzio:2016sbl}%
  \BibitemOpen
  \bibfield  {author} {\bibinfo {author} {\bibfnamefont {L.}~\bibnamefont
  {Di~Luzio}}, \bibinfo {author} {\bibfnamefont {F.}~\bibnamefont {Mescia}}, \
  and\ \bibinfo {author} {\bibfnamefont {E.}~\bibnamefont {Nardi}},\ }\href
  {\doibase 10.1103/PhysRevLett.118.031801} {\bibfield  {journal} {\bibinfo
  {journal} {Phys. Rev. Lett.}\ }\textbf {\bibinfo {volume} {118}},\ \bibinfo
  {pages} {031801} (\bibinfo {year} {2017}{\natexlab{a}})},\ \Eprint
  {http://arxiv.org/abs/1610.07593} {arXiv:1610.07593 [hep-ph]} \BibitemShut
  {NoStop}%
\bibitem [{\citenamefont {Di~Luzio}\ \emph
  {et~al.}(2017{\natexlab{b}})\citenamefont {Di~Luzio}, \citenamefont
  {Mescia},\ and\ \citenamefont {Nardi}}]{DiLuzio:2017pfr}%
  \BibitemOpen
  \bibfield  {author} {\bibinfo {author} {\bibfnamefont {L.}~\bibnamefont
  {Di~Luzio}}, \bibinfo {author} {\bibfnamefont {F.}~\bibnamefont {Mescia}}, \
  and\ \bibinfo {author} {\bibfnamefont {E.}~\bibnamefont {Nardi}},\ }\href
  {\doibase 10.1103/PhysRevD.96.075003} {\bibfield  {journal} {\bibinfo
  {journal} {Phys. Rev.}\ }\textbf {\bibinfo {volume} {D96}},\ \bibinfo {pages}
  {075003} (\bibinfo {year} {2017}{\natexlab{b}})},\ \Eprint
  {http://arxiv.org/abs/1705.05370} {arXiv:1705.05370 [hep-ph]} \BibitemShut
  {NoStop}%
\bibitem [{\citenamefont {Agrawal}\ \emph
  {et~al.}(2018{\natexlab{a}})\citenamefont {Agrawal}, \citenamefont {Fan},
  \citenamefont {Reece},\ and\ \citenamefont {Wang}}]{Agrawal:2017cmd}%
  \BibitemOpen
  \bibfield  {author} {\bibinfo {author} {\bibfnamefont {P.}~\bibnamefont
  {Agrawal}}, \bibinfo {author} {\bibfnamefont {J.}~\bibnamefont {Fan}},
  \bibinfo {author} {\bibfnamefont {M.}~\bibnamefont {Reece}}, \ and\ \bibinfo
  {author} {\bibfnamefont {L.-T.}\ \bibnamefont {Wang}},\ }\href {\doibase
  10.1007/JHEP02(2018)006} {\bibfield  {journal} {\bibinfo  {journal} {JHEP}\
  }\textbf {\bibinfo {volume} {02}},\ \bibinfo {pages} {006} (\bibinfo {year}
  {2018}{\natexlab{a}})},\ \Eprint {http://arxiv.org/abs/1709.06085}
  {arXiv:1709.06085 [hep-ph]} \BibitemShut {NoStop}%
\bibitem [{\citenamefont {Farina}\ \emph {et~al.}(2017)\citenamefont {Farina},
  \citenamefont {Pappadopulo}, \citenamefont {Rompineve},\ and\ \citenamefont
  {Tesi}}]{Farina:2016tgd}%
  \BibitemOpen
  \bibfield  {author} {\bibinfo {author} {\bibfnamefont {M.}~\bibnamefont
  {Farina}}, \bibinfo {author} {\bibfnamefont {D.}~\bibnamefont {Pappadopulo}},
  \bibinfo {author} {\bibfnamefont {F.}~\bibnamefont {Rompineve}}, \ and\
  \bibinfo {author} {\bibfnamefont {A.}~\bibnamefont {Tesi}},\ }\href {\doibase
  10.1007/JHEP01(2017)095} {\bibfield  {journal} {\bibinfo  {journal} {JHEP}\
  }\textbf {\bibinfo {volume} {01}},\ \bibinfo {pages} {095} (\bibinfo {year}
  {2017})},\ \Eprint {http://arxiv.org/abs/1611.09855} {arXiv:1611.09855
  [hep-ph]} \BibitemShut {NoStop}%
\bibitem [{\citenamefont {Craig}\ \emph {et~al.}(2018)\citenamefont {Craig},
  \citenamefont {Hook},\ and\ \citenamefont {Kasko}}]{Craig:2018kne}%
  \BibitemOpen
  \bibfield  {author} {\bibinfo {author} {\bibfnamefont {N.}~\bibnamefont
  {Craig}}, \bibinfo {author} {\bibfnamefont {A.}~\bibnamefont {Hook}}, \ and\
  \bibinfo {author} {\bibfnamefont {S.}~\bibnamefont {Kasko}},\ }\href
  {\doibase 10.1007/JHEP09(2018)028} {\bibfield  {journal} {\bibinfo  {journal}
  {JHEP}\ }\textbf {\bibinfo {volume} {09}},\ \bibinfo {pages} {028} (\bibinfo
  {year} {2018})},\ \Eprint {http://arxiv.org/abs/1805.06538} {arXiv:1805.06538
  [hep-ph]} \BibitemShut {NoStop}%
\bibitem [{\citenamefont {Agrawal}\ \emph
  {et~al.}(2018{\natexlab{b}})\citenamefont {Agrawal}, \citenamefont
  {Marques-Tavares},\ and\ \citenamefont {Xue}}]{Agrawal:2017eqm}%
  \BibitemOpen
  \bibfield  {author} {\bibinfo {author} {\bibfnamefont {P.}~\bibnamefont
  {Agrawal}}, \bibinfo {author} {\bibfnamefont {G.}~\bibnamefont
  {Marques-Tavares}}, \ and\ \bibinfo {author} {\bibfnamefont {W.}~\bibnamefont
  {Xue}},\ }\href {\doibase 10.1007/JHEP03(2018)049} {\bibfield  {journal}
  {\bibinfo  {journal} {JHEP}\ }\textbf {\bibinfo {volume} {03}},\ \bibinfo
  {pages} {049} (\bibinfo {year} {2018}{\natexlab{b}})},\ \Eprint
  {http://arxiv.org/abs/1708.05008} {arXiv:1708.05008 [hep-ph]} \BibitemShut
  {NoStop}%
\bibitem [{\citenamefont {Co}\ \emph {et~al.}(2018)\citenamefont {Co},
  \citenamefont {Hall},\ and\ \citenamefont {Harigaya}}]{Co:2017mop}%
  \BibitemOpen
  \bibfield  {author} {\bibinfo {author} {\bibfnamefont {R.~T.}\ \bibnamefont
  {Co}}, \bibinfo {author} {\bibfnamefont {L.~J.}\ \bibnamefont {Hall}}, \ and\
  \bibinfo {author} {\bibfnamefont {K.}~\bibnamefont {Harigaya}},\ }\href
  {\doibase 10.1103/PhysRevLett.120.211602} {\bibfield  {journal} {\bibinfo
  {journal} {Phys. Rev. Lett.}\ }\textbf {\bibinfo {volume} {120}},\ \bibinfo
  {pages} {211602} (\bibinfo {year} {2018})},\ \Eprint
  {http://arxiv.org/abs/1711.10486} {arXiv:1711.10486 [hep-ph]} \BibitemShut
  {NoStop}%
\bibitem [{\citenamefont {Visinelli}\ and\ \citenamefont
  {Gondolo}(2010)}]{Visinelli:2009kt}%
  \BibitemOpen
  \bibfield  {author} {\bibinfo {author} {\bibfnamefont {L.}~\bibnamefont
  {Visinelli}}\ and\ \bibinfo {author} {\bibfnamefont {P.}~\bibnamefont
  {Gondolo}},\ }\href {\doibase 10.1103/PhysRevD.81.063508} {\bibfield
  {journal} {\bibinfo  {journal} {Phys. Rev.}\ }\textbf {\bibinfo {volume}
  {D81}},\ \bibinfo {pages} {063508} (\bibinfo {year} {2010})},\ \Eprint
  {http://arxiv.org/abs/0912.0015} {arXiv:0912.0015 [astro-ph.CO]} \BibitemShut
  {NoStop}%
\bibitem [{\citenamefont {Ramberg}\ and\ \citenamefont
  {Visinelli}(2019)}]{Ramberg:2019dgi}%
  \BibitemOpen
  \bibfield  {author} {\bibinfo {author} {\bibfnamefont {N.}~\bibnamefont
  {Ramberg}}\ and\ \bibinfo {author} {\bibfnamefont {L.}~\bibnamefont
  {Visinelli}},\ }\href@noop {} {\  (\bibinfo {year} {2019})},\ \Eprint
  {http://arxiv.org/abs/1904.05707} {arXiv:1904.05707 [astro-ph.CO]}
  \BibitemShut {NoStop}%
\bibitem [{\citenamefont {Visinelli}\ and\ \citenamefont
  {Redondo}(2018)}]{Visinelli:2018wza}%
  \BibitemOpen
  \bibfield  {author} {\bibinfo {author} {\bibfnamefont {L.}~\bibnamefont
  {Visinelli}}\ and\ \bibinfo {author} {\bibfnamefont {J.}~\bibnamefont
  {Redondo}},\ }\href@noop {} {\  (\bibinfo {year} {2018})},\ \Eprint
  {http://arxiv.org/abs/1808.01879} {arXiv:1808.01879 [astro-ph.CO]}
  \BibitemShut {NoStop}%
\bibitem [{\citenamefont {Nelson}\ and\ \citenamefont
  {Xiao}(2018)}]{Nelson:2018via}%
  \BibitemOpen
  \bibfield  {author} {\bibinfo {author} {\bibfnamefont {A.~E.}\ \bibnamefont
  {Nelson}}\ and\ \bibinfo {author} {\bibfnamefont {H.}~\bibnamefont {Xiao}},\
  }\href {\doibase 10.1103/PhysRevD.98.063516} {\bibfield  {journal} {\bibinfo
  {journal} {Phys. Rev.}\ }\textbf {\bibinfo {volume} {D98}},\ \bibinfo {pages}
  {063516} (\bibinfo {year} {2018})},\ \Eprint
  {http://arxiv.org/abs/1807.07176} {arXiv:1807.07176 [astro-ph.CO]}
  \BibitemShut {NoStop}%
\bibitem [{\citenamefont {Draper}\ \emph {et~al.}(2018)\citenamefont {Draper},
  \citenamefont {Kozaczuk},\ and\ \citenamefont {Yu}}]{Draper:2018tmh}%
  \BibitemOpen
  \bibfield  {author} {\bibinfo {author} {\bibfnamefont {P.}~\bibnamefont
  {Draper}}, \bibinfo {author} {\bibfnamefont {J.}~\bibnamefont {Kozaczuk}}, \
  and\ \bibinfo {author} {\bibfnamefont {J.-H.}\ \bibnamefont {Yu}},\ }\href
  {\doibase 10.1103/PhysRevD.98.015028} {\bibfield  {journal} {\bibinfo
  {journal} {Phys. Rev.}\ }\textbf {\bibinfo {volume} {D98}},\ \bibinfo {pages}
  {015028} (\bibinfo {year} {2018})},\ \Eprint
  {http://arxiv.org/abs/1803.00015} {arXiv:1803.00015 [hep-ph]} \BibitemShut
  {NoStop}%
\bibitem [{\citenamefont {Kolb}\ and\ \citenamefont
  {Turner}(1990)}]{Kolb:1990vq}%
  \BibitemOpen
  \bibfield  {author} {\bibinfo {author} {\bibfnamefont {E.~W.}\ \bibnamefont
  {Kolb}}\ and\ \bibinfo {author} {\bibfnamefont {M.~S.}\ \bibnamefont
  {Turner}},\ }\href@noop {} {\bibfield  {journal} {\bibinfo  {journal} {Front.
  Phys.}\ }\textbf {\bibinfo {volume} {69}},\ \bibinfo {pages} {1} (\bibinfo
  {year} {1990})}\BibitemShut {NoStop}%
\bibitem [{\citenamefont {Hiramatsu}\ \emph {et~al.}(2012)\citenamefont
  {Hiramatsu}, \citenamefont {Kawasaki}, \citenamefont {Saikawa},\ and\
  \citenamefont {Sekiguchi}}]{Hiramatsu:2012gg}%
  \BibitemOpen
  \bibfield  {author} {\bibinfo {author} {\bibfnamefont {T.}~\bibnamefont
  {Hiramatsu}}, \bibinfo {author} {\bibfnamefont {M.}~\bibnamefont {Kawasaki}},
  \bibinfo {author} {\bibfnamefont {K.}~\bibnamefont {Saikawa}}, \ and\
  \bibinfo {author} {\bibfnamefont {T.}~\bibnamefont {Sekiguchi}},\ }\href
  {\doibase 10.1103/PhysRevD.86.089902, 10.1103/PhysRevD.85.105020} {\bibfield
  {journal} {\bibinfo  {journal} {Phys. Rev.}\ }\textbf {\bibinfo {volume}
  {D85}},\ \bibinfo {pages} {105020} (\bibinfo {year} {2012})},\ \bibinfo
  {note} {[Erratum: Phys. Rev.D86,089902(2012)]},\ \Eprint
  {http://arxiv.org/abs/1202.5851} {arXiv:1202.5851 [hep-ph]} \BibitemShut
  {NoStop}%
\bibitem [{\citenamefont {Gorghetto}\ \emph {et~al.}(2018)\citenamefont
  {Gorghetto}, \citenamefont {Hardy},\ and\ \citenamefont
  {Villadoro}}]{Gorghetto:2018myk}%
  \BibitemOpen
  \bibfield  {author} {\bibinfo {author} {\bibfnamefont {M.}~\bibnamefont
  {Gorghetto}}, \bibinfo {author} {\bibfnamefont {E.}~\bibnamefont {Hardy}}, \
  and\ \bibinfo {author} {\bibfnamefont {G.}~\bibnamefont {Villadoro}},\ }\href
  {\doibase 10.1007/JHEP07(2018)151} {\bibfield  {journal} {\bibinfo  {journal}
  {JHEP}\ }\textbf {\bibinfo {volume} {07}},\ \bibinfo {pages} {151} (\bibinfo
  {year} {2018})},\ \Eprint {http://arxiv.org/abs/1806.04677} {arXiv:1806.04677
  [hep-ph]} \BibitemShut {NoStop}%
\bibitem [{\citenamefont {Klaer}\ and\ \citenamefont
  {Moore}(2017)}]{Klaer:2017ond}%
  \BibitemOpen
  \bibfield  {author} {\bibinfo {author} {\bibfnamefont {V.~B.}\ \bibnamefont
  {Klaer}}\ and\ \bibinfo {author} {\bibfnamefont {G.~D.}\ \bibnamefont
  {Moore}},\ }\href {\doibase 10.1088/1475-7516/2017/11/049} {\bibfield
  {journal} {\bibinfo  {journal} {JCAP}\ }\textbf {\bibinfo {volume} {1711}},\
  \bibinfo {pages} {049} (\bibinfo {year} {2017})},\ \Eprint
  {http://arxiv.org/abs/1708.07521} {arXiv:1708.07521 [hep-ph]} \BibitemShut
  {NoStop}%
\bibitem [{\citenamefont {Jaeckel}\ \emph {et~al.}(2017)\citenamefont
  {Jaeckel}, \citenamefont {Mehta},\ and\ \citenamefont
  {Witkowski}}]{Jaeckel:2016qjp}%
  \BibitemOpen
  \bibfield  {author} {\bibinfo {author} {\bibfnamefont {J.}~\bibnamefont
  {Jaeckel}}, \bibinfo {author} {\bibfnamefont {V.~M.}\ \bibnamefont {Mehta}},
  \ and\ \bibinfo {author} {\bibfnamefont {L.~T.}\ \bibnamefont {Witkowski}},\
  }\href {\doibase 10.1088/1475-7516/2017/01/036} {\bibfield  {journal}
  {\bibinfo  {journal} {JCAP}\ }\textbf {\bibinfo {volume} {1701}},\ \bibinfo
  {pages} {036} (\bibinfo {year} {2017})},\ \Eprint
  {http://arxiv.org/abs/1605.01367} {arXiv:1605.01367 [hep-ph]} \BibitemShut
  {NoStop}%
\bibitem [{\citenamefont {Berges}\ \emph {et~al.}(2019)\citenamefont {Berges},
  \citenamefont {Chatrchyan},\ and\ \citenamefont {Jaeckel}}]{Berges:2019dgr}%
  \BibitemOpen
  \bibfield  {author} {\bibinfo {author} {\bibfnamefont {J.}~\bibnamefont
  {Berges}}, \bibinfo {author} {\bibfnamefont {A.}~\bibnamefont {Chatrchyan}},
  \ and\ \bibinfo {author} {\bibfnamefont {J.}~\bibnamefont {Jaeckel}},\
  }\href@noop {} {\  (\bibinfo {year} {2019})},\ \Eprint
  {http://arxiv.org/abs/1903.03116} {arXiv:1903.03116 [hep-ph]} \BibitemShut
  {NoStop}%
\bibitem [{\citenamefont {Turner}(1986)}]{Turner:1985si}%
  \BibitemOpen
  \bibfield  {author} {\bibinfo {author} {\bibfnamefont {M.~S.}\ \bibnamefont
  {Turner}},\ }\href {\doibase 10.1103/PhysRevD.33.889} {\bibfield  {journal}
  {\bibinfo  {journal} {Phys. Rev.}\ }\textbf {\bibinfo {volume} {D33}},\
  \bibinfo {pages} {889} (\bibinfo {year} {1986})}\BibitemShut {NoStop}%
\bibitem [{\citenamefont {Tanabashi}\ \emph {et~al.}(2018)\citenamefont
  {Tanabashi} \emph {et~al.}}]{Tanabashi:2018oca}%
  \BibitemOpen
  \bibfield  {author} {\bibinfo {author} {\bibfnamefont {M.}~\bibnamefont
  {Tanabashi}} \emph {et~al.} (\bibinfo {collaboration} {Particle Data
  Group}),\ }\href {\doibase 10.1103/PhysRevD.98.030001} {\bibfield  {journal}
  {\bibinfo  {journal} {Phys. Rev.}\ }\textbf {\bibinfo {volume} {D98}},\
  \bibinfo {pages} {030001} (\bibinfo {year} {2018})}\BibitemShut {NoStop}%
\bibitem [{\citenamefont {Banks}\ and\ \citenamefont
  {Dine}(1997)}]{Banks:1996ea}%
  \BibitemOpen
  \bibfield  {author} {\bibinfo {author} {\bibfnamefont {T.}~\bibnamefont
  {Banks}}\ and\ \bibinfo {author} {\bibfnamefont {M.}~\bibnamefont {Dine}},\
  }\href {\doibase 10.1016/S0550-3213(97)00413-6} {\bibfield  {journal}
  {\bibinfo  {journal} {Nucl. Phys.}\ }\textbf {\bibinfo {volume} {B505}},\
  \bibinfo {pages} {445} (\bibinfo {year} {1997})},\ \Eprint
  {http://arxiv.org/abs/hep-th/9608197} {arXiv:hep-th/9608197 [hep-th]}
  \BibitemShut {NoStop}%
\bibitem [{\citenamefont {Kawasaki}\ \emph {et~al.}(2000)\citenamefont
  {Kawasaki}, \citenamefont {Kohri},\ and\ \citenamefont
  {Sugiyama}}]{Kawasaki:2000en}%
  \BibitemOpen
  \bibfield  {author} {\bibinfo {author} {\bibfnamefont {M.}~\bibnamefont
  {Kawasaki}}, \bibinfo {author} {\bibfnamefont {K.}~\bibnamefont {Kohri}}, \
  and\ \bibinfo {author} {\bibfnamefont {N.}~\bibnamefont {Sugiyama}},\ }\href
  {\doibase 10.1103/PhysRevD.62.023506} {\bibfield  {journal} {\bibinfo
  {journal} {Phys. Rev.}\ }\textbf {\bibinfo {volume} {D62}},\ \bibinfo {pages}
  {023506} (\bibinfo {year} {2000})},\ \Eprint
  {http://arxiv.org/abs/astro-ph/0002127} {arXiv:astro-ph/0002127 [astro-ph]}
  \BibitemShut {NoStop}%
\bibitem [{\citenamefont {Hannestad}(2004)}]{Hannestad:2004px}%
  \BibitemOpen
  \bibfield  {author} {\bibinfo {author} {\bibfnamefont {S.}~\bibnamefont
  {Hannestad}},\ }\href {\doibase 10.1103/PhysRevD.70.043506} {\bibfield
  {journal} {\bibinfo  {journal} {Phys. Rev.}\ }\textbf {\bibinfo {volume}
  {D70}},\ \bibinfo {pages} {043506} (\bibinfo {year} {2004})},\ \Eprint
  {http://arxiv.org/abs/astro-ph/0403291} {arXiv:astro-ph/0403291 [astro-ph]}
  \BibitemShut {NoStop}%
\bibitem [{\citenamefont {Joyce}(1997)}]{Joyce:1996cp}%
  \BibitemOpen
  \bibfield  {author} {\bibinfo {author} {\bibfnamefont {M.}~\bibnamefont
  {Joyce}},\ }\href {\doibase 10.1103/PhysRevD.55.1875} {\bibfield  {journal}
  {\bibinfo  {journal} {Phys. Rev.}\ }\textbf {\bibinfo {volume} {D55}},\
  \bibinfo {pages} {1875} (\bibinfo {year} {1997})},\ \Eprint
  {http://arxiv.org/abs/hep-ph/9606223} {arXiv:hep-ph/9606223 [hep-ph]}
  \BibitemShut {NoStop}%
\bibitem [{\citenamefont {Ferreira}\ and\ \citenamefont
  {Joyce}(1998)}]{Ferreira:1997hj}%
  \BibitemOpen
  \bibfield  {author} {\bibinfo {author} {\bibfnamefont {P.~G.}\ \bibnamefont
  {Ferreira}}\ and\ \bibinfo {author} {\bibfnamefont {M.}~\bibnamefont
  {Joyce}},\ }\href {\doibase 10.1103/PhysRevD.58.023503} {\bibfield  {journal}
  {\bibinfo  {journal} {Phys. Rev.}\ }\textbf {\bibinfo {volume} {D58}},\
  \bibinfo {pages} {023503} (\bibinfo {year} {1998})},\ \Eprint
  {http://arxiv.org/abs/astro-ph/9711102} {arXiv:astro-ph/9711102 [astro-ph]}
  \BibitemShut {NoStop}%
\bibitem [{\citenamefont {Barr}\ and\ \citenamefont
  {Seckel}(1992)}]{Barr:1992qq}%
  \BibitemOpen
  \bibfield  {author} {\bibinfo {author} {\bibfnamefont {S.~M.}\ \bibnamefont
  {Barr}}\ and\ \bibinfo {author} {\bibfnamefont {D.}~\bibnamefont {Seckel}},\
  }\href {\doibase 10.1103/PhysRevD.46.539} {\bibfield  {journal} {\bibinfo
  {journal} {Phys. Rev.}\ }\textbf {\bibinfo {volume} {D46}},\ \bibinfo {pages}
  {539} (\bibinfo {year} {1992})}\BibitemShut {NoStop}%
\bibitem [{\citenamefont {Holman}\ \emph
  {et~al.}(1992{\natexlab{a}})\citenamefont {Holman}, \citenamefont {Hsu},
  \citenamefont {Kephart}, \citenamefont {Kolb}, \citenamefont {Watkins},\ and\
  \citenamefont {Widrow}}]{Holman:1992us}%
  \BibitemOpen
  \bibfield  {author} {\bibinfo {author} {\bibfnamefont {R.}~\bibnamefont
  {Holman}}, \bibinfo {author} {\bibfnamefont {S.~D.~H.}\ \bibnamefont {Hsu}},
  \bibinfo {author} {\bibfnamefont {T.~W.}\ \bibnamefont {Kephart}}, \bibinfo
  {author} {\bibfnamefont {E.~W.}\ \bibnamefont {Kolb}}, \bibinfo {author}
  {\bibfnamefont {R.}~\bibnamefont {Watkins}}, \ and\ \bibinfo {author}
  {\bibfnamefont {L.~M.}\ \bibnamefont {Widrow}},\ }\href {\doibase
  10.1016/0370-2693(92)90491-L} {\bibfield  {journal} {\bibinfo  {journal}
  {Phys. Lett.}\ }\textbf {\bibinfo {volume} {B282}},\ \bibinfo {pages} {132}
  (\bibinfo {year} {1992}{\natexlab{a}})},\ \Eprint
  {http://arxiv.org/abs/hep-ph/9203206} {arXiv:hep-ph/9203206 [hep-ph]}
  \BibitemShut {NoStop}%
\bibitem [{\citenamefont {Holman}\ \emph
  {et~al.}(1992{\natexlab{b}})\citenamefont {Holman}, \citenamefont {Hsu},
  \citenamefont {Kolb}, \citenamefont {Watkins},\ and\ \citenamefont
  {Widrow}}]{Holman:1992va}%
  \BibitemOpen
  \bibfield  {author} {\bibinfo {author} {\bibfnamefont {R.}~\bibnamefont
  {Holman}}, \bibinfo {author} {\bibfnamefont {S.~D.~H.}\ \bibnamefont {Hsu}},
  \bibinfo {author} {\bibfnamefont {E.~W.}\ \bibnamefont {Kolb}}, \bibinfo
  {author} {\bibfnamefont {R.}~\bibnamefont {Watkins}}, \ and\ \bibinfo
  {author} {\bibfnamefont {L.~M.}\ \bibnamefont {Widrow}},\ }\href {\doibase
  10.1103/PhysRevLett.69.1489} {\bibfield  {journal} {\bibinfo  {journal}
  {Phys. Rev. Lett.}\ }\textbf {\bibinfo {volume} {69}},\ \bibinfo {pages}
  {1489} (\bibinfo {year} {1992}{\natexlab{b}})}\BibitemShut {NoStop}%
\bibitem [{\citenamefont {Kamionkowski}\ and\ \citenamefont
  {March-Russell}(1992{\natexlab{a}})}]{Kamionkowski:1992ax}%
  \BibitemOpen
  \bibfield  {author} {\bibinfo {author} {\bibfnamefont {M.}~\bibnamefont
  {Kamionkowski}}\ and\ \bibinfo {author} {\bibfnamefont {J.}~\bibnamefont
  {March-Russell}},\ }\href {\doibase 10.1103/PhysRevLett.69.1485} {\bibfield
  {journal} {\bibinfo  {journal} {Phys. Rev. Lett.}\ }\textbf {\bibinfo
  {volume} {69}},\ \bibinfo {pages} {1485} (\bibinfo {year}
  {1992}{\natexlab{a}})},\ \Eprint {http://arxiv.org/abs/hep-th/9201063}
  {arXiv:hep-th/9201063 [hep-th]} \BibitemShut {NoStop}%
\bibitem [{\citenamefont {Kamionkowski}\ and\ \citenamefont
  {March-Russell}(1992{\natexlab{b}})}]{Kamionkowski:1992mf}%
  \BibitemOpen
  \bibfield  {author} {\bibinfo {author} {\bibfnamefont {M.}~\bibnamefont
  {Kamionkowski}}\ and\ \bibinfo {author} {\bibfnamefont {J.}~\bibnamefont
  {March-Russell}},\ }\href {\doibase 10.1016/0370-2693(92)90492-M} {\bibfield
  {journal} {\bibinfo  {journal} {Phys. Lett.}\ }\textbf {\bibinfo {volume}
  {B282}},\ \bibinfo {pages} {137} (\bibinfo {year} {1992}{\natexlab{b}})},\
  \Eprint {http://arxiv.org/abs/hep-th/9202003} {arXiv:hep-th/9202003 [hep-th]}
  \BibitemShut {NoStop}%
\bibitem [{\citenamefont {Aghanim}\ \emph {et~al.}(2018)\citenamefont {Aghanim}
  \emph {et~al.}}]{Aghanim:2018eyx}%
  \BibitemOpen
  \bibfield  {author} {\bibinfo {author} {\bibfnamefont {N.}~\bibnamefont
  {Aghanim}} \emph {et~al.} (\bibinfo {collaboration} {Planck}),\ }\href@noop
  {} {\  (\bibinfo {year} {2018})},\ \Eprint {http://arxiv.org/abs/1807.06209}
  {arXiv:1807.06209 [astro-ph.CO]} \BibitemShut {NoStop}%
\bibitem [{\citenamefont {Feng}\ and\ \citenamefont
  {Shadmi}(2011)}]{Feng:2011ik}%
  \BibitemOpen
  \bibfield  {author} {\bibinfo {author} {\bibfnamefont {J.~L.}\ \bibnamefont
  {Feng}}\ and\ \bibinfo {author} {\bibfnamefont {Y.}~\bibnamefont {Shadmi}},\
  }\href {\doibase 10.1103/PhysRevD.83.095011} {\bibfield  {journal} {\bibinfo
  {journal} {Phys. Rev.}\ }\textbf {\bibinfo {volume} {D83}},\ \bibinfo {pages}
  {095011} (\bibinfo {year} {2011})},\ \Eprint {http://arxiv.org/abs/1102.0282}
  {arXiv:1102.0282 [hep-ph]} \BibitemShut {NoStop}%
\bibitem [{\citenamefont {Cline}\ \emph {et~al.}(2014)\citenamefont {Cline},
  \citenamefont {Liu}, \citenamefont {Moore},\ and\ \citenamefont
  {Xue}}]{Cline:2013zca}%
  \BibitemOpen
  \bibfield  {author} {\bibinfo {author} {\bibfnamefont {J.~M.}\ \bibnamefont
  {Cline}}, \bibinfo {author} {\bibfnamefont {Z.}~\bibnamefont {Liu}}, \bibinfo
  {author} {\bibfnamefont {G.}~\bibnamefont {Moore}}, \ and\ \bibinfo {author}
  {\bibfnamefont {W.}~\bibnamefont {Xue}},\ }\href {\doibase
  10.1103/PhysRevD.90.015023} {\bibfield  {journal} {\bibinfo  {journal} {Phys.
  Rev.}\ }\textbf {\bibinfo {volume} {D90}},\ \bibinfo {pages} {015023}
  (\bibinfo {year} {2014})},\ \Eprint {http://arxiv.org/abs/1312.3325}
  {arXiv:1312.3325 [hep-ph]} \BibitemShut {NoStop}%
\bibitem [{\citenamefont {Boddy}\ \emph {et~al.}(2014)\citenamefont {Boddy},
  \citenamefont {Feng}, \citenamefont {Kaplinghat},\ and\ \citenamefont
  {Tait}}]{Boddy:2014yra}%
  \BibitemOpen
  \bibfield  {author} {\bibinfo {author} {\bibfnamefont {K.~K.}\ \bibnamefont
  {Boddy}}, \bibinfo {author} {\bibfnamefont {J.~L.}\ \bibnamefont {Feng}},
  \bibinfo {author} {\bibfnamefont {M.}~\bibnamefont {Kaplinghat}}, \ and\
  \bibinfo {author} {\bibfnamefont {T.~M.~P.}\ \bibnamefont {Tait}},\ }\href
  {\doibase 10.1103/PhysRevD.89.115017} {\bibfield  {journal} {\bibinfo
  {journal} {Phys. Rev.}\ }\textbf {\bibinfo {volume} {D89}},\ \bibinfo {pages}
  {115017} (\bibinfo {year} {2014})},\ \Eprint {http://arxiv.org/abs/1402.3629}
  {arXiv:1402.3629 [hep-ph]} \BibitemShut {NoStop}%
\bibitem [{\citenamefont {Hochberg}\ \emph {et~al.}(2014)\citenamefont
  {Hochberg}, \citenamefont {Kuflik}, \citenamefont {Volansky},\ and\
  \citenamefont {Wacker}}]{Hochberg:2014dra}%
  \BibitemOpen
  \bibfield  {author} {\bibinfo {author} {\bibfnamefont {Y.}~\bibnamefont
  {Hochberg}}, \bibinfo {author} {\bibfnamefont {E.}~\bibnamefont {Kuflik}},
  \bibinfo {author} {\bibfnamefont {T.}~\bibnamefont {Volansky}}, \ and\
  \bibinfo {author} {\bibfnamefont {J.~G.}\ \bibnamefont {Wacker}},\ }\href
  {\doibase 10.1103/PhysRevLett.113.171301} {\bibfield  {journal} {\bibinfo
  {journal} {Phys. Rev. Lett.}\ }\textbf {\bibinfo {volume} {113}},\ \bibinfo
  {pages} {171301} (\bibinfo {year} {2014})},\ \Eprint
  {http://arxiv.org/abs/1402.5143} {arXiv:1402.5143 [hep-ph]} \BibitemShut
  {NoStop}%
\bibitem [{\citenamefont {Forestell}\ \emph {et~al.}(2018)\citenamefont
  {Forestell}, \citenamefont {Morrissey},\ and\ \citenamefont
  {Sigurdson}}]{Forestell:2017wov}%
  \BibitemOpen
  \bibfield  {author} {\bibinfo {author} {\bibfnamefont {L.}~\bibnamefont
  {Forestell}}, \bibinfo {author} {\bibfnamefont {D.~E.}\ \bibnamefont
  {Morrissey}}, \ and\ \bibinfo {author} {\bibfnamefont {K.}~\bibnamefont
  {Sigurdson}},\ }\href {\doibase 10.1103/PhysRevD.97.075029} {\bibfield
  {journal} {\bibinfo  {journal} {Phys. Rev.}\ }\textbf {\bibinfo {volume}
  {D97}},\ \bibinfo {pages} {075029} (\bibinfo {year} {2018})},\ \Eprint
  {http://arxiv.org/abs/1710.06447} {arXiv:1710.06447 [hep-ph]} \BibitemShut
  {NoStop}%
\bibitem [{\citenamefont {Berlin}\ \emph {et~al.}(2018)\citenamefont {Berlin},
  \citenamefont {Blinov}, \citenamefont {Gori}, \citenamefont {Schuster},\ and\
  \citenamefont {Toro}}]{Berlin:2018tvf}%
  \BibitemOpen
  \bibfield  {author} {\bibinfo {author} {\bibfnamefont {A.}~\bibnamefont
  {Berlin}}, \bibinfo {author} {\bibfnamefont {N.}~\bibnamefont {Blinov}},
  \bibinfo {author} {\bibfnamefont {S.}~\bibnamefont {Gori}}, \bibinfo {author}
  {\bibfnamefont {P.}~\bibnamefont {Schuster}}, \ and\ \bibinfo {author}
  {\bibfnamefont {N.}~\bibnamefont {Toro}},\ }\href {\doibase
  10.1103/PhysRevD.97.055033} {\bibfield  {journal} {\bibinfo  {journal} {Phys.
  Rev.}\ }\textbf {\bibinfo {volume} {D97}},\ \bibinfo {pages} {055033}
  (\bibinfo {year} {2018})},\ \Eprint {http://arxiv.org/abs/1801.05805}
  {arXiv:1801.05805 [hep-ph]} \BibitemShut {NoStop}%
\bibitem [{\citenamefont {Gross}\ \emph {et~al.}(1981)\citenamefont {Gross},
  \citenamefont {Pisarski},\ and\ \citenamefont {Yaffe}}]{Gross:1980br}%
  \BibitemOpen
  \bibfield  {author} {\bibinfo {author} {\bibfnamefont {D.~J.}\ \bibnamefont
  {Gross}}, \bibinfo {author} {\bibfnamefont {R.~D.}\ \bibnamefont {Pisarski}},
  \ and\ \bibinfo {author} {\bibfnamefont {L.~G.}\ \bibnamefont {Yaffe}},\
  }\href {\doibase 10.1103/RevModPhys.53.43} {\bibfield  {journal} {\bibinfo
  {journal} {Rev. Mod. Phys.}\ }\textbf {\bibinfo {volume} {53}},\ \bibinfo
  {pages} {43} (\bibinfo {year} {1981})}\BibitemShut {NoStop}%
\bibitem [{\citenamefont {Borsanyi}\ \emph
  {et~al.}(2016{\natexlab{a}})\citenamefont {Borsanyi} \emph
  {et~al.}}]{borsanyinature}%
  \BibitemOpen
  \bibfield  {author} {\bibinfo {author} {\bibfnamefont {S.}~\bibnamefont
  {Borsanyi}} \emph {et~al.},\ }\href {\doibase 10.1038/nature20115} {\bibfield
   {journal} {\bibinfo  {journal} {Nature}\ }\textbf {\bibinfo {volume}
  {539}},\ \bibinfo {pages} {69} (\bibinfo {year} {2016}{\natexlab{a}})},\
  \Eprint {http://arxiv.org/abs/1606.07494} {arXiv:1606.07494 [hep-lat]}
  \BibitemShut {NoStop}%
\bibitem [{\citenamefont {Borsanyi}\ \emph
  {et~al.}(2016{\natexlab{b}})\citenamefont {Borsanyi}, \citenamefont
  {Dierigl}, \citenamefont {Fodor}, \citenamefont {Katz}, \citenamefont
  {Mages}, \citenamefont {Nogradi}, \citenamefont {Redondo}, \citenamefont
  {Ringwald},\ and\ \citenamefont {Szabo}}]{borsanyia}%
  \BibitemOpen
  \bibfield  {author} {\bibinfo {author} {\bibfnamefont {S.}~\bibnamefont
  {Borsanyi}}, \bibinfo {author} {\bibfnamefont {M.}~\bibnamefont {Dierigl}},
  \bibinfo {author} {\bibfnamefont {Z.}~\bibnamefont {Fodor}}, \bibinfo
  {author} {\bibfnamefont {S.~D.}\ \bibnamefont {Katz}}, \bibinfo {author}
  {\bibfnamefont {S.~W.}\ \bibnamefont {Mages}}, \bibinfo {author}
  {\bibfnamefont {D.}~\bibnamefont {Nogradi}}, \bibinfo {author} {\bibfnamefont
  {J.}~\bibnamefont {Redondo}}, \bibinfo {author} {\bibfnamefont
  {A.}~\bibnamefont {Ringwald}}, \ and\ \bibinfo {author} {\bibfnamefont
  {K.~K.}\ \bibnamefont {Szabo}},\ }\href {\doibase
  10.1016/j.physletb.2015.11.020} {\bibfield  {journal} {\bibinfo  {journal}
  {Phys. Lett.}\ }\textbf {\bibinfo {volume} {B752}},\ \bibinfo {pages} {175}
  (\bibinfo {year} {2016}{\natexlab{b}})},\ \Eprint
  {http://arxiv.org/abs/1508.06917} {arXiv:1508.06917 [hep-lat]} \BibitemShut
  {NoStop}%
\bibitem [{\citenamefont {Dine}\ \emph {et~al.}(2017)\citenamefont {Dine},
  \citenamefont {Draper}, \citenamefont {Stephenson-Haskins},\ and\
  \citenamefont {Xu}}]{Dine:2017swf}%
  \BibitemOpen
  \bibfield  {author} {\bibinfo {author} {\bibfnamefont {M.}~\bibnamefont
  {Dine}}, \bibinfo {author} {\bibfnamefont {P.}~\bibnamefont {Draper}},
  \bibinfo {author} {\bibfnamefont {L.}~\bibnamefont {Stephenson-Haskins}}, \
  and\ \bibinfo {author} {\bibfnamefont {D.}~\bibnamefont {Xu}},\ }\href
  {\doibase 10.1103/PhysRevD.96.095001} {\bibfield  {journal} {\bibinfo
  {journal} {Phys. Rev.}\ }\textbf {\bibinfo {volume} {D96}},\ \bibinfo {pages}
  {095001} (\bibinfo {year} {2017})},\ \Eprint
  {http://arxiv.org/abs/1705.00676} {arXiv:1705.00676 [hep-ph]} \BibitemShut
  {NoStop}%
\bibitem [{\citenamefont {Adshead}\ \emph {et~al.}(2016)\citenamefont
  {Adshead}, \citenamefont {Cui},\ and\ \citenamefont
  {Shelton}}]{Adshead:2016xxj}%
  \BibitemOpen
  \bibfield  {author} {\bibinfo {author} {\bibfnamefont {P.}~\bibnamefont
  {Adshead}}, \bibinfo {author} {\bibfnamefont {Y.}~\bibnamefont {Cui}}, \ and\
  \bibinfo {author} {\bibfnamefont {J.}~\bibnamefont {Shelton}},\ }\href
  {\doibase 10.1007/JHEP06(2016)016} {\bibfield  {journal} {\bibinfo  {journal}
  {JHEP}\ }\textbf {\bibinfo {volume} {06}},\ \bibinfo {pages} {016} (\bibinfo
  {year} {2016})},\ \Eprint {http://arxiv.org/abs/1604.02458} {arXiv:1604.02458
  [hep-ph]} \BibitemShut {NoStop}%
\bibitem [{\citenamefont {Anastassopoulos}\ \emph {et~al.}(2017)\citenamefont
  {Anastassopoulos} \emph {et~al.}}]{Anastassopoulos:2017ftl}%
  \BibitemOpen
  \bibfield  {author} {\bibinfo {author} {\bibfnamefont {V.}~\bibnamefont
  {Anastassopoulos}} \emph {et~al.} (\bibinfo {collaboration} {CAST}),\ }\href
  {\doibase 10.1038/nphys4109} {\bibfield  {journal} {\bibinfo  {journal}
  {Nature Phys.}\ }\textbf {\bibinfo {volume} {13}},\ \bibinfo {pages} {584}
  (\bibinfo {year} {2017})},\ \Eprint {http://arxiv.org/abs/1705.02290}
  {arXiv:1705.02290 [hep-ex]} \BibitemShut {NoStop}%
\bibitem [{\citenamefont {Cadamuro}\ and\ \citenamefont
  {Redondo}(2012)}]{Cadamuro:2011fd}%
  \BibitemOpen
  \bibfield  {author} {\bibinfo {author} {\bibfnamefont {D.}~\bibnamefont
  {Cadamuro}}\ and\ \bibinfo {author} {\bibfnamefont {J.}~\bibnamefont
  {Redondo}},\ }\href {\doibase 10.1088/1475-7516/2012/02/032} {\bibfield
  {journal} {\bibinfo  {journal} {JCAP}\ }\textbf {\bibinfo {volume} {1202}},\
  \bibinfo {pages} {032} (\bibinfo {year} {2012})},\ \Eprint
  {http://arxiv.org/abs/1110.2895} {arXiv:1110.2895 [hep-ph]} \BibitemShut
  {NoStop}%
\bibitem [{\citenamefont {Friedland}\ \emph {et~al.}(2013)\citenamefont
  {Friedland}, \citenamefont {Giannotti},\ and\ \citenamefont
  {Wise}}]{Friedland:2012hj}%
  \BibitemOpen
  \bibfield  {author} {\bibinfo {author} {\bibfnamefont {A.}~\bibnamefont
  {Friedland}}, \bibinfo {author} {\bibfnamefont {M.}~\bibnamefont
  {Giannotti}}, \ and\ \bibinfo {author} {\bibfnamefont {M.}~\bibnamefont
  {Wise}},\ }\href {\doibase 10.1103/PhysRevLett.110.061101} {\bibfield
  {journal} {\bibinfo  {journal} {Phys. Rev. Lett.}\ }\textbf {\bibinfo
  {volume} {110}},\ \bibinfo {pages} {061101} (\bibinfo {year} {2013})},\
  \Eprint {http://arxiv.org/abs/1210.1271} {arXiv:1210.1271 [hep-ph]}
  \BibitemShut {NoStop}%
\bibitem [{\citenamefont {Raffelt}(1996)}]{Raffelt:1996wa}%
  \BibitemOpen
  \bibfield  {author} {\bibinfo {author} {\bibfnamefont {G.~G.}\ \bibnamefont
  {Raffelt}},\ }\href
  {http://wwwth.mpp.mpg.de/members/raffelt/mypapers/199613.pdf} {\emph
  {\bibinfo {title} {{Stars as laboratories for fundamental physics}}}}\
  (\bibinfo {year} {1996})\BibitemShut {NoStop}%
\bibitem [{\citenamefont {Dolan}\ \emph {et~al.}(2017)\citenamefont {Dolan},
  \citenamefont {Ferber}, \citenamefont {Hearty}, \citenamefont {Kahlhoefer},\
  and\ \citenamefont {Schmidt-Hoberg}}]{Dolan:2017osp}%
  \BibitemOpen
  \bibfield  {author} {\bibinfo {author} {\bibfnamefont {M.~J.}\ \bibnamefont
  {Dolan}}, \bibinfo {author} {\bibfnamefont {T.}~\bibnamefont {Ferber}},
  \bibinfo {author} {\bibfnamefont {C.}~\bibnamefont {Hearty}}, \bibinfo
  {author} {\bibfnamefont {F.}~\bibnamefont {Kahlhoefer}}, \ and\ \bibinfo
  {author} {\bibfnamefont {K.}~\bibnamefont {Schmidt-Hoberg}},\ }\href
  {\doibase 10.1007/JHEP12(2017)094} {\bibfield  {journal} {\bibinfo  {journal}
  {JHEP}\ }\textbf {\bibinfo {volume} {12}},\ \bibinfo {pages} {094} (\bibinfo
  {year} {2017})},\ \Eprint {http://arxiv.org/abs/1709.00009} {arXiv:1709.00009
  [hep-ph]} \BibitemShut {NoStop}%
\bibitem [{\citenamefont {Lee}(2018)}]{Lee:2018lcj}%
  \BibitemOpen
  \bibfield  {author} {\bibinfo {author} {\bibfnamefont {J.~S.}\ \bibnamefont
  {Lee}},\ }\href@noop {} {\  (\bibinfo {year} {2018})},\ \Eprint
  {http://arxiv.org/abs/1808.10136} {arXiv:1808.10136 [hep-ph]} \BibitemShut
  {NoStop}%
\bibitem [{\citenamefont {Payez}\ \emph {et~al.}(2015)\citenamefont {Payez},
  \citenamefont {Evoli}, \citenamefont {Fischer}, \citenamefont {Giannotti},
  \citenamefont {Mirizzi},\ and\ \citenamefont {Ringwald}}]{Payez:2014xsa}%
  \BibitemOpen
  \bibfield  {author} {\bibinfo {author} {\bibfnamefont {A.}~\bibnamefont
  {Payez}}, \bibinfo {author} {\bibfnamefont {C.}~\bibnamefont {Evoli}},
  \bibinfo {author} {\bibfnamefont {T.}~\bibnamefont {Fischer}}, \bibinfo
  {author} {\bibfnamefont {M.}~\bibnamefont {Giannotti}}, \bibinfo {author}
  {\bibfnamefont {A.}~\bibnamefont {Mirizzi}}, \ and\ \bibinfo {author}
  {\bibfnamefont {A.}~\bibnamefont {Ringwald}},\ }\href {\doibase
  10.1088/1475-7516/2015/02/006} {\bibfield  {journal} {\bibinfo  {journal}
  {JCAP}\ }\textbf {\bibinfo {volume} {1502}},\ \bibinfo {pages} {006}
  (\bibinfo {year} {2015})},\ \Eprint {http://arxiv.org/abs/1410.3747}
  {arXiv:1410.3747 [astro-ph.HE]} \BibitemShut {NoStop}%
\bibitem [{\citenamefont {Masso}\ and\ \citenamefont
  {Toldra}(1997)}]{Masso:1997ru}%
  \BibitemOpen
  \bibfield  {author} {\bibinfo {author} {\bibfnamefont {E.}~\bibnamefont
  {Masso}}\ and\ \bibinfo {author} {\bibfnamefont {R.}~\bibnamefont {Toldra}},\
  }\href {\doibase 10.1103/PhysRevD.55.7967} {\bibfield  {journal} {\bibinfo
  {journal} {Phys. Rev.}\ }\textbf {\bibinfo {volume} {D55}},\ \bibinfo {pages}
  {7967} (\bibinfo {year} {1997})},\ \Eprint
  {http://arxiv.org/abs/hep-ph/9702275} {arXiv:hep-ph/9702275 [hep-ph]}
  \BibitemShut {NoStop}%
\bibitem [{\citenamefont {Overduin}\ and\ \citenamefont
  {Wesson}(2004)}]{Overduin:2004sz}%
  \BibitemOpen
  \bibfield  {author} {\bibinfo {author} {\bibfnamefont {J.~M.}\ \bibnamefont
  {Overduin}}\ and\ \bibinfo {author} {\bibfnamefont {P.~S.}\ \bibnamefont
  {Wesson}},\ }\href {\doibase 10.1016/j.physrep.2004.07.006} {\bibfield
  {journal} {\bibinfo  {journal} {Phys. Rept.}\ }\textbf {\bibinfo {volume}
  {402}},\ \bibinfo {pages} {267} (\bibinfo {year} {2004})},\ \Eprint
  {http://arxiv.org/abs/astro-ph/0407207} {arXiv:astro-ph/0407207 [astro-ph]}
  \BibitemShut {NoStop}%
\bibitem [{\citenamefont {Abramowski}\ \emph {et~al.}(2013)\citenamefont
  {Abramowski} \emph {et~al.}}]{Abramowski:2013oea}%
  \BibitemOpen
  \bibfield  {author} {\bibinfo {author} {\bibfnamefont {A.}~\bibnamefont
  {Abramowski}} \emph {et~al.} (\bibinfo {collaboration} {H.E.S.S.}),\ }\href
  {\doibase 10.1103/PhysRevD.88.102003} {\bibfield  {journal} {\bibinfo
  {journal} {Phys. Rev.}\ }\textbf {\bibinfo {volume} {D88}},\ \bibinfo {pages}
  {102003} (\bibinfo {year} {2013})},\ \Eprint {http://arxiv.org/abs/1311.3148}
  {arXiv:1311.3148 [astro-ph.HE]} \BibitemShut {NoStop}%
\bibitem [{\citenamefont {Ajello}\ \emph {et~al.}(2016)\citenamefont {Ajello}
  \emph {et~al.}}]{TheFermi-LAT:2016zue}%
  \BibitemOpen
  \bibfield  {author} {\bibinfo {author} {\bibfnamefont {M.}~\bibnamefont
  {Ajello}} \emph {et~al.} (\bibinfo {collaboration} {Fermi-LAT}),\ }\href
  {\doibase 10.1103/PhysRevLett.116.161101} {\bibfield  {journal} {\bibinfo
  {journal} {Phys. Rev. Lett.}\ }\textbf {\bibinfo {volume} {116}},\ \bibinfo
  {pages} {161101} (\bibinfo {year} {2016})},\ \Eprint
  {http://arxiv.org/abs/1603.06978} {arXiv:1603.06978 [astro-ph.HE]}
  \BibitemShut {NoStop}%
\bibitem [{\citenamefont {Chen}\ and\ \citenamefont
  {Conlon}(2018)}]{Chen:2017mjf}%
  \BibitemOpen
  \bibfield  {author} {\bibinfo {author} {\bibfnamefont {L.}~\bibnamefont
  {Chen}}\ and\ \bibinfo {author} {\bibfnamefont {J.~P.}\ \bibnamefont
  {Conlon}},\ }\href {\doibase 10.1093/mnras/sty1591} {\bibfield  {journal}
  {\bibinfo  {journal} {Mon. Not. Roy. Astron. Soc.}\ }\textbf {\bibinfo
  {volume} {479}},\ \bibinfo {pages} {2243} (\bibinfo {year} {2018})},\ \Eprint
  {http://arxiv.org/abs/1712.08313} {arXiv:1712.08313 [astro-ph.HE]}
  \BibitemShut {NoStop}%
\bibitem [{\citenamefont {Arvanitaki}\ \emph {et~al.}(2015)\citenamefont
  {Arvanitaki}, \citenamefont {Baryakhtar},\ and\ \citenamefont
  {Huang}}]{Arvanitaki:2014wva}%
  \BibitemOpen
  \bibfield  {author} {\bibinfo {author} {\bibfnamefont {A.}~\bibnamefont
  {Arvanitaki}}, \bibinfo {author} {\bibfnamefont {M.}~\bibnamefont
  {Baryakhtar}}, \ and\ \bibinfo {author} {\bibfnamefont {X.}~\bibnamefont
  {Huang}},\ }\href {\doibase 10.1103/PhysRevD.91.084011} {\bibfield  {journal}
  {\bibinfo  {journal} {Phys. Rev.}\ }\textbf {\bibinfo {volume} {D91}},\
  \bibinfo {pages} {084011} (\bibinfo {year} {2015})},\ \Eprint
  {http://arxiv.org/abs/1411.2263} {arXiv:1411.2263 [hep-ph]} \BibitemShut
  {NoStop}%
\bibitem [{\citenamefont {Boutan}\ \emph {et~al.}(2018)\citenamefont {Boutan}
  \emph {et~al.}}]{Boutan:2018uoc}%
  \BibitemOpen
  \bibfield  {author} {\bibinfo {author} {\bibfnamefont {C.}~\bibnamefont
  {Boutan}} \emph {et~al.} (\bibinfo {collaboration} {ADMX}),\ }\href {\doibase
  10.1103/PhysRevLett.121.261302} {\bibfield  {journal} {\bibinfo  {journal}
  {Phys. Rev. Lett.}\ }\textbf {\bibinfo {volume} {121}},\ \bibinfo {pages}
  {261302} (\bibinfo {year} {2018})},\ \Eprint
  {http://arxiv.org/abs/1901.00920} {arXiv:1901.00920 [hep-ex]} \BibitemShut
  {NoStop}%
\bibitem [{\citenamefont {Hagmann}\ \emph {et~al.}(1990)\citenamefont
  {Hagmann}, \citenamefont {Sikivie}, \citenamefont {Sullivan},\ and\
  \citenamefont {Tanner}}]{Hagmann:1990tj}%
  \BibitemOpen
  \bibfield  {author} {\bibinfo {author} {\bibfnamefont {C.}~\bibnamefont
  {Hagmann}}, \bibinfo {author} {\bibfnamefont {P.}~\bibnamefont {Sikivie}},
  \bibinfo {author} {\bibfnamefont {N.~S.}\ \bibnamefont {Sullivan}}, \ and\
  \bibinfo {author} {\bibfnamefont {D.~B.}\ \bibnamefont {Tanner}},\ }\href
  {\doibase 10.1103/PhysRevD.42.1297} {\bibfield  {journal} {\bibinfo
  {journal} {Phys. Rev.}\ }\textbf {\bibinfo {volume} {D42}},\ \bibinfo {pages}
  {1297} (\bibinfo {year} {1990})}\BibitemShut {NoStop}%
\bibitem [{\citenamefont {De~Panfilis}\ \emph {et~al.}(1987)\citenamefont
  {De~Panfilis}, \citenamefont {Melissinos}, \citenamefont {Moskowitz},
  \citenamefont {Rogers}, \citenamefont {Semertzidis}, \citenamefont {Wuensch},
  \citenamefont {Halama}, \citenamefont {Prodell}, \citenamefont {Fowler},\
  and\ \citenamefont {Nezrick}}]{DePanfilis:1987dk}%
  \BibitemOpen
  \bibfield  {author} {\bibinfo {author} {\bibfnamefont {S.}~\bibnamefont
  {De~Panfilis}}, \bibinfo {author} {\bibfnamefont {A.~C.}\ \bibnamefont
  {Melissinos}}, \bibinfo {author} {\bibfnamefont {B.~E.}\ \bibnamefont
  {Moskowitz}}, \bibinfo {author} {\bibfnamefont {J.~T.}\ \bibnamefont
  {Rogers}}, \bibinfo {author} {\bibfnamefont {Y.~K.}\ \bibnamefont
  {Semertzidis}}, \bibinfo {author} {\bibfnamefont {W.}~\bibnamefont
  {Wuensch}}, \bibinfo {author} {\bibfnamefont {H.~J.}\ \bibnamefont {Halama}},
  \bibinfo {author} {\bibfnamefont {A.~G.}\ \bibnamefont {Prodell}}, \bibinfo
  {author} {\bibfnamefont {W.~B.}\ \bibnamefont {Fowler}}, \ and\ \bibinfo
  {author} {\bibfnamefont {F.~A.}\ \bibnamefont {Nezrick}},\ }\href {\doibase
  10.1103/PhysRevLett.59.839} {\bibfield  {journal} {\bibinfo  {journal} {Phys.
  Rev. Lett.}\ }\textbf {\bibinfo {volume} {59}},\ \bibinfo {pages} {839}
  (\bibinfo {year} {1987})}\BibitemShut {NoStop}%
\bibitem [{\citenamefont {Wuensch}\ \emph {et~al.}(1989)\citenamefont
  {Wuensch}, \citenamefont {De~Panfilis-Wuensch}, \citenamefont {Semertzidis},
  \citenamefont {Rogers}, \citenamefont {Melissinos}, \citenamefont {Halama},
  \citenamefont {Moskowitz}, \citenamefont {Prodell}, \citenamefont {Fowler},\
  and\ \citenamefont {Nezrick}}]{Wuensch:1989sa}%
  \BibitemOpen
  \bibfield  {author} {\bibinfo {author} {\bibfnamefont {W.}~\bibnamefont
  {Wuensch}}, \bibinfo {author} {\bibfnamefont {S.}~\bibnamefont
  {De~Panfilis-Wuensch}}, \bibinfo {author} {\bibfnamefont {Y.~K.}\
  \bibnamefont {Semertzidis}}, \bibinfo {author} {\bibfnamefont {J.~T.}\
  \bibnamefont {Rogers}}, \bibinfo {author} {\bibfnamefont {A.~C.}\
  \bibnamefont {Melissinos}}, \bibinfo {author} {\bibfnamefont {H.~J.}\
  \bibnamefont {Halama}}, \bibinfo {author} {\bibfnamefont {B.~E.}\
  \bibnamefont {Moskowitz}}, \bibinfo {author} {\bibfnamefont {A.~G.}\
  \bibnamefont {Prodell}}, \bibinfo {author} {\bibfnamefont {W.~B.}\
  \bibnamefont {Fowler}}, \ and\ \bibinfo {author} {\bibfnamefont {F.~A.}\
  \bibnamefont {Nezrick}},\ }\href {\doibase 10.1103/PhysRevD.40.3153}
  {\bibfield  {journal} {\bibinfo  {journal} {Phys. Rev.}\ }\textbf {\bibinfo
  {volume} {D40}},\ \bibinfo {pages} {3153} (\bibinfo {year}
  {1989})}\BibitemShut {NoStop}%
\bibitem [{\citenamefont {Shokair}\ \emph {et~al.}(2014)\citenamefont {Shokair}
  \emph {et~al.}}]{Shokair:2014rna}%
  \BibitemOpen
  \bibfield  {author} {\bibinfo {author} {\bibfnamefont {T.~M.}\ \bibnamefont
  {Shokair}} \emph {et~al.},\ }\href {\doibase 10.1142/S0217751X14430040}
  {\bibfield  {journal} {\bibinfo  {journal} {Int. J. Mod. Phys.}\ }\textbf
  {\bibinfo {volume} {A29}},\ \bibinfo {pages} {1443004} (\bibinfo {year}
  {2014})},\ \Eprint {http://arxiv.org/abs/1405.3685} {arXiv:1405.3685
  [physics.ins-det]} \BibitemShut {NoStop}%
\bibitem [{\citenamefont {Petrakou}(2017)}]{Petrakou:2017epq}%
  \BibitemOpen
  \bibfield  {author} {\bibinfo {author} {\bibfnamefont {E.}~\bibnamefont
  {Petrakou}} (\bibinfo {collaboration} {CAPP/IBS}),\ }\bibfield  {booktitle}
  {\emph {\bibinfo {booktitle} {{Proceedings, 5th International Conference on
  New Frontiers in Physics: Kolymbari, Crete, Greece, July 6-14, 2016}}},\
  }\href {\doibase 10.1051/epjconf/201716401012} {\bibfield  {journal}
  {\bibinfo  {journal} {EPJ Web Conf.}\ }\textbf {\bibinfo {volume} {164}},\
  \bibinfo {pages} {01012} (\bibinfo {year} {2017})},\ \Eprint
  {http://arxiv.org/abs/1702.03664} {arXiv:1702.03664 [physics.ins-det]}
  \BibitemShut {NoStop}%
\bibitem [{\citenamefont {Alesini}\ \emph {et~al.}(2017)\citenamefont
  {Alesini}, \citenamefont {Babusci}, \citenamefont {Di~Gioacchino},
  \citenamefont {Gatti}, \citenamefont {Lamanna},\ and\ \citenamefont
  {Ligi}}]{Alesini:2017ifp}%
  \BibitemOpen
  \bibfield  {author} {\bibinfo {author} {\bibfnamefont {D.}~\bibnamefont
  {Alesini}}, \bibinfo {author} {\bibfnamefont {D.}~\bibnamefont {Babusci}},
  \bibinfo {author} {\bibfnamefont {D.}~\bibnamefont {Di~Gioacchino}}, \bibinfo
  {author} {\bibfnamefont {C.}~\bibnamefont {Gatti}}, \bibinfo {author}
  {\bibfnamefont {G.}~\bibnamefont {Lamanna}}, \ and\ \bibinfo {author}
  {\bibfnamefont {C.}~\bibnamefont {Ligi}},\ }\href@noop {} {\  (\bibinfo
  {year} {2017})},\ \Eprint {http://arxiv.org/abs/1707.06010} {arXiv:1707.06010
  [physics.ins-det]} \BibitemShut {NoStop}%
\bibitem [{\citenamefont {Gatti}\ \emph {et~al.}(2018)\citenamefont {Gatti}
  \emph {et~al.}}]{Gatti:2018ojx}%
  \BibitemOpen
  \bibfield  {author} {\bibinfo {author} {\bibfnamefont {C.}~\bibnamefont
  {Gatti}} \emph {et~al.},\ }in\ \href@noop {} {\emph {\bibinfo {booktitle}
  {{14th Patras Workshop on Axions, WIMPs and WISPs (AXION-WIMP 2018) (PATRAS
  2018) Hamburg, Germany, June 18-22, 2018}}}}\ (\bibinfo {year} {2018})\
  \Eprint {http://arxiv.org/abs/1811.06754} {arXiv:1811.06754
  [physics.ins-det]} \BibitemShut {NoStop}%
\bibitem [{\citenamefont {Irastorza}\ \emph {et~al.}(2013)\citenamefont
  {Irastorza} \emph {et~al.}}]{Irastorza:2013dav}%
  \BibitemOpen
  \bibfield  {author} {\bibinfo {author} {\bibfnamefont {I.}~\bibnamefont
  {Irastorza}} \emph {et~al.} (\bibinfo {collaboration} {IAXO}),\ }\href@noop
  {} {\  (\bibinfo {year} {2013})}\BibitemShut {NoStop}%
\bibitem [{\citenamefont {B{\"{a}}hre}\ \emph {et~al.}(2013)\citenamefont
  {B{\"{a}}hre} \emph {et~al.}}]{Bahre:2013ywa}%
  \BibitemOpen
  \bibfield  {author} {\bibinfo {author} {\bibfnamefont {R.}~\bibnamefont
  {B{\"{a}}hre}} \emph {et~al.},\ }\href {\doibase
  10.1088/1748-0221/8/09/T09001} {\bibfield  {journal} {\bibinfo  {journal}
  {JINST}\ }\textbf {\bibinfo {volume} {8}},\ \bibinfo {pages} {T09001}
  (\bibinfo {year} {2013})},\ \Eprint {http://arxiv.org/abs/1302.5647}
  {arXiv:1302.5647 [physics.ins-det]} \BibitemShut {NoStop}%
\bibitem [{\citenamefont {Hook}\ \emph {et~al.}(2018)\citenamefont {Hook},
  \citenamefont {Kahn}, \citenamefont {Safdi},\ and\ \citenamefont
  {Sun}}]{Hook:2018iia}%
  \BibitemOpen
  \bibfield  {author} {\bibinfo {author} {\bibfnamefont {A.}~\bibnamefont
  {Hook}}, \bibinfo {author} {\bibfnamefont {Y.}~\bibnamefont {Kahn}}, \bibinfo
  {author} {\bibfnamefont {B.~R.}\ \bibnamefont {Safdi}}, \ and\ \bibinfo
  {author} {\bibfnamefont {Z.}~\bibnamefont {Sun}},\ }\href {\doibase
  10.1103/PhysRevLett.121.241102} {\bibfield  {journal} {\bibinfo  {journal}
  {Phys. Rev. Lett.}\ }\textbf {\bibinfo {volume} {121}},\ \bibinfo {pages}
  {241102} (\bibinfo {year} {2018})},\ \Eprint
  {http://arxiv.org/abs/1804.03145} {arXiv:1804.03145 [hep-ph]} \BibitemShut
  {NoStop}%
\bibitem [{\citenamefont {Creque-Sarbinowski}\ and\ \citenamefont
  {Kamionkowski}(2018)}]{Creque-Sarbinowski:2018ebl}%
  \BibitemOpen
  \bibfield  {author} {\bibinfo {author} {\bibfnamefont {C.}~\bibnamefont
  {Creque-Sarbinowski}}\ and\ \bibinfo {author} {\bibfnamefont
  {M.}~\bibnamefont {Kamionkowski}},\ }\href {\doibase
  10.1103/PhysRevD.98.063524} {\bibfield  {journal} {\bibinfo  {journal} {Phys.
  Rev.}\ }\textbf {\bibinfo {volume} {D98}},\ \bibinfo {pages} {063524}
  (\bibinfo {year} {2018})},\ \Eprint {http://arxiv.org/abs/1806.11119}
  {arXiv:1806.11119 [astro-ph.CO]} \BibitemShut {NoStop}%
\bibitem [{\citenamefont {McAllister}\ \emph {et~al.}(2018)\citenamefont
  {McAllister}, \citenamefont {Goryachev}, \citenamefont {Bourhill},
  \citenamefont {Ivanov},\ and\ \citenamefont {Tobar}}]{McAllister:2018ndu}%
  \BibitemOpen
  \bibfield  {author} {\bibinfo {author} {\bibfnamefont {B.~T.}\ \bibnamefont
  {McAllister}}, \bibinfo {author} {\bibfnamefont {M.}~\bibnamefont
  {Goryachev}}, \bibinfo {author} {\bibfnamefont {J.}~\bibnamefont {Bourhill}},
  \bibinfo {author} {\bibfnamefont {E.~N.}\ \bibnamefont {Ivanov}}, \ and\
  \bibinfo {author} {\bibfnamefont {M.~E.}\ \bibnamefont {Tobar}},\ }\href@noop
  {} {\  (\bibinfo {year} {2018})},\ \Eprint {http://arxiv.org/abs/1803.07755}
  {arXiv:1803.07755 [physics.ins-det]} \BibitemShut {NoStop}%
\bibitem [{\citenamefont {Ouellet}\ and\ \citenamefont
  {Bogorad}(2018)}]{Ouellet:2018nfr}%
  \BibitemOpen
  \bibfield  {author} {\bibinfo {author} {\bibfnamefont {J.}~\bibnamefont
  {Ouellet}}\ and\ \bibinfo {author} {\bibfnamefont {Z.}~\bibnamefont
  {Bogorad}},\ }\href@noop {} {\  (\bibinfo {year} {2018})},\ \Eprint
  {http://arxiv.org/abs/1809.10709} {arXiv:1809.10709 [hep-ph]} \BibitemShut
  {NoStop}%
\bibitem [{\citenamefont {Beutter}\ \emph {et~al.}(2018)\citenamefont
  {Beutter}, \citenamefont {Pargner}, \citenamefont {Schwetz},\ and\
  \citenamefont {Todarello}}]{Beutter:2018xfx}%
  \BibitemOpen
  \bibfield  {author} {\bibinfo {author} {\bibfnamefont {M.}~\bibnamefont
  {Beutter}}, \bibinfo {author} {\bibfnamefont {A.}~\bibnamefont {Pargner}},
  \bibinfo {author} {\bibfnamefont {T.}~\bibnamefont {Schwetz}}, \ and\
  \bibinfo {author} {\bibfnamefont {E.}~\bibnamefont {Todarello}},\ }\href@noop
  {} {\  (\bibinfo {year} {2018})},\ \Eprint {http://arxiv.org/abs/1812.05487}
  {arXiv:1812.05487 [hep-ph]} \BibitemShut {NoStop}%
\bibitem [{\citenamefont {Kim}\ \emph {et~al.}(2018)\citenamefont {Kim},
  \citenamefont {Kim}, \citenamefont {Jung}, \citenamefont {Kim}, \citenamefont
  {Shin},\ and\ \citenamefont {Semertzidis}}]{Kim:2018sci}%
  \BibitemOpen
  \bibfield  {author} {\bibinfo {author} {\bibfnamefont {Y.}~\bibnamefont
  {Kim}}, \bibinfo {author} {\bibfnamefont {D.}~\bibnamefont {Kim}}, \bibinfo
  {author} {\bibfnamefont {J.}~\bibnamefont {Jung}}, \bibinfo {author}
  {\bibfnamefont {J.}~\bibnamefont {Kim}}, \bibinfo {author} {\bibfnamefont
  {Y.~C.}\ \bibnamefont {Shin}}, \ and\ \bibinfo {author} {\bibfnamefont
  {Y.~K.}\ \bibnamefont {Semertzidis}},\ }\href@noop {} {\  (\bibinfo {year}
  {2018})},\ \Eprint {http://arxiv.org/abs/1810.02459} {arXiv:1810.02459
  [hep-ph]} \BibitemShut {NoStop}%
\bibitem [{\citenamefont {Goryachev}\ \emph {et~al.}(2018)\citenamefont
  {Goryachev}, \citenamefont {Mcallister},\ and\ \citenamefont
  {Tobar}}]{Goryachev:2018vjt}%
  \BibitemOpen
  \bibfield  {author} {\bibinfo {author} {\bibfnamefont {M.}~\bibnamefont
  {Goryachev}}, \bibinfo {author} {\bibfnamefont {B.}~\bibnamefont
  {Mcallister}}, \ and\ \bibinfo {author} {\bibfnamefont {M.~E.}\ \bibnamefont
  {Tobar}},\ }\href@noop {} {\  (\bibinfo {year} {2018})},\ \Eprint
  {http://arxiv.org/abs/1806.07141} {arXiv:1806.07141 [physics.ins-det]}
  \BibitemShut {NoStop}%
\bibitem [{\citenamefont {Marsh}\ \emph {et~al.}(2018)\citenamefont {Marsh},
  \citenamefont {Fong}, \citenamefont {Lentz}, \citenamefont {Smejkal},\ and\
  \citenamefont {Ali}}]{Marsh:2018dlj}%
  \BibitemOpen
  \bibfield  {author} {\bibinfo {author} {\bibfnamefont {D.~J.~E.}\
  \bibnamefont {Marsh}}, \bibinfo {author} {\bibfnamefont {K.-C.}\ \bibnamefont
  {Fong}}, \bibinfo {author} {\bibfnamefont {E.~W.}\ \bibnamefont {Lentz}},
  \bibinfo {author} {\bibfnamefont {L.}~\bibnamefont {Smejkal}}, \ and\
  \bibinfo {author} {\bibfnamefont {M.~N.}\ \bibnamefont {Ali}},\ }\href@noop
  {} {\  (\bibinfo {year} {2018})},\ \Eprint {http://arxiv.org/abs/1807.08810}
  {arXiv:1807.08810 [hep-ph]} \BibitemShut {NoStop}%
\bibitem [{\citenamefont {Bogorad}\ \emph {et~al.}(2019)\citenamefont
  {Bogorad}, \citenamefont {Hook}, \citenamefont {Kahn},\ and\ \citenamefont
  {Soreq}}]{Bogorad:2019pbu}%
  \BibitemOpen
  \bibfield  {author} {\bibinfo {author} {\bibfnamefont {Z.}~\bibnamefont
  {Bogorad}}, \bibinfo {author} {\bibfnamefont {A.}~\bibnamefont {Hook}},
  \bibinfo {author} {\bibfnamefont {Y.}~\bibnamefont {Kahn}}, \ and\ \bibinfo
  {author} {\bibfnamefont {Y.}~\bibnamefont {Soreq}},\ }\href@noop {} {\
  (\bibinfo {year} {2019})},\ \Eprint {http://arxiv.org/abs/1902.01418}
  {arXiv:1902.01418 [hep-ph]} \BibitemShut {NoStop}%
\bibitem [{\citenamefont {Janish}\ \emph {et~al.}(2019)\citenamefont {Janish},
  \citenamefont {Narayan}, \citenamefont {Rajendran},\ and\ \citenamefont
  {Riggins}}]{Janish:2019dpr}%
  \BibitemOpen
  \bibfield  {author} {\bibinfo {author} {\bibfnamefont {R.}~\bibnamefont
  {Janish}}, \bibinfo {author} {\bibfnamefont {V.}~\bibnamefont {Narayan}},
  \bibinfo {author} {\bibfnamefont {S.}~\bibnamefont {Rajendran}}, \ and\
  \bibinfo {author} {\bibfnamefont {P.}~\bibnamefont {Riggins}},\ }\href@noop
  {} {\  (\bibinfo {year} {2019})},\ \Eprint {http://arxiv.org/abs/1904.07245}
  {arXiv:1904.07245 [hep-ph]} \BibitemShut {NoStop}%
\bibitem [{\citenamefont {Edwards}\ \emph {et~al.}(2019)\citenamefont
  {Edwards}, \citenamefont {Chianese}, \citenamefont {Kavanagh}, \citenamefont
  {Nissanke},\ and\ \citenamefont {Weniger}}]{Edwards:2019tzf}%
  \BibitemOpen
  \bibfield  {author} {\bibinfo {author} {\bibfnamefont {T.~D.~P.}\
  \bibnamefont {Edwards}}, \bibinfo {author} {\bibfnamefont {M.}~\bibnamefont
  {Chianese}}, \bibinfo {author} {\bibfnamefont {B.~J.}\ \bibnamefont
  {Kavanagh}}, \bibinfo {author} {\bibfnamefont {S.~M.}\ \bibnamefont
  {Nissanke}}, \ and\ \bibinfo {author} {\bibfnamefont {C.}~\bibnamefont
  {Weniger}},\ }\href@noop {} {\  (\bibinfo {year} {2019})},\ \Eprint
  {http://arxiv.org/abs/1905.04686} {arXiv:1905.04686 [hep-ph]} \BibitemShut
  {NoStop}%
\bibitem [{\citenamefont {DeRocco}\ and\ \citenamefont
  {Hook}(2018)}]{DeRocco:2018jwe}%
  \BibitemOpen
  \bibfield  {author} {\bibinfo {author} {\bibfnamefont {W.}~\bibnamefont
  {DeRocco}}\ and\ \bibinfo {author} {\bibfnamefont {A.}~\bibnamefont {Hook}},\
  }\href {\doibase 10.1103/PhysRevD.98.035021} {\bibfield  {journal} {\bibinfo
  {journal} {Phys. Rev.}\ }\textbf {\bibinfo {volume} {D98}},\ \bibinfo {pages}
  {035021} (\bibinfo {year} {2018})},\ \Eprint
  {http://arxiv.org/abs/1802.07273} {arXiv:1802.07273 [hep-ph]} \BibitemShut
  {NoStop}%
\bibitem [{\citenamefont {Liu}\ \emph {et~al.}(2018)\citenamefont {Liu},
  \citenamefont {Elwood}, \citenamefont {Evans},\ and\ \citenamefont
  {Thaler}}]{Liu:2018icu}%
  \BibitemOpen
  \bibfield  {author} {\bibinfo {author} {\bibfnamefont {H.}~\bibnamefont
  {Liu}}, \bibinfo {author} {\bibfnamefont {B.~D.}\ \bibnamefont {Elwood}},
  \bibinfo {author} {\bibfnamefont {M.}~\bibnamefont {Evans}}, \ and\ \bibinfo
  {author} {\bibfnamefont {J.}~\bibnamefont {Thaler}},\ }\href@noop {} {\
  (\bibinfo {year} {2018})},\ \Eprint {http://arxiv.org/abs/1809.01656}
  {arXiv:1809.01656 [hep-ph]} \BibitemShut {NoStop}%
\bibitem [{\citenamefont {Obata}\ \emph {et~al.}(2018)\citenamefont {Obata},
  \citenamefont {Fujita},\ and\ \citenamefont {Michimura}}]{Obata:2018vvr}%
  \BibitemOpen
  \bibfield  {author} {\bibinfo {author} {\bibfnamefont {I.}~\bibnamefont
  {Obata}}, \bibinfo {author} {\bibfnamefont {T.}~\bibnamefont {Fujita}}, \
  and\ \bibinfo {author} {\bibfnamefont {Y.}~\bibnamefont {Michimura}},\ }\href
  {\doibase 10.1103/PhysRevLett.121.161301} {\bibfield  {journal} {\bibinfo
  {journal} {Phys. Rev. Lett.}\ }\textbf {\bibinfo {volume} {121}},\ \bibinfo
  {pages} {161301} (\bibinfo {year} {2018})},\ \Eprint
  {http://arxiv.org/abs/1805.11753} {arXiv:1805.11753 [astro-ph.CO]}
  \BibitemShut {NoStop}%
\bibitem [{\citenamefont {Blinov}\ \emph {et~al.}()\citenamefont {Blinov},
  \citenamefont {Dolan}, \citenamefont {Draper},\ and\ \citenamefont
  {Kozaczuk}}]{ALPminiclusters}%
  \BibitemOpen
  \bibfield  {author} {\bibinfo {author} {\bibfnamefont {N.}~\bibnamefont
  {Blinov}}, \bibinfo {author} {\bibfnamefont {M.~J.}\ \bibnamefont {Dolan}},
  \bibinfo {author} {\bibfnamefont {P.}~\bibnamefont {Draper}}, \ and\ \bibinfo
  {author} {\bibfnamefont {J.}~\bibnamefont {Kozaczuk}},\ }\bibfield
  {booktitle} {\emph {\bibinfo {booktitle} {to appear}},\ }\href@noop {} {\
  }\Eprint {http://arxiv.org/abs/19xx.xxxxx} {arXiv:19xx.xxxxx} \BibitemShut
  {NoStop}%
\end{thebibliography}%
\end{document}